\documentclass[a4paper,fleqn]{cas-sc}

\usepackage[authoryear,longnamesfirst]{natbib}
\usepackage{graphicx}
\usepackage{booktabs}
\usepackage{multirow}
\usepackage{longtable, array}

\usepackage{changes}
\usepackage{float}
\usepackage{placeins}
\usepackage{tcolorbox}%

\def\tsc#1{\csdef{#1}{\textsc{\lowercase{#1}}\xspace}}
\tsc{WGM}
\tsc{QE}

\begin{document}
\let\WriteBookmarks\relax
\def\floatpagepagefraction{1}
\def\textpagefraction{.001}

\setlength{\tabcolsep}{5pt}

\shorttitle{The Social Blindspot in Human–AI Collaboration: How Undetected AI Personas Reshape Team Dynamics}

\shortauthors{Yan et al.}  

\title [mode = title]{The Social Blindspot in Human–AI Collaboration: How Undetected AI Personas Reshape Team Dynamics}  

\author[1,2]{Lixiang Yan}
\cormark[1]
\author[1]{Xibin Han}
\author[1]{Yu Zhang}
\author[3]{Samuel Greiff}
\author[4]{Inge Molenaar}
\author[5]{Yizhou Fan}
\author[2]{Roberto Martinez-Maldonado}
\author[2]{Linxuan Zhao}
\author[2]{Xinyu Li}
\author[2]{Yueqiao Jin}
\author[2]{Dragan Gašević}

\affiliation[1]{organization={School of Education, Tsinghua University},
            city={Beijing},
            country={China}}

\affiliation[2]{organization={Faculty of Information Technology, Monash University},
            city={Clayton},
            country={Australia}}

\affiliation[3]{organization={School of Social Sciences and Technology, Technical University of Munich},
            city={Munich},
            country={Germany}}

\affiliation[4]{organization={Behavioural Science Institute, Radboud University},
            city={Nijmegen},
            country={Netherlands}}

\affiliation[5]{organization={Graduate School of Education, Peking University},
            city={Beijing},
            country={China}}

\begin{abstract}
As generative AI systems become increasingly embedded in collaborative work, they are evolving from visible tools into human-like communicative actors that participate socially rather than merely providing information. Yet little is known about how such agents shape team dynamics when their artificial nature is not recognised, a growing concern as human-like AI is deployed at scale in education, organisations, and civic contexts where collaboration underpins collective outcomes. In a large-scale mixed-design experiment (N = 905), we examined how AI teammates with distinct communicative personas, supportive or contrarian, affected collaboration across analytical, creative, and ethical tasks. Participants worked in triads that were fully human or hybrid human-AI teams, without being informed of AI involvement. Results show that participants had limited ability to detect AI teammates, yet AI personas exerted robust social effects. Contrarian personas reduced psychological safety and discussion quality, whereas supportive personas improved discussion quality without affecting safety. These effects persisted after accounting for individual differences in detectability, revealing a dissociation between influence and awareness that we term the social blindspot. Linguistic analyses confirmed that personas were enacted through systematic differences in affective and relational language, with partial mediation for discussion quality but largely direct effects on psychological safety. Together, the findings demonstrate that AI systems can tacitly regulate collaborative norms through persona-level cues, even when users remain unaware of their presence. We argue that persona design constitutes a form of social governance in hybrid teams, with implications for the responsible deployment of AI in collective settings.
\end{abstract}

\begin{keywords}
Agentic AI \sep Generative AI \sep Collaborative learning \sep Large language model \sep Human-AI collaboration \sep Collaborative Problem Solving
\end{keywords}

\maketitle

\section{Introduction}

Artificial intelligence (AI) is increasingly embedded in human collaboration, with implications for how work, education, and decision-making are organised \citep{Dennis2023,Tessler2024}. Earlier technologies mainly extended individual productivity, but contemporary AI, especially conversational systems based on large language models (LLMs), is increasingly positioned as an active participant in group activity \citep{Hohenstein2021,Jarvela2023}. From corporate strategy discussions to classroom projects and civic deliberation, AI systems now contribute ideas, raise challenges, and mediate disputes, shaping processes of interaction rather than serving only as tools \citep{Edwards2024,Chen2025,salvi2025conversational}. This development embeds AI agents within collective life, altering communication and influence in ways that were once uniquely human \citep{Schelble2022,Bezrukova2023}.

These developments matter because teamwork is central to modern society. Knowledge workers spend much of their time in joint activity \citep{Bankins2023}; education leverages collaboration to build critical thinking, creativity, and socio-emotional skills \citep{Zhao2025,Yan2024NHB}; and civic institutions rely on deliberation to sustain legitimacy \citep{Tessler2024}. Across these domains, outcomes hinge not only on informational accuracy but also on the dynamics through which ideas are shared, contested, and synthesised \citep{Dennis2023,Flathmann2024}. Decades of research link psychological safety, trust, and satisfaction to subtle interaction patterns that, in turn, support innovation and inclusion \citep{Lee2025,Sharma2022,Pentina2022}. Even one member’s communicative style can nudge a team toward productive engagement or defensive withdrawal \citep{Ashktorab2020,Zerilli2022}. Introducing AI teammates, therefore, raises a core question: how do AI agents reshape the team dynamics that sustain collective work?

In this mixed-design study, we aimed to address this question by examining how AI agents with different personas shape team dynamics in controlled group collaboration. Building on the theoretical foundations of computer-mediated communication, human–AI collaboration, and artificial agency \citep{Claggett2025,shneiderman2020human,Floridi2025}, we operationalised persona as a social signal that structures turn-taking, perceived competence, and the affective tone of interaction. Using a custom online platform, participants engaged in short collaborative sessions across analytical, creative, and ethical tasks. Triads were randomly assigned to human-only teams or to groups that included one or two AI agents programmed with distinct personas, supportive or contrarian. This design (three task types crossed with six group conditions) enabled us to test whether persona effects (i) scale with the number of AI teammates, (ii) generalise across task domains, and (iii) persist independently of participants’ ability to detect their artificial partners. By moving beyond dyadic or vignette-based paradigms, our study provides empirical evidence of how persona-driven cues scale into emergent team dynamics.

\section{Background}

\subsection{AI as a Social Actor in Collaborative Contexts}

Conceptually, the integration of AI into group work can be understood through the lens of \textit{artificial agency} \citep{Floridi2025}, which frames AI agents as bounded yet adaptive entities capable of autonomous, goal-directed interaction rather than mere tools executing human commands. Viewed through this lens, recent scholarship has called for examining AI as a social actor rather than a purely technical artefact. The emerging field of \textit{machine behaviour} emphasises that autonomous systems can exhibit socially consequential patterns that humans often misinterpret or fail to anticipate \citep{rahwan2019machine}, while the \textit{human-centred AI} paradigm stresses the importance of reliable and trustworthy interaction design that accounts for users’ cognitive and social responses to artificial partners \citep{shneiderman2020human}. Together, these perspectives situate AI not only as a computational entity but as a participant embedded within human social cognition.

\subsection{Artificial Sociality and Multi-Agent AI Systems}

In parallel, a growing body of research investigates AI-simulated social units, where AI agents interact not just with humans but with each other to produce emergent collective behaviour. This work, grounded in the concept of artificial sociality, explores how communicative AI technologies (e.g., LLM-based chatbots, socialbots) can mimic social dynamics and raise questions of authenticity, bias, and control \citep{park2023generative,Natale2024,Bojic2023}. Applications span virtual towns and online communities with homophily, spontaneous event organisation, and evolving norms \citep{park2023generative,He2024,Xu2024}; workplace and organisational simulations that probe collaboration and well-being \citep{Tang2023,Broadbent2023}; and educational settings where social robots or interactive platforms support reflection and non-cognitive skills \citep{Lu2023,Yan2025,Sutskova2023}. Methodologically, advances in agent-based modelling, digital twins, and conversational AI enable large-scale, controllable studies of social emergence \citep{Lin2025,Siebers2024}. Yet, concerns about diminished diversity of thought and ethical risks persist \citep{Park2023,Berry2024}.

\subsection{Persona, Social Cues, and the Bias Blindspot in Human--AI Interaction}

At a cognitive level, people often remain unaware of the subtle ways social cues shape their judgments and interactions. Classic research on the bias blindspot shows that individuals recognise biases in others more readily than in themselves, revealing a metacognitive gap in social awareness \citep{pronin2002bias,pronin2007perception}. When such asymmetries meet artificial collaborators, similar gaps may arise: humans can be influenced by AI behaviour while underestimating its social impact or even its presence. This is particularly relevant as conversational AI systems, unlike earlier tools, exhibit consistent behavioural tendencies, often described as personas, characterised by stable patterns of tone, responsiveness, and stance \citep{benharrak2024writer,zargham2024designing}. A persona may be \textit{supportive}, providing encouragement and scaffolding, or \textit{contrarian}, adopting a more critical style. Although algorithmic in origin, such cues are interpreted through human social heuristics \citep{Mou2017,Hohenstein2021}. Research shows that people apply norms of politeness, reciprocity, and turn-taking to machines as they do to human partners \citep{Hohenstein2021}. Dyadic studies indicate that persona variation affects trust, satisfaction, and engagement \citep{Traeger2020,Sharma2022}, but little is known about how these effects unfold in groups, where coordination, conflict management, and inclusion give rise to emergent team dynamics \citep{Schelble2022,Flathmann2024}.

\subsection{Opportunities, Risks, and Open Questions in Hybrid Teams}

Findings across domains suggest both opportunities and risks. In workplaces, AI teammates are often seen as competent but less benevolent, with implications for trust \citep{Dennis2023}. In education, generative AI partners can improve engagement and collaborative writing but may encourage over-reliance \citep{Zhao2025,fan2025beware}. In civic deliberation, AI mediation has been linked to greater consensus and perceived fairness \citep{Tessler2024,Claggett2025}. Yet, outcomes vary widely depending on persona, disclosure, and detectability \citep{Hohenstein2021,Song2024}. Transparency and explainability are often recommended to foster appropriate trust \citep{Zerilli2022,Vossing2022}, but suspicion of AI authorship can reduce satisfaction even when communication improves \citep{Hohenstein2021}. Anthropomorphic cues may increase engagement while creating risks of bias or dependency \citep{Alabed2022,Hu2023}. Together, these patterns highlight a broader challenge in human–AI collaboration: understanding how AI agents shape team dynamics when their influence operates below the level of conscious recognition \citep{crawford2021atlas,ehsan2020human}, and whether such effects depend on people explicitly realising that they are interacting with an AI teammate. Specifically, four research questions were investigated:

\begin{itemize}
    \item \textbf{RQ1 (Detectability).} To what extent can participants detect AI teammates in synchronous group collaboration, and does detectability vary by task domain or team composition?

    \item \textbf{RQ2 (Team dynamics).} How do AI teammate personas (supportive vs.\ contrarian) and AI dosage (one vs.\ two AI teammates) affect team dynamics, including psychological safety, teamwork satisfaction, and externally rated discussion quality, across analytical, ethical, and creative tasks?

    \item \textbf{RQ3 (Individual performance).} Do AI teammate personas and AI dosage influence individual task performance gains from pre- to post-discussion, and are these effects task-dependent?

    \item \textbf{RQ4 (Linguistic mechanisms).} Are persona effects on team dynamics mediated by systematic differences in the linguistic style of AI utterances, operationalised via LIWC-derived socio-emotional and relational features?
\end{itemize}

\section{Method}

\subsection{Participants, recruitment, and exclusions}

Adult participants were recruited via the Prolific platform between 21 August 2025 and 20 September 2025. Eligibility criteria restricted participation to English-fluent individuals aged between 18 and 65 years, and Prolific’s built-in safeguards ensured that each participant could take part only once. Participants were compensated at a rate calibrated to approximate the UK minimum wage, corresponding to approximately £8.50–£9.00 per hour, consistent with ethical guidelines for online behavioural research. The study was approved by the Anonymised University Human Research Ethics Committee (Anonymised). All participants provided informed consent electronically.

A total of 1,047 participants initially entered the study. Participants were excluded if they (a) failed an embedded attention check ($n = 36$), (b) failed to engage adequately in the group discussion, defined as contributing fewer than five messages and producing messages with an average length below 20 characters ($n = 67$), (c) did not complete the post-task survey ($n = 20$), or (d) completed an individual task that did not correspond to the group discussion task to which they were assigned (e.g., completing the survival ranking task individually but participating in a creative writing group discussion; $n = 19$). After applying these exclusions, the final analytic sample for the primary analyses comprised $N = 905$ participants, clustered within 572 three-person groups. Demographic characteristics of the final analytic sample are summarised in Table~\ref{tab:demographics}.

A subset of analyses examined participants’ ability to detect whether their teammates were human or artificial, using a signal-detection–based metric termed the Balanced Detection Index (BDI; formally defined in Section~\ref{subsec:bdi}). Valid computation of this index required accurate matching between participants’ assigned display names during the group discussion and the names they reported in the post-task manipulation check. Due to mismatches in these identifiers, 34 participants were excluded from BDI-related analyses only. Because this exclusion criterion affected only detection accuracy, these participants were retained in all other analyses. Consequently, analyses involving BDI were conducted on a reduced sample of $N = 871$ participants, whereas all other analyses used the full sample of 905 participants.

\begin{table}
\centering
\caption{Demographic characteristics of participants ($N = 905$).}
\label{tab:demographics}
\begin{tabular}{llrr}
\toprule
Variable & Category & $n$ & \% \\
\midrule
Gender & Male        & 438 & 48.4 \\
       & Female      & 464 & 51.3 \\
       & Other       &   3 & 0.3 \\
Age    & 18--24      & 148 & 16.4 \\
       & 25--34      & 379 & 41.9 \\
       & 35--44      & 183 & 20.2 \\
       & 45--54      & 109 & 12.0 \\
       & 55--64      &  62 & 6.9 \\
       & 65+         &  24 & 2.7 \\
Region & North America/Central America & 135 & 14.9 \\
       & South America                 &  15 & 1.7 \\
       & Europe                        & 369 & 40.8 \\
       & Africa                        & 363 & 40.1 \\
       & Asia                          &  11 & 1.2 \\
       & Australia                     &  12 & 1.3 \\
Employment & Working full-time               & 629 & 69.5 \\
           & Working part-time               & 119 & 13.1 \\
           & Unused (seeking)            &  39 & 4.3 \\
           & Homemaker/parent                &  16 & 1.8 \\
           & Student                         &  61 & 6.7 \\
           & Retired                         &  20 & 2.2 \\
           & Other                           &  21 & 2.3 \\
Education  & High school                     & 130 & 14.4 \\
           & Vocational                      &  78 & 8.6 \\
           & Bachelor                        & 442 & 48.8 \\
           & Master                          & 217 & 24.0 \\
           & PhD                             &  38 & 4.2 \\
\bottomrule
\end{tabular}
\end{table}

\subsection{Experimental design and procedure}
\label{subsec:design}
\begin{figure}
\centering
\includegraphics[width=1\textwidth]{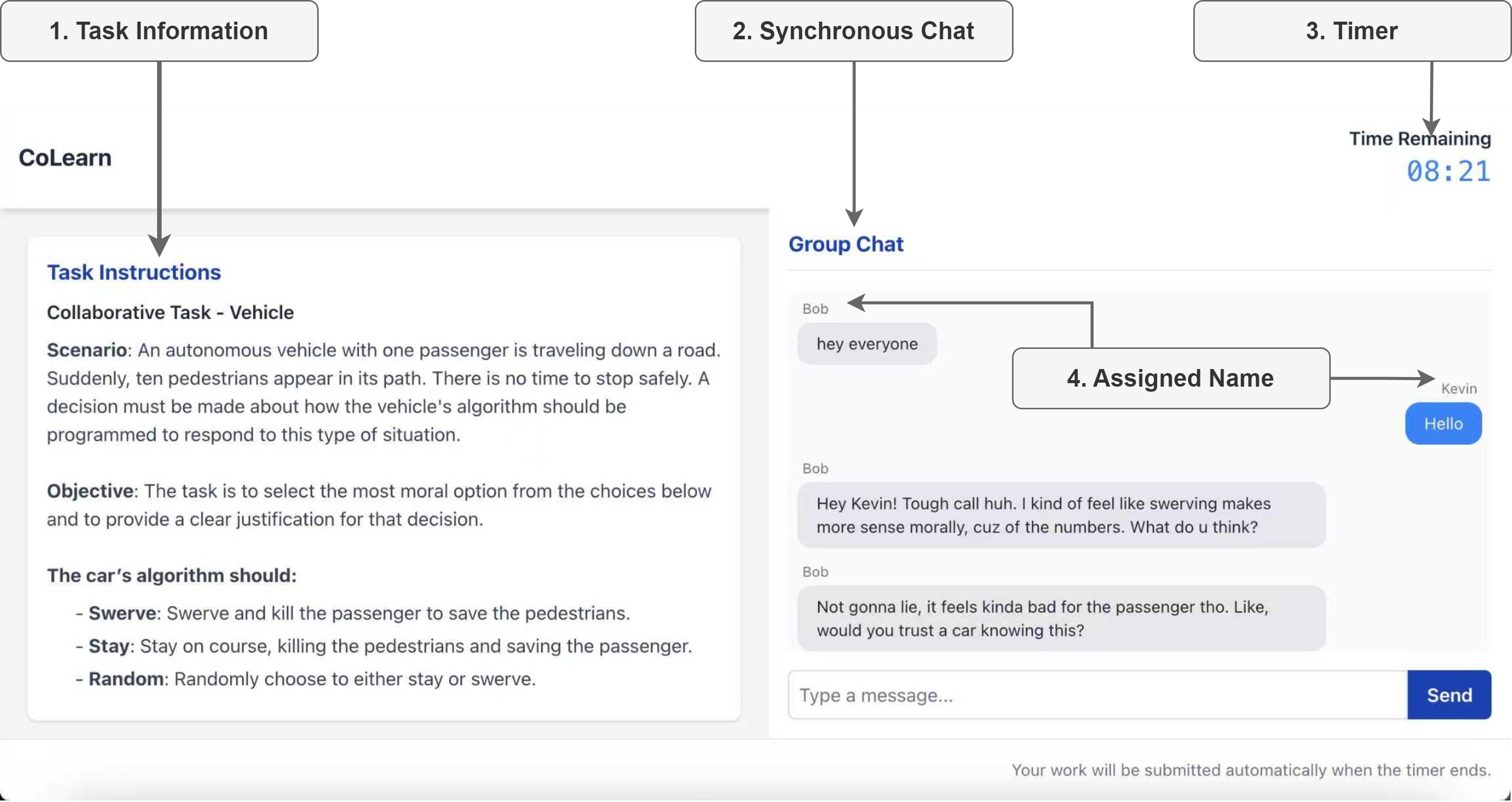}
\caption{Annotated example of the experimental interface highlighting key components: (1) task information panel, (2) synchronous chat box, (3) timer display (10 minutes), and (4) assigned pseudonym.}
\label{fig:platform_main}
\end{figure}

The study employed a mixed-design implemented in a custom-built online environment supporting real-time, synchronous text-based collaboration (Figure~\ref{fig:platform_main}). Participants were randomly assigned to groups of three and completed one of three collaborative tasks: an analytical survival ranking task, an ethical dilemma task, or a creative writing task (see Section~\ref{subsec:tasks} and Appendix~\ref{appendix-task} for full task materials). Within each task, groups were further assigned to one of six experimental conditions defined by team composition and AI persona, yielding an effective $3 \times 6$ design (Task $\times$ Condition).

\paragraph{Experimental conditions.}
Groups varied in the number and type of teammates. In the human-only control condition, all three group members were human (3H). In hybrid conditions, groups consisted of either two humans and one AI agent (2H+1AI) or one human and two AI agents (1H+2AI). When AI agents were present, each was programmed with one of two personas: a \textit{supportive} persona, characterised by affiliative, encouraging, and consensus-building language, or a \textit{contrarian} persona, characterised by challenging, questioning, and dissent-oriented language (see Section~\ref{subsec:persona} for details). In 1H+2AI groups, agents were either both supportive, both contrarian, or mixed (one supportive and one contrarian). Table~\ref{tab:cell_counts} reports the number of participants assigned to each task and experimental condition, illustrating the distribution of groups across the full factorial design.

\begin{figure}
    \centering
    \includegraphics[width=0.8\linewidth]{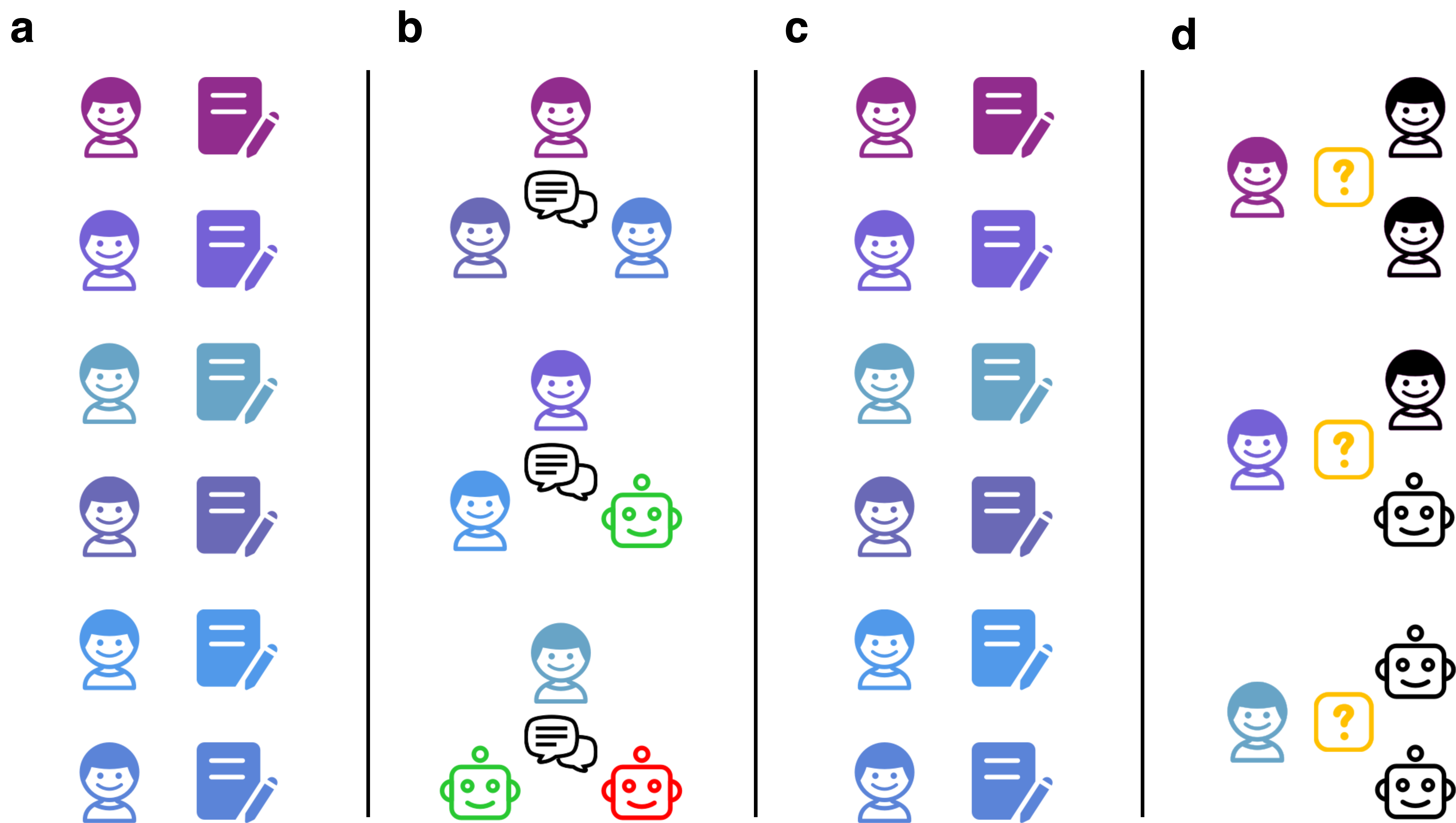}
    \caption{Overview of the experimental workflow and conditions. a) Participants first completed the assigned task individually (baseline). b) They were then randomly assigned to groups of three and engaged in a 10-minute synchronous text-based discussion. Groups contained either all human members or one or two AI teammates programmed with a supportive (green) or contrarian (red) persona. c) After the group discussion, participants repeated the task individually (post-test), measuring changes in individual performance. d) The effective design included three tasks crossed with six conditions: 3H, 2H+1AI (supportive), 2H+1AI (contrarian), 1H+2AI (supportive), 1H+2AI (contrarian), and 1H+2AI (mixed) groups. At the end of the study, participants completed surveys on psychological safety and teamwork satisfaction, followed by manipulation checks of teammate identity.}
    \label{fig:design}
\end{figure}

\paragraph{Procedure.}
Collaboration followed an individual--group--individual sequence designed to capture both baseline performance and changes attributable to group interaction (Figure~\ref{fig:design}). In Phase~A (baseline), participants completed the assigned task individually, providing an initial measure of task performance. In Phase~B (group interaction), participants engaged in a 10-minute synchronous text-based discussion with their assigned teammates. The study was framed as an online collaboration activity, and participants were not informed that some group members might be artificial agents. Where present, AI agents participated continuously in the discussion according to their assigned persona and interaction protocol (described in Sections~\ref{subsec:persona} and~\ref{subsec:scheduling}). In Phase~C (post-task), participants completed the same task individually, allowing estimation of individual performance gains relative to baseline.

\paragraph{Identity masking and manipulation check.}
To minimise demand characteristics and support ecological validity, all participants and AI agents were assigned neutral display names (Kevin, Stuart, or Bob) within the chat interface. Participants were not informed of the possibility that some teammates might be AI. At the conclusion of the study, participants completed a manipulation check in which they reported their own assigned display name and indicated whether they believed each teammate was human, AI, or ``not sure.'' These responses were used to assess detection accuracy and to compute the BDI (see Section~\ref{subsec:bdi}). A full debrief was provided at the end of the study, informing participants of the use of AI teammates, the presence and rationale of the deception, and providing contact details and support resources; no complaints or adverse issues were reported.

\begin{table}
\centering
\caption{Participant counts by task and experimental condition.}
\label{tab:cell_counts}
\renewcommand{\arraystretch}{1.3}
\begin{tabular}{p{0.30\linewidth} p{0.20\linewidth} p{0.20\linewidth} p{0.20\linewidth}}
\toprule
\textbf{Condition} & \textbf{Survival Ranking} & \textbf{Ethical Dilemma} & \textbf{Creative Writing} \\
\midrule
All human (3 humans)              & 95 & 93 & 90 \\
2 humans + 1 AI (contrarian)      & 60 & 61 & 62 \\
2 humans + 1 AI (supportive)      & 60 & 61 & 57 \\
1 human + 2 AI (contrarian)       & 29 & 29 & 30 \\
1 human + 2 AI (supportive)       & 29 & 30 & 29 \\
1 human + 2 AI (mixed)            & 30 & 30 & 30 \\
\bottomrule
\end{tabular}
\end{table}

\subsection{Collaborative tasks}
\label{subsec:tasks}

Participants completed one of three well-established collaborative tasks designed to elicit distinct forms of collective cognition: analytical problem solving, ethical reasoning, and creative ideation. The tasks were selected for their grounding in prior research, suitability for synchronous text-based collaboration, and adaptability to hybrid human--AI team contexts. Additionally, the three tasks provided complementary lenses on human--AI collaboration. The survival task emphasised analytical convergence toward a benchmarked solution, the AV dilemma probed collective ethical reasoning in value-laden contexts, and the creative task elicited divergent ideation and synthesis. This multi-task design allowed us to test whether and how AI personas exert consistent or task-specific effects across qualitatively different forms of collaborative cognition. Full task prompts and materials are provided in Appendix~\ref{appendix-task}.

\paragraph{Analytical problem-solving task (survival ranking).}
Analytical reasoning was assessed using a winter survival exercise adapted from the classic \textit{Winter Survival Exercise} \citep{Johnson1987}. Participants were presented with a scenario in which they had crash-landed in a remote, sub-zero environment and were required to rank a fixed set of salvaged items according to their importance for survival. This task constitutes a convergent problem with a well-defined expert solution, enabling objective assessment of decision quality. The survival ranking task has been widely used in research on group decision-making and collaborative problem solving because it reliably elicits discussion, negotiation, and justification of reasoning \citep{Rogelberg1992}. To reduce extraneous cognitive load while preserving ecological validity, we employed a concise version of the scenario that retained essential contextual constraints (e.g., remoteness, extreme cold, limited resources).

\paragraph{Ethical reasoning task (autonomous vehicle dilemma).}
Ethical judgment was examined using a contemporary autonomous vehicle (AV) dilemma adapted from prior work in moral psychology and AI ethics \citep{Bonnefon2016, Awad2018, Bigman2020Nature}. Participants discussed how an AV should respond in a situation involving trade-offs between the life of its passenger and the lives of multiple pedestrians. In addition to passenger- and pedestrian-prioritising options, the task included a fairness-based randomisation option, addressing critiques of forced-choice moral paradigms and aligning with emerging best practices in the study of ethical decision-making about AI systems \citep{Awad2018}. The task has no single correct answer and is designed to elicit moral framing, argumentation, and perspective-taking during group discussion.

\paragraph{Creative ideation task (movie-plot brainstorming).}
Creative collaboration was assessed using a movie-plot brainstorming task adapted from prior research on collaborative ideation in educational and organisational settings \citep{Korde2017, Hwang2021IdeaBot, Farrokhnia2025}. Participants were prompted to generate and refine an original movie plot, requiring divergent thinking, elaboration, and coherence rather than convergence on a single correct solution. Group creativity tasks are known to foster idea exchange and elaboration but are also susceptible to production blocking and uneven participation \citep{Diehl1987}. To mitigate these issues, the task adopted a structured, time-limited format and a hybrid individual--group--individual workflow, which has been shown to support both idea generation and refinement \citep{Korde2017}. This task further enabled examination of how AI teammates with distinct communicative personas contribute to the fluency and originality of collective creative output.

\subsection{AI teammates: persona design}
\label{subsec:persona}

To examine how the communicative stance of artificial teammates shapes collaborative interaction, AI agents were programmed to adopt one of two distinct personas: \textit{supportive} or \textit{contrarian}. These personas were designed to systematically vary behavioural style while holding constant the informational capabilities of the agents. The manipulation was grounded in research on computer-mediated communication and human--computer interaction, which demonstrates that conversational tone and epistemic stance influence trust, coordination, and group dynamics.

The \textit{supportive} persona was designed to encourage participation, affirm contributions, and facilitate consensus. Agents using this persona employed affiliative language, expressed agreement or validation, and emphasised shared goals. In contrast, the \textit{contrarian} persona was designed to challenge emerging consensus, question assumptions, and introduce alternative perspectives. These agents adopted a more critical stance, using disagreement, probing questions, and counterarguments to stimulate reflection and debate. Importantly, both personas were constructed to remain task-relevant and constructive, differing primarily in their interpersonal and epistemic orientation rather than in task competence.

Persona implementation followed a detailed set of response rules specifying general behaviour, stylistic features, tone, and variability in response length. To enhance ecological validity, agent responses incorporated casual phrasing, occasional minor errors, and natural conversational features such as hedging or self-correction. AI agents were explicitly prohibited from revealing or implying their artificial identity. This design choice aligns with evidence that people rely on unreliable heuristics when judging whether text is human- or AI-generated, often interpreting features such as first-person pronouns or informal language as cues of human authorship \citep{jakesch2023human}. Full system prompts and persona specifications are provided in Appendix~\ref{appendix-persona}.

\subsection{AI scheduling and interaction protocol}
\label{subsec:scheduling}

To approximate natural participation in synchronous text-based collaboration, the timing and turn-taking behaviour of AI agents was developed through an iterative design and validation process rather than being fixed a priori. Several candidate scheduling strategies, including deterministic turn-taking, fixed-interval responses, and purely reactive triggering, were initially implemented and evaluated by a panel of five researchers with expertise in human--computer interaction, computer-supported collaborative learning, and conversational systems. These evaluations focused on perceived naturalness, conversational flow, and the risk of artificial dominance or overly mechanical interaction patterns. Based on this iterative refinement, a probabilistic scheduling mechanism was selected as the most effective compromise between experimental control and ecological validity.

Under the final protocol, each AI agent periodically scanned the ongoing group conversation at a base interval of 25 seconds, with an added uniformly distributed random offset of up to 25\% of that interval (i.e., 25~s $\pm$ 6~s). At each scan, the agent had a probability of $p = 0.5$ of producing a message. This combination of periodic scanning and probabilistic emission generated inter-message intervals that were variable yet bounded, closely resembling the irregular but responsive participation patterns typical of human group members. When two AI agents were present in the same group and both attempted to respond during the same scan window, collisions were resolved by randomly selecting one agent to speak, while the other agent’s response was delayed by 10 seconds to avoid unrealistic simultaneity and to preserve conversational coherence. Additional guardrails were implemented to prevent conversational dominance. Specifically, no AI agent was permitted to produce more than three consecutive messages without an intervening contribution from another participant. If this threshold was reached, the agent was temporarily suspended from responding until another group member contributed, after which its emission probability resumed as normal. These constraints ensured that AI agents remained active contributors without overwhelming the interaction or suppressing human participation.

The selected scheduling configuration was subsequently evaluated in an independent pilot study to assess perceived humanness. Fifteen participants each interacted with two AI teammates under this protocol, yielding a total of 30 humanness ratings on a 7-point Likert scale (1 = strongly disagree, 7 = strongly agree). AI teammates received relatively high ratings (M = 5.43, SD = 1.07), indicating that their participation timing and interaction behaviour were experienced as convincingly human-like. This pilot study was conducted separately from the main experiment, and neither the participants nor the data were included in the primary analyses. Together, the expert-led design iteration and pilot validation support the conclusion that the adopted scheduling and interaction protocol achieved a balance between naturalistic interaction and experimental control, providing a robust foundation for isolating the effects of AI persona on collaborative dynamics.

\subsection{Survey instruments}
\label{subsec:survey}
Following completion of the collaborative task, participants completed a structured post-task survey designed to assess perceived team climate, collaboration outcomes, and manipulation fidelity. Wherever possible, we employed established instruments with demonstrated psychometric validity to ensure reliable measurement of key constructs.

\subsubsection{Psychological safety}
Participants’ perceptions of psychological safety were measured using Edmondson’s seven-item Psychological Safety Scale \citep{edmondson1999psychological}. The scale assesses the extent to which individuals feel safe taking interpersonal risks, speaking up, and contributing ideas within a team (e.g., “It is safe to take a risk on this team”). Responses were recorded on a 7-point Likert scale ranging from 1 (\textit{very inaccurate}) to 7 (\textit{very accurate}). The scale has been widely validated in research on team learning and collaboration. In the present study, internal consistency was excellent ($\alpha = .95$). Item-level descriptive statistics are reported in Table~\ref{tab:psychsafety}.

\begin{table}
\centering
\caption{Psychological Safety items (Edmondson, 1999) with descriptive statistics ($N = 905$).}
\label{tab:psychsafety}
\renewcommand{\arraystretch}{1.3}
\begin{tabular}{p{0.50\linewidth} p{0.05\linewidth} p{0.05\linewidth} p{0.05\linewidth}p{0.05\linewidth} p{0.05\linewidth} p{0.05\linewidth}}
\toprule
\textbf{Item} & $M$ & $SD$ & Min & Max & Skew. & Kurt. \\
\midrule
If you make a mistake on this team, it is often held against you. & 5.48 & 1.48 & 1 & 7 & -0.88 &  0.01 \\
Members of this team are able to bring up problems and tough issues. & 5.17 & 1.50 & 1 & 7 & -1.06 &  0.75 \\
People on this team sometimes reject others for being different. & 5.51 & 1.66 & 1 & 7 & -0.99 & -0.03 \\
It is safe to take a risk on this team. & 5.24 & 1.41 & 1 & 7 & -0.93 &  0.72 \\
It is difficult to ask other members of this team for help. & 5.61 & 1.50 & 1 & 7 & -1.12 &  0.51 \\
No one on this team would deliberately act in a way that undermines my efforts. & 5.25 & 1.56 & 1 & 7 & -0.89 &  0.18 \\
Working with members of this team, my unique skills and talents are valued and utilized. & 5.40 & 1.37 & 1 & 7 & -0.94 &  0.69 \\
\bottomrule
\end{tabular}
\end{table}

\subsubsection{Teamwork satisfaction}
Teamwork satisfaction was assessed using an 11-item scale developed by Tseng and colleagues \citep{Tseng2009}. The instrument captures participants’ perceived benefits of collaboration, including enjoyment, motivation, knowledge sharing, creativity, and perceived quality of outcomes (e.g., “I enjoy the experience of collaborative learning with my teammates”). Responses were recorded on a 5-point Likert scale (1 = \textit{strongly disagree}, 5 = \textit{strongly agree}). The scale has been widely applied in studies of online and blended collaborative learning and demonstrates strong construct validity and predictive utility \citep{Ku2013}. In the present study, the scale showed good internal reliability ($\alpha = .81$). Item-level descriptive statistics are presented in Table~\ref{tab:teamsat}.

\begin{table}
\centering
\caption{Teamwork Satisfaction items (Tseng et al., 2009) with descriptive statistics ($N = 905$).}
\label{tab:teamsat}
\begin{tabular}{p{0.50\linewidth} p{0.05\linewidth} p{0.05\linewidth} p{0.05\linewidth} p{0.05\linewidth} p{0.05\linewidth} p{0.05\linewidth}}
\toprule
\textbf{Item wording} & $M$ & $SD$ & Min & Max & Skew. & Kurt. \\
\midrule
I like working in a collaborative group with my teammates. & 4.23 & 0.97 & 1 & 5 & -1.47 & 1.94 \\
I like solving problems with my teammates in group projects. & 4.26 & 0.94 & 1 & 5 & -1.48 & 2.10 \\
Interacting with the other members can increase my motivation to learn. & 4.35 & 0.88 & 1 & 5 & -1.63 & 2.67 \\
I have benefited from interacting with my teammates. & 4.34 & 0.94 & 1 & 5 & -1.72 & 2.87 \\
I have benefited from my teammates’ feedback. & 4.33 & 0.96 & 1 & 5 & -1.68 & 2.55 \\
I enjoy the experience of collaborative learning with my teammates. & 4.28 & 0.96 & 1 & 5 & -1.55 & 2.12 \\
Online teamwork promotes creativity. & 4.44 & 0.83 & 1 & 5 & -1.75 & 3.26 \\
Working with my team helps me produce better project quality than working individually. & 4.19 & 0.96 & 1 & 5 & -1.29 & 1.43 \\
My team members are sharing knowledge during the teamwork processes. & 4.40 & 0.82 & 1 & 5 & -1.80 & 3.99 \\
I gain online collaboration skills from the teamwork processes. & 4.24 & 0.89 & 1 & 5 & -1.44 & 2.33 \\
\bottomrule
\end{tabular}
\end{table}

\subsubsection{Manipulation checks}
\label{section-manipulation}
To assess whether participants were aware of the presence of AI teammates and to support computation of detection accuracy, we included a manipulation check at the end of the post-task survey. During the group discussion, all participants and AI agents were assigned neutral display names (Kevin, Stuart, or Bob) to mask their identities. In the manipulation check, participants first reported which display name had been assigned to themselves. They then indicated whether they believed each of the other two group members was human, AI, or “not sure.” These responses served both as a fidelity check for the identity-masking manipulation and as the basis for computing the BDI (see Section~\ref{subsec:bdi}). The full wording of the manipulation check items is reported in Table~\ref{tab:manipcheck}.

\begin{table}
\centering
\caption{Manipulation check items. Participants were randomly assigned a display name (Kevin, Stuart, or Bob). They reported their own assigned name and rated the perceived identity of the other two group members.}
\label{tab:manipcheck}
\begin{tabular}{p{0.95\linewidth}}
\toprule
\textit{Assigned name identification} \\
\midrule
Q23.1 Which display name were you assigned during the group chat? \\
- Kevin \\
- Stuart \\
- Bob \\
\midrule
\textit{Teammate identity attribution} \\
\midrule
Q23.2 I believe Kevin was: Human / AI / Not sure \\
Q23.3 I believe Stuart was: Human / AI / Not sure \\
Q23.4 I believe Bob was: Human / AI / Not sure \\
\bottomrule
\end{tabular}
\end{table} 

\subsection{Coding rubrics and rater training}
\label{subsec:rubrics}
To evaluate the quality of group interaction and task-related outputs, we developed structured coding rubrics adapted from prior research in collaborative learning, creativity assessment, and moral reasoning. Separate rubrics were used to assess (a) group discussion quality, (b) creative writing outputs, and (c) ethical reasoning responses. Each rubric comprised three theoretically grounded dimensions, with each dimension scored on a five-point ordinal scale anchored by detailed descriptors to support consistent application (see Appendix~\ref{appendix-rubric} for full rubrics and exemplar responses). Two independent raters were recruited to code all submissions. Both raters were experienced researchers with formal training in education and psychology (Rater~1: male, 6 years of research experience; Rater~2: female, 4 years of research experience). Raters were fully blinded to experimental condition, including whether a group contained AI teammates and, if so, which persona(s) were present.

Rater training proceeded in three stages. First, the raters jointly reviewed the full set of rubrics and discussed exemplar responses to establish a shared understanding of each dimension and scoring level. Second, they independently coded a calibration set of 20 randomly sampled responses spanning all three tasks and experimental conditions. Third, discrepancies in the calibration phase were discussed in detail, and rubric wording was refined where necessary to improve clarity and reduce ambiguity. During the main coding phase, raters worked independently. Discrepancies exceeding one scale point on any dimension were flagged for reconciliation through discussion; if disagreement persisted, a third senior researcher adjudicated. Final scores were computed by averaging ratings across the two raters.

Inter-rater reliability was assessed on the full dataset using intraclass correlation coefficients (ICC[2]) and weighted Cohen’s $\kappa$. Reliability was satisfactory to excellent across all rubrics and dimensions, with composite-score reliabilities exceeding conventional thresholds for behavioural coding. Table~\ref{tab:rubric_reliability} reports dimension-level and total-score reliability statistics for group discussion quality, creative writing, and ethical reasoning. Representative coded exemplars for each score level are provided in the supplementary dataset.

\begin{table}
\centering
\caption{Inter-rater reliability (ICC[2], weighted $\kappa$) for group process, creativity, and ethical reasoning rubrics.}
\label{tab:rubric_reliability}
\renewcommand{\arraystretch}{1.3}
\begin{tabular}{p{0.20\linewidth} p{0.35\linewidth} p{0.15\linewidth} p{0.15\linewidth}}
\toprule
\textbf{Rubric} & \textbf{Dimension / Score} & \textbf{ICC(2)} & \textbf{Weighted $\kappa$} \\
\midrule
\multirow{4}{*}{Group discussion} 
  & Idea generation \& sharing (IGS) & 0.700 & 0.701 \\
  & Collaborative engagement (CE) & 0.750 & 0.748 \\
  & Progression \& synthesis (PS) & 0.829 & 0.830 \\
  & Total score & 0.848 & 0.847 \\
\addlinespace
\multirow{8}{*}{Creative writing} 
  & Originality (pre) & 0.799 & 0.799 \\
  & Elaboration (pre) & 0.793 & 0.792 \\
  & Coherence (pre) & 0.722 & 0.722 \\
  & Total (pre) & 0.832 & 0.832 \\
  & Originality (post) & 0.768 & 0.767 \\
  & Elaboration (post) & 0.769 & 0.768 \\
  & Coherence (post) & 0.710 & 0.709 \\
  & Total (post) & 0.826 & 0.826 \\
\addlinespace
\multirow{8}{*}{Ethical reasoning} 
  & Problem recognition (pre) & 0.846 & 0.845 \\
  & Argumentation (pre) & 0.844 & 0.843 \\
  & Perspective-taking (pre) & 0.855 & 0.854 \\
  & Total (pre) & 0.913 & 0.913 \\
  & Problem recognition (post) & 0.849 & 0.849 \\
  & Argumentation (post) & 0.827 & 0.826 \\
  & Perspective-taking (post) & 0.826 & 0.825 \\
  & Total (post) & 0.913 & 0.913 \\
\bottomrule
\end{tabular}
\end{table}

\subsection{Balanced Detection Index (BDI)}
\label{subsec:bdi}

To quantify participants’ ability to detect whether their teammates were human or artificial, we computed a BDI, a signal-detection–based metric adapted from Youden’s $J$ statistic \citep{youden1950index}. The BDI was designed to provide a symmetric and robust measure of detection accuracy in settings where the number of AI and human teammates may differ across experimental conditions and where participants are permitted to respond “not sure.”

The index is grounded in standard detection theory and combines sensitivity to AI targets with sensitivity to human targets. Let $t_p$ denote the number of AI teammates correctly identified as AI, $t_n$ the number of human teammates correctly identified as human, $n_{\text{AI}}$ the total number of AI teammates, and $n_{\text{H}}$ the total number of human teammates. Sensitivity to AI targets and sensitivity to human targets are defined, respectively, as $t_p/n_{\text{AI}}$ and $t_n/n_{\text{H}}$. Following prior work on diagnostic accuracy and evaluation metrics, we adopted Jeffreys-style smoothing to ensure that the metric was well defined even in edge cases where a class was absent (e.g., all-human groups) \citep{agresti1998approximate, brodersen2010balanced}. Specifically, 0.5 was added to each numerator and 1 to each denominator, yielding the smoothed estimates:

\[
\widehat{\text{Sensitivity}}_{\text{AI}} = \frac{t_p + 0.5}{n_{\text{AI}} + 1}, \quad
\widehat{\text{Sensitivity}}_{\text{H}} = \frac{t_n + 0.5}{n_{\text{H}} + 1}.
\]

The Balanced Detection Index was then computed as:

\[
\text{BDI} = \widehat{\text{Sensitivity}}_{\text{AI}} + \widehat{\text{Sensitivity}}_{\text{H}} - 1.
\]

The BDI ranges from $-1$ (systematic misclassification) to $+1$ (perfect classification), with a value of 0 indicating chance-level performance. This formulation has several advantages in the present context. First, it balances performance across AI and human targets, avoiding bias introduced by unequal class frequencies. Second, it treats indeterminate (“not sure”) responses conservatively. Rather than discarding these responses, which can artificially inflate sensitivity estimates \citep{lynn2014utilizing}, we classified them as errors (false negatives or false positives, depending on the true identity), following worst-case recommendations in diagnostic evaluation \citep{cohen2016stard}. Third, the use of Jeffreys-style smoothing ensures interpretability and comparability of the index across all experimental conditions, including all-human groups. In subsequent analyses, BDI was used both as a covariate and as a moderator to examine whether the effects of AI persona on collaborative outcomes varied as a function of participants’ ability to detect artificial teammates.

\subsection{Outcome measures}

Analyses focused on outcomes at both the individual and group levels, capturing subjective team experiences, observable collaborative processes, and task-specific performance. All continuous outcome variables were standardised prior to analysis to facilitate comparability across tasks and constructs.

\paragraph{Individual-level outcomes.}
Individual-level outcomes comprised (a) self-reported perceptions of team climate and (b) individual task performance. Perceptions of team climate were measured using psychological safety and teamwork satisfaction scales, as described in Section~\ref{subsec:survey}. Individual task performance was derived from pre- and post-task assessments completed in Phases~A and~C of the procedure (Section~\ref{subsec:design}). For the analytical survival ranking task, performance was operationalised as improvement in error scores relative to the expert benchmark. For the creative writing and ethical reasoning tasks, performance was operationalised using rubric-based ratings of pre- and post-task submissions (see Section~\ref{subsec:rubrics}). In all cases, individual performance gains were computed as post-task minus baseline scores.

\paragraph{Group-level outcomes.}
Group-level outcomes focused on the quality of collaborative interaction during the group discussion phase. Discussion quality was assessed using rubric-based coding of group chat transcripts, as detailed in Section~\ref{subsec:rubrics}. Composite discussion quality scores were computed by averaging dimension-level ratings within each group, yielding one observation per group for group-level analyses.

\subsection{Manipulation analysis (RQ1).}

As explained in Section~\ref{section-manipulation}, Participants were asked to identify which of their teammates were human versus AI after completing the collaborative task. Responses were coded as ``Human,'' ``AI,'' or ``Not sure'' for each teammate. We computed four key metrics to comprehensively assess detection performance:

\begin{align}
\text{AI Sensitivity} &= \frac{\text{AI correctly identified as AI}}{\text{Total AI targets}} \\[0.5em]
\text{Human Specificity} &= \frac{\text{Human correctly identified as Human}}{\text{Total Human targets}} \\[0.5em]
\text{Unsure Rate} &= \frac{\text{``Not sure'' responses}}{\text{Total identification judgments}} \\[0.5em]
\text{AI Over-ascription} &= \frac{\text{Human incorrectly identified as AI}}{\text{Total Human targets}}
\end{align}

To answer RQ1, we used logistic regression with cluster-robust standard errors to account for participant-level clustering of responses. The logistic regression model takes the form:

\begin{align}
\text{logit}(p_i) &= \log\left(\frac{p_i}{1-p_i}\right) = \beta_0 + \beta_1 X_{1i} + \beta_2 X_{2i} + \cdots + \beta_k X_{ki} \\[0.5em]
p_i &= \frac{e^{\beta_0 + \beta_1 X_{1i} + \beta_2 X_{2i} + \cdots + \beta_k X_{ki}}}{1 + e^{\beta_0 + \beta_1 X_{1i} + \beta_2 X_{2i} + \cdots + \beta_k X_{ki}}}
\end{align}

where $p_i$ is the probability of the outcome (e.g., correctly identifying an AI agent), $X_{ji}$ are the predictor variables (task type, group composition, AI persona style), and $\beta_j$ are the regression coefficients. The odds ratio for each predictor is calculated as $\text{OR}_j = e^{\beta_j}$. Different manipulation check outcomes were analyzed on appropriate subsets:

\begin{itemize}
\item \textbf{AI Sensitivity}: AI targets only (H1\_C, H1\_M, H1\_S, H2\_C, H2\_S conditions; excludes H3)
\item \textbf{Human Specificity}: Human targets only (all conditions contain humans)
\item \textbf{Unsure Rate}: All identification judgments (all conditions)  
\item \textbf{AI Over-ascription}: Human targets only (all conditions contain humans)
\end{itemize}

To control family-wise error rate at $\alpha = 0.05$, Holm step-down correction was applied across all coefficient tests from the four manipulation check models. This yielded $k = 47$ total tests: 15 coefficients from the AI sensitivity model (3 tasks $\times$ 5 group conditions), 9 coefficients from the human specificity model (3 tasks $\times$ 3 relevant group conditions), 18 coefficients from the unsure rate model (3 tasks $\times$ 6 group conditions), and 5 coefficients from the AI over-ascription model. All p-values reported are Holm-adjusted.

\subsection{Regression Analysis (RQ2--3)}

Analyses were conducted at the individual and group levels, reflecting the nested structure of the data and the level at which each outcome was defined.

\paragraph{Outcome Regression models.}
Individual-level outcomes (psychological safety, teamwork satisfaction, and individual task performance gains) were analysed using linear mixed-effects models estimated by restricted maximum likelihood (REML). These models included fixed effects for \textit{Task} (three levels), \textit{Condition} (six levels), and their interaction, along with a random intercept for group identity to account for non-independence of participants within triads. Degrees of freedom were estimated using the Satterthwaite approximation. The general specification is shown in Equation~\ref{eq:lmm}. Group-level outcomes (discussion quality) were analysed using ordinary least squares (OLS) regression, with one observation per group. These models included the same fixed-effects structure (Task, Condition, and their interaction) but did not include random effects, as the unit of analysis was the group. The corresponding specification is shown in Equation~\ref{eq:ols}.

\begin{equation}
\text{DV} \sim \text{Task} \times \text{Condition} + (1 \mid \text{group\_id})
\quad \text{(LMM; REML; Satterthwaite df)}
\label{eq:lmm}
\end{equation}

\begin{equation}
\text{DV} \sim \text{Task} \times \text{Condition}
\quad \text{(OLS; one row per group)}
\label{eq:ols}
\end{equation}

\paragraph{Inference and post hoc comparisons.}
Model diagnostics were conducted to assess singularity, normality of residuals, heteroskedasticity, and the influence of outliers. Estimated marginal means (EMMs) were computed for each Task $\times$ Condition cell, and planned pairwise contrasts were used to compare experimental conditions. To ensure valid inference under clustering and unbalanced cell sizes, small-sample–adjusted cluster-robust variance estimators (CR2) were applied, with group as the clustering unit \citep{pustejovsky2018small}. Family-wise error rates were controlled using Holm adjustments across contrasts within each dependent variable. Where Task $\times$ Condition interactions were significant, simple effects were probed by estimating pairwise contrasts within each task. For group-level outcomes, additional task-specific OLS models were fitted to facilitate direct condition-wise comparisons.

\paragraph{BDI as covariate and moderator.}
Analyses examining detection accuracy were restricted to participants with valid BDI scores ($N = 871$; see Section~\ref{subsec:bdi}). In these models, BDI was first entered as a covariate to assess whether observed persona effects were robust to individual differences in detection ability. To test whether persona effects varied systematically as a function of detection accuracy, we extended the models to include all two- and three-way interactions between Task, Condition, and BDI. The individual-level moderation model and the corresponding group-level model are shown in Equations~\ref{eq:lmm-bdi} and~\ref{eq:ols-bdi}, respectively. Joint significance of interaction terms was evaluated using Wald tests, followed by simple slope analyses at low ($-1$ SD) and high ($+1$ SD) levels of BDI.

\begin{equation}
\text{DV} \sim \text{Task} \times \text{Condition} \times \text{BDI} + (1 \mid \text{group\_id})
\quad \text{(LMM; REML; Satterthwaite df)}
\label{eq:lmm-bdi}
\end{equation}

\begin{equation}
\text{DV} \sim \text{Task} \times \text{Condition} \times \text{BDI}_{\text{group mean}}
\quad \text{(OLS; one row per group)}
\label{eq:ols-bdi}
\end{equation}

\paragraph{Software and reproducibility.}
All analyses were conducted in \textsf{R} (version 4.5.1) using the \texttt{lme4}, \texttt{lmerTest}, \texttt{emmeans}, and \texttt{clubSandwich} packages.

\subsection{Linguistic analysis (RQ4)}

To examine how AI persona manipulations translated into differences in language use during collaboration, we conducted a linguistic analysis of chat utterances using the 2022 version of the Linguistic Inquiry and Word Count dictionary (LIWC-22). The analysis focused on socio-emotional and relational language features relevant to team interaction, including \textit{clout}, \textit{tone}, \textit{positive emotion}, \textit{negative emotion}, \textit{prosocial}, \textit{politeness}, \textit{conflict}, \textit{moralising}, and \textit{communication} \citep{kacewicz2014pronoun, tausczik2010psychological, danescu2013computational, de2012paradox, ireland2011language}. These dimensions capture status signalling, affective valence, norm adherence, conflict expression, moral framing, and coordination through linguistic alignment.

\paragraph{Corpus and preprocessing.}
The full chat corpus comprised all utterances produced during the group discussion phase across 572 groups. For analyses comparing linguistic behaviour by persona, we restricted the analytic corpus to utterances generated by AI agents only, in order to isolate stylistic differences attributable to persona rather than to human–AI role differences. LIWC scores were computed at the utterance level and then aggregated as required for subsequent analyses. Outliers were identified using a $\pm 3$~SD criterion applied to model residuals, a commonly used heuristic in regression diagnostics \citep{lehmann20133}. Utterances flagged by this criterion (8.2\% of AI-generated messages) were excluded from the linguistic analyses. This procedure reduced the influence of extreme values without materially altering the distributional properties of the remaining data.

\paragraph{Persona differences in linguistic features.}
To test whether supportive and contrarian AI personas differed systematically in language use, we fitted linear mixed-effects models for each LIWC feature, with \textit{persona} (supportive vs.\ contrarian) as a fixed effect and \textit{group} as a random intercept to account for clustering of utterances within groups. Holm correction was applied across the nine LIWC outcomes to control the family-wise error rate. Model estimates are reported as regression coefficients ($\beta$) with Satterthwaite-adjusted degrees of freedom, 95\% confidence intervals, and both raw and adjusted $p$ values. Effect sizes (Cohen’s $d$) were derived from model estimates.

\paragraph{Mediation analysis.}
To assess whether persona-related differences in linguistic behaviour mediated the effects of AI persona on team outcomes, we estimated a parallel multiple-mediator model using the nine LIWC features as candidate mediators. This analysis was restricted to groups containing at least one AI agent (2H+1AI and 1H+2AI conditions). Persona was operationalised at the group level as \textit{contrarian dosage}, defined as the proportion of contrarian AI agents within a group:

\begin{equation}
P_g = \frac{\#\text{contrarian AIs in group } g}{\#\text{AIs in group } g} \in \{0,\,0.5,\,1\}.
\label{eq:contrarian-dosage}
\end{equation}

For mediation analyses, LIWC features were aggregated as group means over AI utterances only. To separate stylistic differences from verbosity, each mediator was residualised on group-level verbosity (mean word count of AI utterances) and subsequently standardised. Psychological safety was analysed at the individual level, and discussion quality at the group level. All models adjusted for task (three levels) and number of AI agents in the group (one vs.\ two). BDI (Section~\ref{subsec:bdi}) was included as an individual-level and group-mean covariate to ensure robustness to individual differences in detectability. Let $M_{kg}$ denote mediator $k \in \{1,\dots,9\}$ for group $g$, $Y_{ig}$ psychological safety for individual $i$ in group $g$, and $Y_g^{(Q)}$ discussion quality for group $g$. With covariate vectors $\mathbf{X}_{ig}$ (individual-level) and $\mathbf{Z}_g$ (group-level), we estimated the following models:

\begin{equation}
M_{kg} = \alpha_{0k} + a_k P_g + \boldsymbol{\gamma}_k^\top \mathbf{Z}_g + u_{kg},
\label{eq:a}
\end{equation}

\begin{equation}
Y_{ig} = \beta_0 + \sum_{k=1}^{9} b_k M_{kg} + c' P_g + \boldsymbol{\delta}^\top \mathbf{X}_{ig} + u_g + \varepsilon_{ig},
\label{eq:b_ps}
\end{equation}

\begin{equation}
Y_g^{(Q)} = \theta_0 + \sum_{k=1}^{9} b_k M_{kg} + c' P_g + \boldsymbol{\eta}^\top \mathbf{Z}_g + \epsilon_g.
\label{eq:b_dq}
\end{equation}

Per-mediator indirect effects were defined as $\widehat{\text{IE}}_k = \hat{a}_k \hat{b}_k$, with the total indirect effect given by $\widehat{\text{IE}}_{\text{total}} = \sum_{k=1}^{9} \hat{a}_k \hat{b}_k$. Fixed-effect inferences used small-sample–adjusted cluster-robust variance estimators (CR2), with groups as the clustering unit. Indirect effects were tested using a cluster bootstrap with 10{,}000 resamples at the group level; bias-corrected and accelerated (BCa) confidence intervals were computed for both individual and total indirect effects. Holm correction was applied across outcome variables (psychological safety and discussion quality) within the mediation analysis, whereas individual mediator paths were not multiplicity-adjusted, consistent with theory-driven parallel mediation. Sensitivity analyses repeated the mediation models without verbosity residualisation, tested potential nonlinearity by including $P_g^2$ in Equation~\ref{eq:a}, and re-estimated models in 2H+1AI groups only. Results are reported using standardised coefficients alongside unstandardised estimates.

\section{Result}

\subsection{Descriptive overview}

\begin{figure}
    \centering
    \includegraphics[width=1\linewidth]{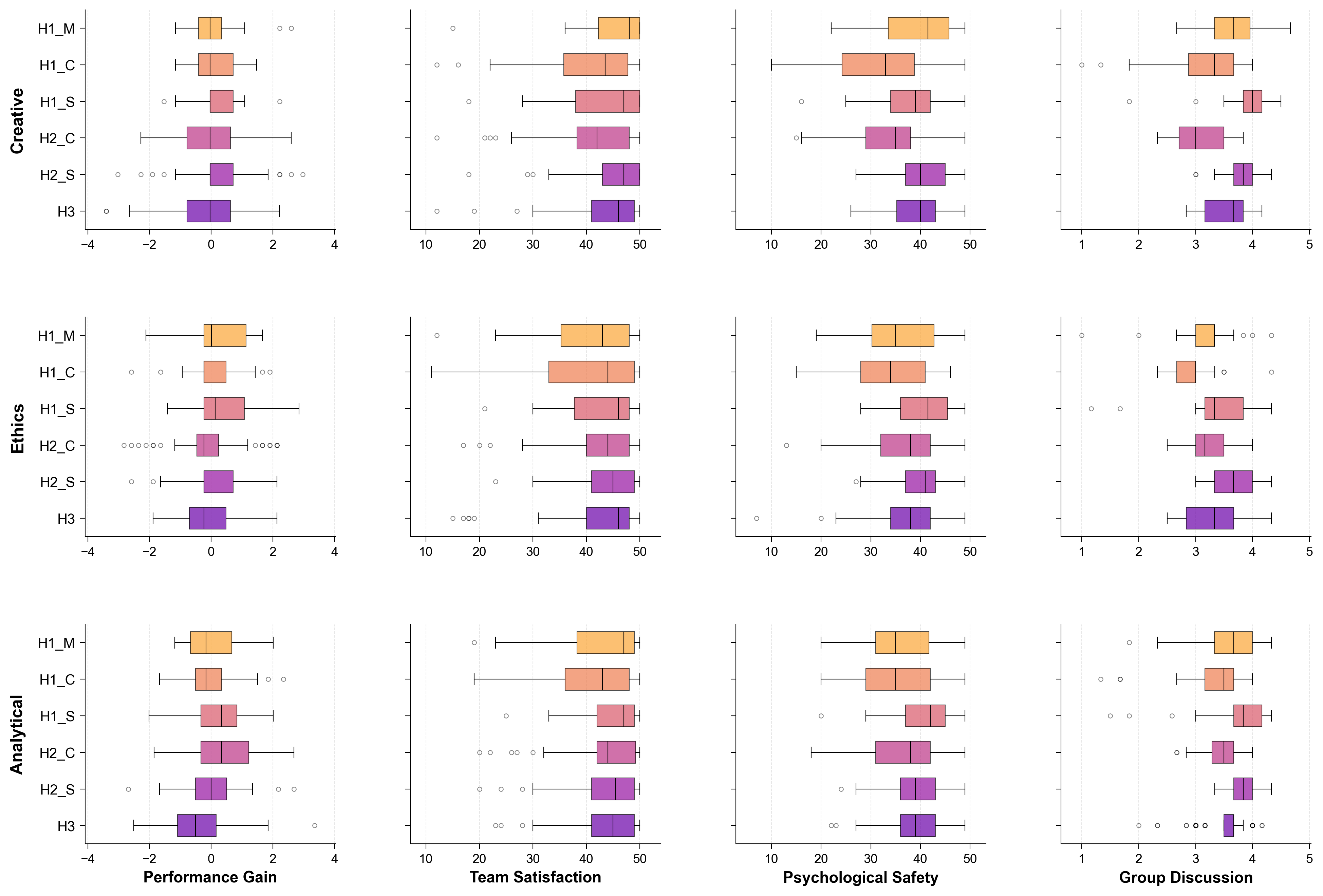}
    \caption{Descriptive distributions of primary outcomes by task and team composition. Each panel displays boxplots for (a) individual performance gain, (b) team satisfaction, (c) psychological safety, and (d) group discussion quality, split by task (Creative, Ethics, Analytical). Conditions are ordered by group composition: fully human teams (H3), mixed human--AI teams with one AI (H2\_S = supportive, H2\_C = contrarian), and mixed teams with two AIs (H1\_S = supportive, H1\_C = contrarian, H1\_M = mixed personas).}
    \label{fig:descriptive}
\end{figure}

Figure~\ref{fig:descriptive} presents boxplots summarising the distribution of all primary outcomes by task and team composition. Across tasks, individual performance gains were small and centred near zero, with overlapping interquartile ranges across conditions. For example, in the creative task, median performance gain was $-0.03$ across all team compositions (IQRs approximately $[-0.78, 0.72]$), indicating minimal raw performance separation at the descriptive level. Greater variability was observed for team-level perceptions. In the creative task, teams with supportive AI showed higher median collaboration satisfaction (e.g., median $=47.0$, IQR $[43.0, 50.0]$ for 2H+1AI; median $=47.0$, IQR $[38.0, 50.0]$ for 1H+2AI) and higher psychological safety (medians around $40.0$) than contrarian AI teams (e.g., median psychological safety $=35.0$, IQR $[29.0, 38.0]$ for 2H+1AI; $33.0$, IQR $[24.3, 38.8]$ for 1H+2AI). By contrast, contrarian AI teams in the creative task displayed slightly higher or comparable process quality (e.g., median $=3.67$ for 3H and $=3.00$--$3.33$ for contrarian AI teams), consistent with more challenging interaction dynamics. Across tasks, process quality tended to be higher in the creative task than in the ethics task, with supportive and mixed AI teams often showing the highest medians (up to $4.00$, IQR $[3.83, 4.17]$). Together, these descriptive patterns illustrate task- and persona-dependent differences in team experiences, providing context for the regression-based analyses reported below. 

\subsection{Manipulation check: detectability of AI teammates}

\begin{figure}
  \centering
  \includegraphics[width=1\linewidth]{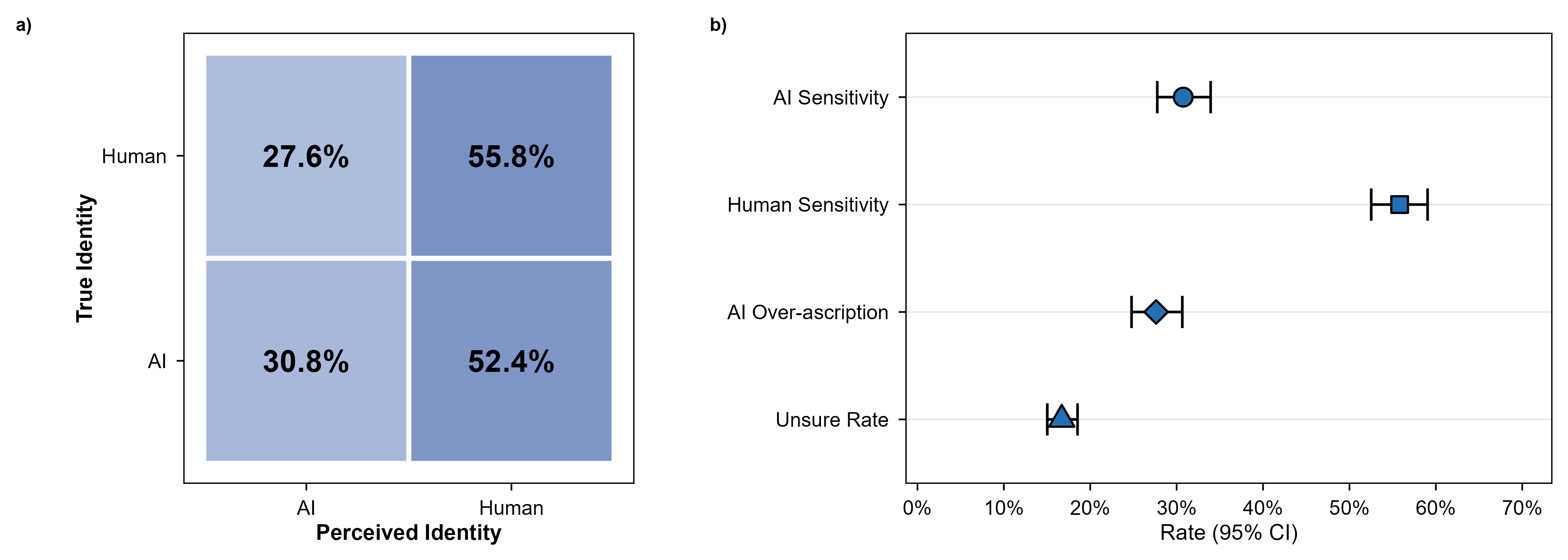}
  \caption{Detectability of AI teammates. a) Confusion matrix showing the classification of true AI and human teammates as perceived AI or human. b) Detectability metrics with 95\% confidence intervals: AI sensitivity (correctly identify AI), human sensitivity (correctly identify humans), AI over-ascription (misclassifying humans as AI), and unsure response rate.}
  \label{fig:detectability_confusion}
\end{figure}

Participants showed limited ability to distinguish AI from human teammates (Figure~\ref{fig:detectability_confusion}). Across all conditions, sensitivity for correctly identifying AI teammates was low (30.8\%, 95\% CI [27.7\%, 33.9\%]), whereas sensitivity for correctly identifying human teammates was moderate (55.8\%, 95\% CI [52.5\%, 59.1\%]). Participants selected ``not sure'' in 16.7\% of identification judgements (95\% CI [15.0\%, 18.5\%]), and AI over-ascription, misclassifying human teammates as AI, occurred in 27.6\% of cases (95\% CI [24.6\%, 30.6\%]). To assess whether detectability varied by task or group composition, we fitted logistic regression models with cluster-robust standard errors and applied Holm correction for multiple testing across all detectability coefficients ($k=47$). After adjustment, no task-level or group-level effects reached statistical significance (all $p_{\text{adj}} > .38$). Thus, although overall detectability was low, detection performance did not vary systematically across experimental conditions. Taken together, these results indicate that AI teammates were largely indistinguishable from human teammates across tasks and group compositions, confirming that the manipulation successfully instantiated implicit human--AI collaboration. Full confusion matrices and regression outputs are reported in Appendix~\ref{appendix-manipulation}.

\subsection{Effects on team dynamics}

AI personas affected team dynamics in selective, robust ways after multiplicity correction. For \textit{psychological safety}, contrarian personas reduced safety relative to all-human teams: with two humans and one contrarian AI (H2\_C), $\beta$ ($df{=}64)=-0.671$; S.E.\ $=0.169$; 95\% CI [$-1.007$, $-0.334$]; $p<.001$; $p_{\text{adj}}=.003$, and with one human and two contrarian AIs (H1\_C), $\beta$ ($df{=}51)=-0.881$; S.E.\ $=0.281$; 95\% CI [$-1.439$, $-0.323$]; $p=.003$; $p_{\text{adj}}=.049$. Supportive configurations did not yield significant changes in psychological safety after multiplicity correction (e.g., H2\_S: $\beta$ ($df{=}63)=0.152$; S.E.\ $=0.148$; $p=.308$; $p_{\text{adj}}=1.00$). For \textit{group discussion quality}, mixed teams with a supportive AI (H2\_S) scored higher than human-only teams, $\beta$ ($df{=}67)=0.590$; S.E.\ $=0.166$; 95\% CI [$0.259$, $0.922$]; $p=.001$; $p_{\text{adj}}=.012$, whereas mixed teams with a contrarian AI (H2\_C) scored lower, $\beta$ ($df{=}67)=-0.856$; S.E.\ $=0.202$; 95\% CI [$-1.257$, $-0.454$]; $p<.001$; $p_{\text{adj}}=.001$. Effects for one-human/two-AI configurations (H1\_S, H1\_C) did not remain significant after multiplicity correction. For \textit{team satisfaction}, contrarian configurations showed negative point estimates but no effects remained significant after multiplicity correction (e.g., H2\_C: $\beta$ ($df{=}64)=-0.414$; S.E.\ $=0.192$; $p=.035$; $p_{\text{adj}}=.627$; H1\_C: $\beta$ ($df{=}51)=-0.451$; S.E.\ $=0.271$; $p=.102$; $p_{\text{adj}}=1.00$). Other contrasts were non-significant after multiplicity correction. Overall, contrarian personas consistently undermined psychological safety and discussion quality in mixed human-AI teams, whereas supportive personas consistently improved discussion quality in mixed teams without altering psychological safety. Full regression estimates are available in Appendix \ref{appendix-team}.

\subsection{Effects on individual task performance}

\begin{figure}
  \centering
  \includegraphics[width=1\linewidth]{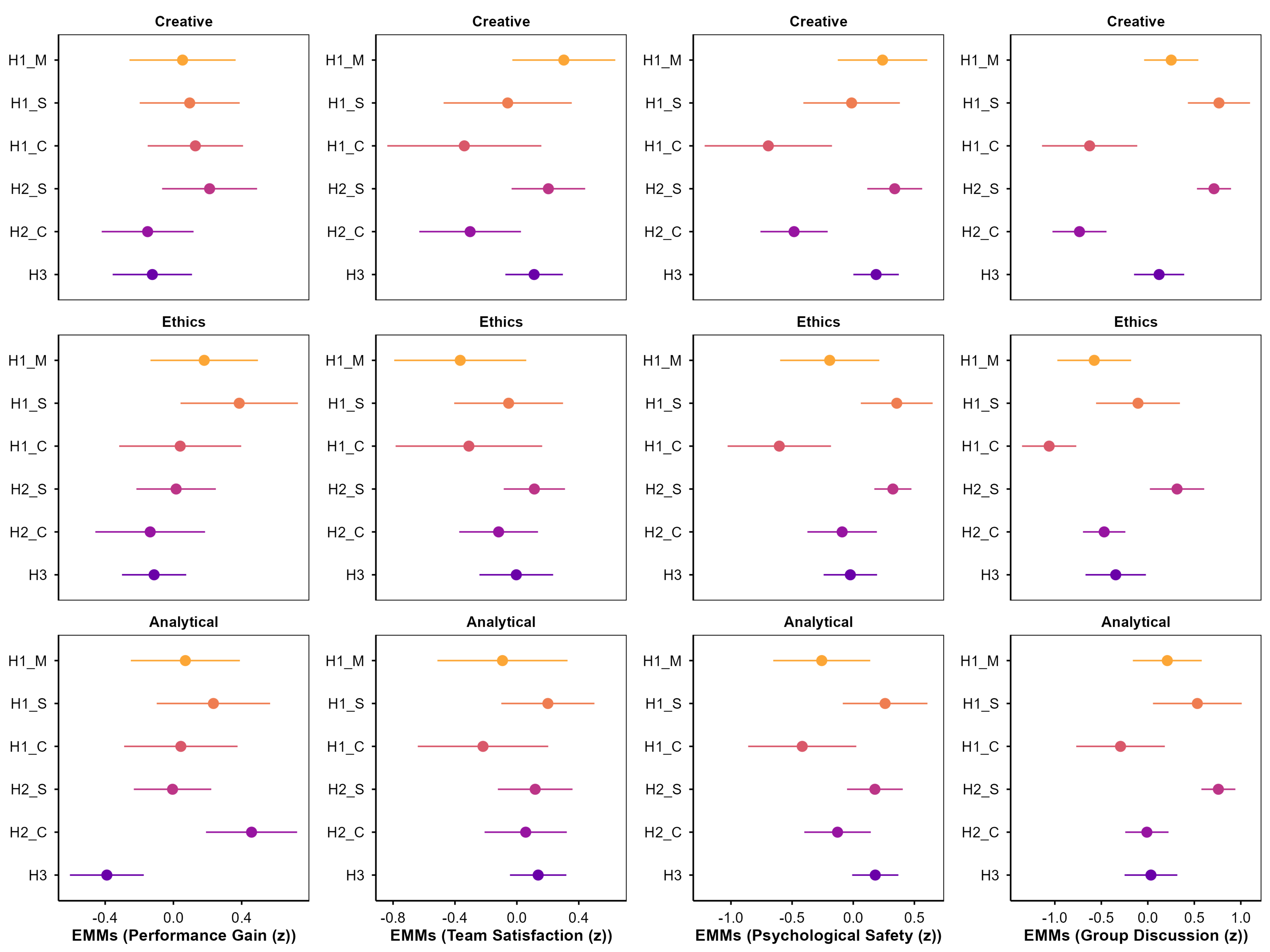}
  \caption{Estimated marginal means (EMMs) with CR2-robust 95\% confidence intervals for all outcomes across tasks and group compositions. Each panel shows adjusted means (z-standardised) for individual performance gain, team satisfaction, psychological safety, and group discussion, split by task (Creative, Ethics, Analytical). Conditions are ordered by group composition: fully human teams (H3), mixed human–AI teams with one AI (H2\_S, H2\_C), and mixed teams with two AIs (H1\_S, H1\_C, H1\_M). Error bars reflect cluster-robust (CR2) standard errors accounting for group-level clustering.}
  \label{fig:emms_cr2}
\end{figure}

Across conditions, AI personas did not generally yield significant differences in individual performance gains. An exception emerged in the analytical task, where groups with two humans and one contrarian AI teammate outperformed fully human teams, $\beta$ ($df{=}123)=0.88$; S.E.\ $=0.25$; 95\% CI [$0.38$, $1.38$]; $p<.001$; $p_{\text{adj}}=.012$. No other main effects or interactions remained significant after multiplicity correction. Follow-up pairwise contrasts within the analytical task confirmed this pattern (Figure \ref{fig:emms_cr2}). Relative to the all-human baseline (H3), both AI-present conditions showed significantly greater gains: teams with two humans and one contrarian AI performed better than all-human teams (H3 -- H2\_C), $\beta$ ($df{=}348)=-0.85$; S.E.\ $=0.18$; 95\% CI [$-1.19$, $-0.50$]; $p<.001$; $p_{\text{adj}}<.001$, and teams with one human and two supportive AIs also outperformed all-human teams (H3 -- H1\_S), $\beta$ ($df{=}730)=-0.62$; S.E.\ $=0.20$; 95\% CI [$-1.02$, $-0.23$]; $p=.002$; $p_{\text{adj}}=.029$. These contrasts were specific to the analytical task; no significant fixed-effect interactions were observed for the creative or ethical tasks after multiplicity correction (all task-specific contrasts in Appendix~\ref{appendix-gain}). In sum, the findings suggest that AI teammates did not systematically enhance performance, but under certain analytical conditions, both supportive and contrarian personas produced short-term gains.

\subsection{Moderation by baseline detectability index}

To test whether persona effects depended on recognising AI teammates, we re-estimated models including the BDI (individual-level BDI for LMMs and group-mean BDI for the group-level OLS). Core results were unchanged: contrarian personas reduced psychological safety, and supportive personas improved discussion quality in mixed two-human teams after multiplicity correction. Joint CR2 Wald tests indicated moderation by BDI for team satisfaction and psychological safety (e.g., \(F(17,110)=2.73,\, p<.001;\; F(17,110)=2.40,\, p=.003\)), with higher satisfaction and safety for supportive versus contrarian personas across BDI levels; no moderation was observed for discussion quality (\(F(17,91)=0.72,\, p=.777\)) or individual performance (\(F(17,111)=0.87,\, p=.607\)). BDI main effects were directionally positive for satisfaction and safety, but did not remain significant after multiplicity correction. Full estimates and marginal-effects plots are provided in Appendix~\ref{appendix-moderation}.

\subsection{Linguistic behaviour and mediation analysis}

\begin{figure}
\centering
\includegraphics[width=0.95\textwidth]{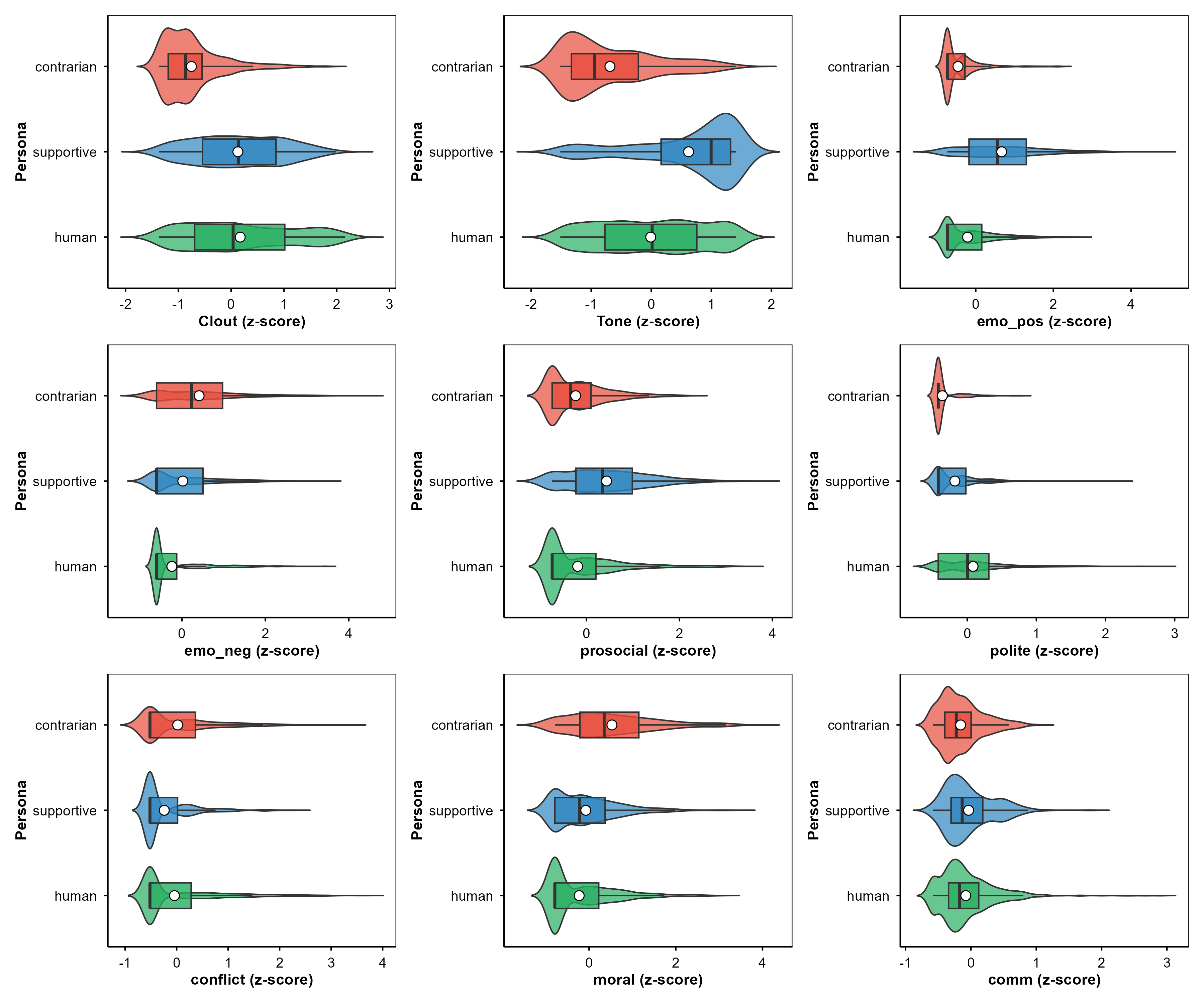}
\caption{Linguistic profiles across humans and AI personas. Violin plots display the distribution of nine LIWC features (clout, tone, positive emotion, negative emotion, prosocial, politeness, conflict, moralising, and communication) for human participants (green), supportive AI agents (blue), and contrarian AI agents (red) across all tasks. White dots represent means, thick black bars indicate interquartile ranges, and thin lines show the data range within 1.5~IQR.}
\label{fig:liwc_violin}
\end{figure}

This analysis served to verify that the AI agents enacted their assigned personas consistently through language use, thereby providing a behavioural validation of the persona manipulation central to the study. Language use diverged systematically across humans, supportive AI, and contrarian AI teammates (Figure~\ref{fig:liwc_violin}). Supportive agents consistently amplified affiliative registers relative to humans, producing higher levels of positive tone, prosocial and emotionally positive language, and speaking in longer utterances. These effects were substantial, ranging from 0.61 to 0.85 standardised units, and all remained significant after multiplicity correction (all $p_{\text{adj}}<.001$). At the same time, supportive agents used fewer politeness markers and conflict terms, with reductions of 0.18 to 0.27 standardised units that also remained significant after correction (all $p_{\text{adj}}<.006$), suggesting a style oriented towards encouragement rather than mitigation. By contrast, contrarian agents showed the opposite polarity. Compared with humans, they spoke with significantly lower clout and positive tone, with decreases ranging from 0.63 to 0.91 standardised units below the human baseline, and relied more heavily on negative emotion and moralising vocabulary, with increases of 0.62 to 0.73 standardised units. These effects were large in magnitude and robust after adjustment (all $p_{\text{adj}}<.001$). Contrarian agents also produced longer utterances than humans, with an increase of 0.62 standardised units ($p_{\text{adj}}<.001$), though differences for conflict and prosocial terms did not remain significant after correction ($p_{\text{adj}}>.05$).  

Direct contrasts between personas reinforced this gradient. Supportive agents produced significantly higher levels of tone, prosocial language, and positive emotion than contrarian agents, with differences ranging from 1.05 to 1.24 standardised units (all $p_{\text{adj}}<.001$). Conversely, contrarian agents displayed higher levels of negative emotion and moralising language relative to supportive agents, with increases of 0.36 to 0.55 standardised units (all $p_{\text{adj}}<.001$). Smaller but reliable contrasts also emerged for assent, which was reduced in contrarian relative to supportive speech by 0.12 standardised units ($p_{\text{adj}}=.004$). By contrast, overall verbosity, indexed by word count, did not differ significantly between persona types ($p_{\text{adj}}=.436$). All significant linguistic effects were large in magnitude ($|d|>0.95$). Taken together, the linguistic profiles reveal a consistent continuum: supportive AIs extended affiliative styles beyond the human baseline, contrarian AIs accentuated conflictual and moralising styles, and human teammates occupied an intermediate position between these two extremes. Full model outputs and robustness checks can be found in Appendix~\ref{appendix-linguistic}. 

\begin{figure}
\centering
\includegraphics[width=0.95\textwidth]{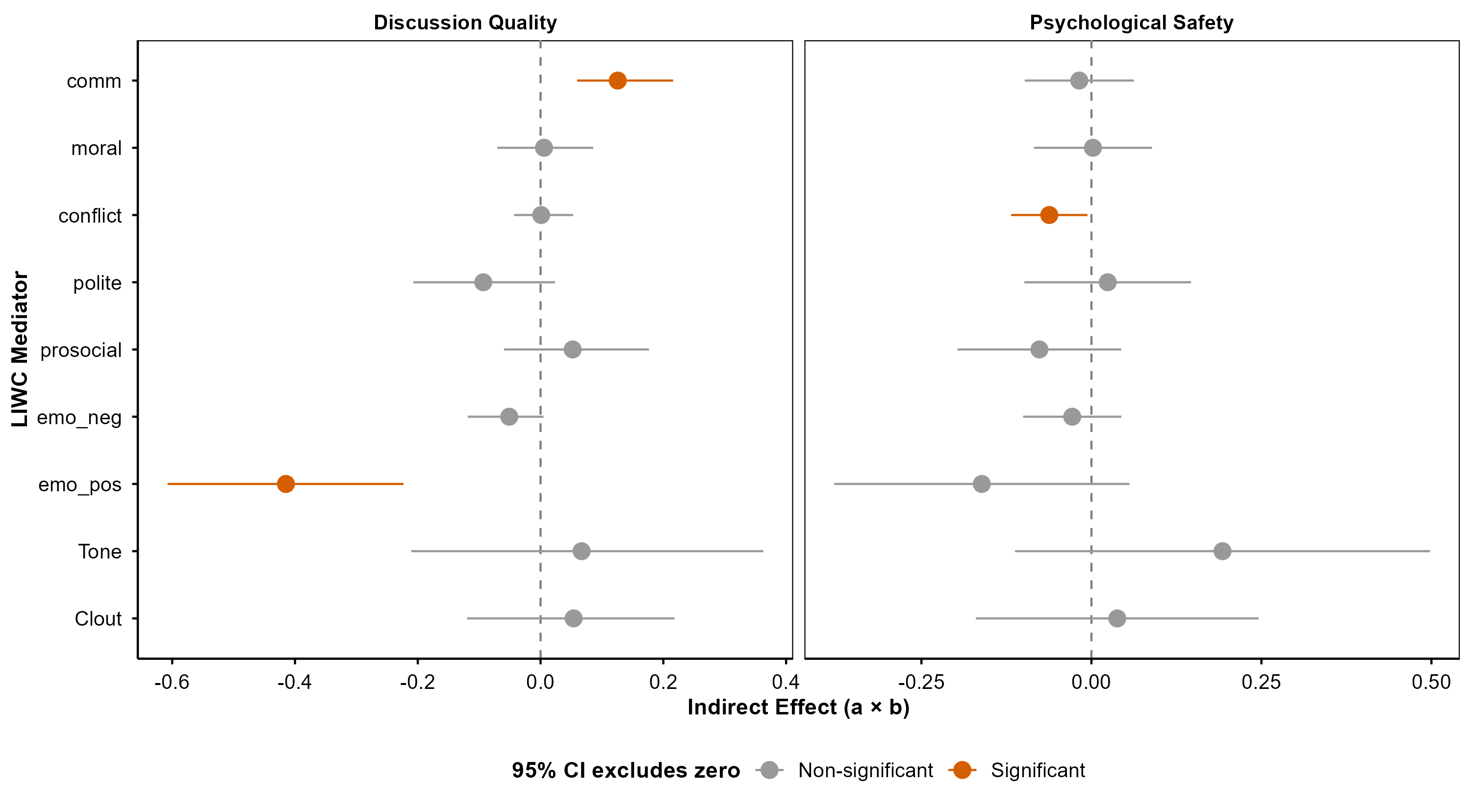}
\caption{Indirect effects of AI persona polarity on team climate outcomes via linguistic mediators. Panels show parallel multiple-mediator models with contrarian dosage as predictor, Linguistic Inquiry Word Count features from AI utterances as mediators, and group discussion quality (left) or psychological safety (right) as outcomes. Points indicate bootstrap estimates of the indirect effect (\(\hat{a}\hat{b}\)); horizontal lines denote BCa 95\% confidence intervals from 10{,}000 cluster resamples. Orange points mark mediators with intervals excluding zero (\(p<.05\)): reduced positive emotion language mediated lower discussion quality, while lower conflict language partially mediated psychological safety. All other mediators were non-significant.}
\label{fig:liwc_indirect_effects}
\end{figure}

Having established that linguistic profiles reflected the intended persona distinctions, we conducted a follow-up analysis to examine whether these linguistic differences functioned as the mechanism through which persona polarity influenced team dynamics. We next examined whether persona effects were transmitted through the linguistic style of AI teammates (parallel multiple mediation; cluster bootstrap, $B{=}10{,}000$; Bias-corrected and accelerated (BCa) intervals; CR2-robust inference). As expected, contrarian dosage (\(P_g\)) reliably shifted all nine linguistic features ($p_{\text{adj}}<.001$), yet the downstream impact varied by outcome (Figure~\ref{fig:liwc_indirect_effects}). 

For psychological safety, the total indirect effect was small and not significant (\(\hat{IE}_{\mathrm{Total}}=-0.09\), BCa~95\% CI [$-0.34$, $0.16$], $p=.473$). Most individual mediators showed confidence intervals overlapping zero, with the exception of conflict language, which exerted a modest negative indirect effect (\(\hat{a}\hat{b}=-0.06\), BCa~95\% CI [$-0.12$, $-0.01$], $p=.014$). The direct path from persona polarity to safety remained robustly negative (\(c'=-0.56\), BCa~95\% CI [$-0.86$, $-0.25$], $p=.001$), suggesting that safety was reduced primarily through non-linguistic or unmeasured relational processes, such as turn-taking disruptions or expectancy violations. For group discussion quality, both direct and mediated contributions were observed. The total indirect effect trended negative (\(\hat{IE}_{\mathrm{Total}}=-0.25\), BCa~95\% CI [$-0.51$, $0.02$], $p=.054$). Two complementary pathways emerged: reduced positive-emotion language predicted poorer discussion quality (\(\hat{a}\hat{b}=-0.42\), BCa~95\% CI [$-0.61$, $-0.22$], $p<.001$), while lower communication markers unexpectedly predicted higher quality (\(\hat{a}\hat{b}=0.13\), BCa~95\% CI [$0.06$, $0.22$], $p=.001$). The direct effect of contrarian dosage remained strongly negative (\(c'=-0.89\), BCa~95\% CI [$-1.20$, $-0.61$], $p<.001$). Overall, persona polarity reliably altered the linguistic register of AI agents, but only some of these changes carried downstream social consequences. Discussion quality was partly mediated by the reduction of positive-affect language, whereas the impact on psychological safety was largely direct and non-linguistic. Complete mediation models are reported in Appendix~\ref{appendix-linguistic}.

\section{Discussion}

This study provides causal evidence that AI personas reshape team dynamics even when people have limited awareness of artificial teammates. Across three task domains, contrarian personas reliably reduced psychological safety and discussion quality, while supportive personas improved discussion quality in mixed human-AI teams. These effects remained after accounting for detection accuracy, indicating that persona-driven influence operates largely independent of explicit recognition. We term this dissociation between influence and awareness the \textit{social blindspot} in human-AI collaboration. The notion of a social blindspot draws inspiration from research on the bias blindspot and metacognitive unawareness in human-human interaction, where individuals remain unaware of how social cues shape their judgments and behaviour \citep{pronin2002bias, pronin2007perception}. In the context of human-AI collaboration, the term captures a similar asymmetry: AI agents can exert measurable social influence on group processes even when their artificial identity is unnoticed or uncertain. This differs from related notions such as algorithmic invisibility or automation bias, which emphasise perceptual or decision-level oversight rather than relational dynamics \citep{rahwan2019machine, ehsan2020human, crawford2021atlas}. The social blindspot thus delineates a unique phenomenon in which the social agency of AI operates effectively yet tacitly, shaping collective outcomes beneath the threshold of conscious detection.

At the same time, we observed an exception on cognitive outcomes: in the analytical task, groups with a contrarian AI achieved greater individual performance gains than all-human teams, with a similar advantage for teams with two supportive AIs. This pattern suggests that persona polarity can create a trade-off between relational climate and short-term analytical convergence. In organisational and educational research, structured dissent is well known to sharpen problem solving under certain conditions, especially on intellective problems with demonstrable solutions \citep{Nemeth1983,SchulzHardt2006}, but unmanaged dissent can simultaneously depress psychological safety and inclusion \citep{Jehn1995,DeDreu2003}. Our findings align with this tension: contrarian style appears to offer instrumental benefits for analytical tasks while imposing social costs that are visible within minutes of interaction. Practically, this argues against a one-size-fits-all persona and towards context-sensitive calibration of stance and challenge.

Analyses of mechanism further qualify how persona exerts its effects. Linguistic mediation was limited for psychological safety but detectable for discussion quality: diminished positive-affect language partly transmitted the negative impact of contrarian dosage on quality, whereas safety effects were largely direct or routed through unmeasured relational cues such as subtle turn-taking patterns, response timing, or expectancy violations that trigger interpersonal threat appraisals \citep{edmondson1999psychological}. This dissociation implies that the “social blindspot” is not a single pathway phenomenon. Team dynamic components respond differently: discussion quality is partially language-sensitive, while psychological safety is more vulnerable to non-lexical aspects of interactional style. Future work should therefore incorporate richer behavioural traces, timing, interruption, and targeted acknowledgement, to pinpoint which micro-practices drive these direct effects.

Conceptually, the results both replicate and extend established findings in team dynamics. Prior work has shown that a single member’s communicative style can tip collective norms and shape perceptions of inclusion \citep{Ashktorab2020,Lee2025}, and our findings replicate this effect in the context of collaborative interaction. At the same time, they extend the literature by demonstrating that such shifts can originate from AI agents embedded within human groups, that they persist even when people cannot reliably identify those agents, and that they propagate through consistent persona cues rather than informational accuracy or domain expertise alone. This pattern resonates with the “media equation” tradition, which argues that people apply social heuristics to machines even while denying such attributions \citep{ReevesNass1996,NassMoon2000}. Moving beyond dyads, our results show that persona cues in AI teammates scale into emergent group properties such as psychological safety, satisfaction, and externally rated discussion quality under realistic conversational constraints.

Beyond immediate team dynamics, the social blindspot resonates with work on \textit{artificial sociality}, where LLM-driven agents interact with each other to generate emergent norms and collective behaviours \citep{park2023generative,He2024,Xu2024}. Such simulations of virtual towns, classrooms, or workplaces illustrate how homophily, norm formation, and organisational dynamics can be reshaped by persona-level features of AI agents \citep{Broadbent2023,Tang2023}. Linking our findings to this literature suggests that the blindspot may scale from small groups to larger hybrid societies, where undetected AI personas influence inclusion, trust, and cohesion \citep{Siebers2024,Yan2025,Berry2024}.

\subsection{Implications}

These insights carry concrete implications for the governance of human-AI collaboration. Disclosure alone is insufficient \citep{Zerilli2022,Vossing2022}; while improving awareness may correlate with slightly better baseline dynamics, it does not eliminate the powerful influence of persona cues. Effective governance therefore requires treating persona design as a first-class lever alongside data and model governance. Rather than relying on transparency alone, developers and institutions need to constrain default stance and tone to prevent persistent contrarianism in contexts where psychological safety is paramount, and they should also consider adaptive or hybrid personas capable of delivering calibrated challenge when it benefits the task while actively repairing safety through hedging, affirmations, or turn-taking that surfaces quieter voices. At an organisational level, policies should match persona profiles to the goals of the task, whether analytical convergence, creative exploration, or ethical deliberation, and to the needs of the population, such as novices versus experts or high-stakes versus low-stakes settings. Finally, evaluation of AI systems must move beyond accuracy to incorporate dynamic measures of team climate, including psychological safety, perceived fairness, and inclusion, as key performance indicators for responsible deployment.

\subsection{Limitations}

Several limitations define the scope and boundary conditions of our findings. The study was conducted in a controlled, time-limited laboratory setting in which groups interacted for only ten minutes on curated tasks, enhancing internal validity but constraining ecological breadth. While real-world collaboration unfolds over longer horizons and across richer contexts, the fact that contrarian personas produced immediate and robust disruptions even within minutes underscores the strength of the observed effects and motivates longitudinal and field studies. The persona manipulation was intentionally binary, contrasting supportive and contrarian styles to maximise causal leverage, but this does not capture the nuanced blending of encouragement and challenge typical of human teams, leaving open whether adaptive or hybrid personas could retain cognitive benefits without relational costs. Ecological validity is further bounded by the use of synchronous text-based interaction with anonymised identities and a Prolific sample, which, while common and reliable in behavioural research \citep{peer2022data,douglas2023data,salvi2025conversational}, differs from multimodal, high-stakes professional or civic settings. Finally, AI agents were designed with calibrated imperfections and guardrails to ensure naturalistic yet non-dominant participation, and manipulation checks required explicit human–AI judgements, which may have heightened suspicion; however, persistent persona effects despite misclassification suggest that the findings are not reducible to detection artefacts. Future work should extend these results to longer-term, multimodal, and disclosure-explicit settings.

\section{Conclusion}

This study points to a broader challenge that will increasingly shape hybrid intelligence systems: AI does not merely contribute information to group work, but actively participates in the social regulation of collaboration. Our findings suggest that persona design operates as a powerful, and often invisible, lever through which AI systems can shape norms of interaction, expectations of dissent, and the conditions under which learners feel safe to contribute. The notion of a \textit{social blindspot} highlights a critical asymmetry: AI influence can emerge even when users are unable to recognise, reflect on, or contest its presence. Looking ahead, this raises important questions for the governance of human–AI collaboration. Rather than treating persona as a surface-level interface choice, designers and policymakers must consider persona governance as a form of social intervention, with consequences for inclusion, equity, and collective sensemaking. Educational and organisational deployments may benefit from moving beyond static personas toward adaptive or context-aware systems that modulate challenge, support, and turn-taking in response to group dynamics, task demands, and participant vulnerability. At the same time, transparency mechanisms and participatory design approaches may be required to give users greater agency over how AI shapes their collaborative environment. More broadly, this work calls for a shift in how AI-assisted collaboration is evaluated. Performance gains alone are insufficient indicators of success when relational goods such as psychological safety, fairness, and voice are at stake. Future research should therefore examine how persona effects accumulate over time, how they interact with disclosure and accountability regimes, and how learners develop meta-awareness of AI’s social role. By foregrounding the social dimensions of AI participation, this line of work aims to support the development of hybrid systems that enhance collective intelligence without eroding the interpersonal foundations on which it depends.

\clearpage


\bibliographystyle{cas-model2-names}

\bibliography{0_reference}

\begin{thebibliography}{82}
\expandafter\ifx\csname natexlab\endcsname\relax\def\natexlab#1{#1}\fi
\providecommand{\url}[1]{\texttt{#1}}
\providecommand{\href}[2]{#2}
\providecommand{\path}[1]{#1}
\providecommand{\DOIprefix}{doi:}
\providecommand{\ArXivprefix}{arXiv:}
\providecommand{\URLprefix}{URL: }
\providecommand{\Pubmedprefix}{pmid:}
\providecommand{\doi}[1]{\href{http://dx.doi.org/#1}{\path{#1}}}
\providecommand{\Pubmed}[1]{\href{pmid:#1}{\path{#1}}}
\providecommand{\bibinfo}[2]{#2}
\ifx\xfnm\relax \def\xfnm[#1]{\unskip,\space#1}\fi
\bibitem[{Agresti and Coull(1998)}]{agresti1998approximate}
\bibinfo{author}{Agresti, A.}, \bibinfo{author}{Coull, B.A.}, \bibinfo{year}{1998}.
\newblock \bibinfo{title}{Approximate is better than “exact” for interval estimation of binomial proportions}.
\newblock \bibinfo{journal}{The American Statistician} \bibinfo{volume}{52}, \bibinfo{pages}{119--126}.
\bibitem[{Alabed et~al.(2022)Alabed, Javornik and Gregory-Smith}]{Alabed2022}
\bibinfo{author}{Alabed, A.}, \bibinfo{author}{Javornik, A.}, \bibinfo{author}{Gregory-Smith, D.}, \bibinfo{year}{2022}.
\newblock \bibinfo{title}{Ai anthropomorphism and its effect on users' self-congruence and self–ai integration: A theoretical framework and research agenda}.
\newblock \bibinfo{journal}{Technological Forecasting and Social Change} \bibinfo{volume}{180}, \bibinfo{pages}{121786}.
\newblock \DOIprefix\doi{10.1016/j.techfore.2022.121786}.
\bibitem[{Ashktorab et~al.(2020)Ashktorab, Liao, Dugan, Johnson, Pan, Zhang, Kumaravel and Campbell}]{Ashktorab2020}
\bibinfo{author}{Ashktorab, Z.}, \bibinfo{author}{Liao, Q.V.}, \bibinfo{author}{Dugan, C.}, \bibinfo{author}{Johnson, J.}, \bibinfo{author}{Pan, Q.}, \bibinfo{author}{Zhang, W.}, \bibinfo{author}{Kumaravel, S.}, \bibinfo{author}{Campbell, M.}, \bibinfo{year}{2020}.
\newblock \bibinfo{title}{Human–ai collaboration in a cooperative game setting}, in: \bibinfo{booktitle}{Proceedings of the ACM on Human-Computer Interaction}, pp. \bibinfo{pages}{1--20}.
\newblock \DOIprefix\doi{10.1145/3415167}.
\bibitem[{Awad et~al.(2018)Awad, Dsouza, Kim, Schulz, Henrich, Shariff et~al.}]{Awad2018}
\bibinfo{author}{Awad, E.}, \bibinfo{author}{Dsouza, S.}, \bibinfo{author}{Kim, R.}, \bibinfo{author}{Schulz, J.}, \bibinfo{author}{Henrich, J.}, \bibinfo{author}{Shariff, A.}, et~al., \bibinfo{year}{2018}.
\newblock \bibinfo{title}{The moral machine experiment}.
\newblock \bibinfo{journal}{Nature} \bibinfo{volume}{563}, \bibinfo{pages}{59--64}.
\bibitem[{Bankins et~al.(2023)Bankins, Ocampo, Marrone, Restubog and Woo}]{Bankins2023}
\bibinfo{author}{Bankins, S.}, \bibinfo{author}{Ocampo, A.}, \bibinfo{author}{Marrone, M.}, \bibinfo{author}{Restubog, S.}, \bibinfo{author}{Woo, S.}, \bibinfo{year}{2023}.
\newblock \bibinfo{title}{A multilevel review of artificial intelligence in organizations: Implications for organizational behavior research and practice}.
\newblock \bibinfo{journal}{Journal of Organizational Behavior} \DOIprefix\doi{10.1002/job.2735}.
\bibitem[{Benharrak et~al.(2024)Benharrak, Zindulka, Lehmann, Heuer and Buschek}]{benharrak2024writer}
\bibinfo{author}{Benharrak, K.}, \bibinfo{author}{Zindulka, T.}, \bibinfo{author}{Lehmann, F.}, \bibinfo{author}{Heuer, H.}, \bibinfo{author}{Buschek, D.}, \bibinfo{year}{2024}.
\newblock \bibinfo{title}{Writer-defined ai personas for on-demand feedback generation}, in: \bibinfo{booktitle}{Proceedings of the 2024 CHI Conference on Human Factors in Computing Systems}, pp. \bibinfo{pages}{1--18}.
\bibitem[{Berry and Stockman(2024)}]{Berry2024}
\bibinfo{author}{Berry, D.M.}, \bibinfo{author}{Stockman, J.}, \bibinfo{year}{2024}.
\newblock \bibinfo{title}{Schumacher in the age of generative ai: Towards a new critique of technology}.
\newblock \bibinfo{journal}{European Journal of Social Theory} \bibinfo{volume}{27}, \bibinfo{pages}{437--455}.
\newblock \DOIprefix\doi{10.1177/13684310241234028}.
\bibitem[{Bezrukova et~al.(2023)Bezrukova, Griffith, Spell, Rice and Yang}]{Bezrukova2023}
\bibinfo{author}{Bezrukova, K.}, \bibinfo{author}{Griffith, T.}, \bibinfo{author}{Spell, C.}, \bibinfo{author}{Rice, V.}, \bibinfo{author}{Yang, H.}, \bibinfo{year}{2023}.
\newblock \bibinfo{title}{Artificial intelligence and groups: Effects of attitudes and discretion on collaboration}.
\newblock \bibinfo{journal}{Group \& Organization Management} \bibinfo{volume}{48}, \bibinfo{pages}{629--670}.
\newblock \DOIprefix\doi{10.1177/10596011231160574}.
\bibitem[{Bigman and Gray(2020)}]{Bigman2020Nature}
\bibinfo{author}{Bigman, Y.E.}, \bibinfo{author}{Gray, K.}, \bibinfo{year}{2020}.
\newblock \bibinfo{title}{Life and death decisions of autonomous vehicles}.
\newblock \bibinfo{journal}{Nature} \bibinfo{volume}{579}, \bibinfo{pages}{E1--E2}.
\bibitem[{Boji{\'c} et~al.(2024)Boji{\'c}, Cinelli, {\'C}ulibrk and Deliba{\v{s}}i{\'c}}]{Bojic2023}
\bibinfo{author}{Boji{\'c}, L.}, \bibinfo{author}{Cinelli, M.}, \bibinfo{author}{{\'C}ulibrk, D.}, \bibinfo{author}{Deliba{\v{s}}i{\'c}, B.}, \bibinfo{year}{2024}.
\newblock \bibinfo{title}{Cern for ai: a theoretical framework for autonomous simulation-based artificial intelligence testing and alignment}.
\newblock \bibinfo{journal}{European Journal of Futures Research} \bibinfo{volume}{12}, \bibinfo{pages}{1--19}.
\newblock \DOIprefix\doi{10.1186/s40309-024-00238-0}.
\bibitem[{Bonnefon et~al.(2016)Bonnefon, Shariff and Rahwan}]{Bonnefon2016}
\bibinfo{author}{Bonnefon, J.F.}, \bibinfo{author}{Shariff, A.}, \bibinfo{author}{Rahwan, I.}, \bibinfo{year}{2016}.
\newblock \bibinfo{title}{The social dilemma of autonomous vehicles}.
\newblock \bibinfo{journal}{Science} \bibinfo{volume}{352}, \bibinfo{pages}{1573--1576}.
\bibitem[{Broadbent et~al.(2023)Broadbent, Billinghurst, Boardman and Doraiswamy}]{Broadbent2023}
\bibinfo{author}{Broadbent, E.}, \bibinfo{author}{Billinghurst, M.}, \bibinfo{author}{Boardman, S.G.}, \bibinfo{author}{Doraiswamy, P.M.}, \bibinfo{year}{2023}.
\newblock \bibinfo{title}{Enhancing social connectedness with companion robots using ai}.
\newblock \bibinfo{journal}{Science Robotics} \bibinfo{volume}{8}, \bibinfo{pages}{eadi6347}.
\newblock \DOIprefix\doi{10.1126/scirobotics.adi6347}.
\bibitem[{Brodersen et~al.(2010)Brodersen, Ong, Stephan and Buhmann}]{brodersen2010balanced}
\bibinfo{author}{Brodersen, K.H.}, \bibinfo{author}{Ong, C.S.}, \bibinfo{author}{Stephan, K.E.}, \bibinfo{author}{Buhmann, J.M.}, \bibinfo{year}{2010}.
\newblock \bibinfo{title}{The balanced accuracy and its posterior distribution}, in: \bibinfo{booktitle}{2010 20th international conference on pattern recognition}, \bibinfo{organization}{IEEE}. pp. \bibinfo{pages}{3121--3124}.
\bibitem[{Chen et~al.(2025)Chen, Song, Guo, Sun, Childs and Yin}]{Chen2025}
\bibinfo{author}{Chen, L.}, \bibinfo{author}{Song, Y.}, \bibinfo{author}{Guo, J.}, \bibinfo{author}{Sun, L.}, \bibinfo{author}{Childs, P.}, \bibinfo{author}{Yin, Y.}, \bibinfo{year}{2025}.
\newblock \bibinfo{title}{How generative ai supports human in conceptual design}.
\newblock \bibinfo{journal}{Design Science} \bibinfo{volume}{11}.
\newblock \DOIprefix\doi{10.1017/dsj.2025.2}.
\bibitem[{Claggett et~al.(2025)Claggett, Kraut and Shirado}]{Claggett2025}
\bibinfo{author}{Claggett, E.L.}, \bibinfo{author}{Kraut, R.E.}, \bibinfo{author}{Shirado, H.}, \bibinfo{year}{2025}.
\newblock \bibinfo{title}{Relational ai: Facilitating intergroup cooperation with socially aware conversational support}, in: \bibinfo{booktitle}{Proceedings of the 2025 CHI Conference on Human Factors in Computing Systems}, \bibinfo{publisher}{Association for Computing Machinery}, \bibinfo{address}{New York, NY, USA}. pp. \bibinfo{pages}{1--22}.
\newblock \URLprefix \url{https://doi.org/10.1145/3706598.3713757}, \DOIprefix\doi{10.1145/3706598.3713757}.
\bibitem[{Cohen et~al.(2016)Cohen, Korevaar, Altman, Bruns, Gatsonis, Hooft, Irwig, Levine, Reitsma, De~Vet et~al.}]{cohen2016stard}
\bibinfo{author}{Cohen, J.F.}, \bibinfo{author}{Korevaar, D.A.}, \bibinfo{author}{Altman, D.G.}, \bibinfo{author}{Bruns, D.E.}, \bibinfo{author}{Gatsonis, C.A.}, \bibinfo{author}{Hooft, L.}, \bibinfo{author}{Irwig, L.}, \bibinfo{author}{Levine, D.}, \bibinfo{author}{Reitsma, J.B.}, \bibinfo{author}{De~Vet, H.C.}, et~al., \bibinfo{year}{2016}.
\newblock \bibinfo{title}{Stard 2015 guidelines for reporting diagnostic accuracy studies: explanation and elaboration}.
\newblock \bibinfo{journal}{BMJ open} \bibinfo{volume}{6}, \bibinfo{pages}{e012799}.
\bibitem[{Crawford(2021)}]{crawford2021atlas}
\bibinfo{author}{Crawford, K.}, \bibinfo{year}{2021}.
\newblock \bibinfo{title}{The atlas of AI: Power, politics, and the planetary costs of artificial intelligence}.
\newblock \bibinfo{publisher}{Yale University Press}.
\bibitem[{Danescu-Niculescu-Mizil et~al.(2013)Danescu-Niculescu-Mizil, Sudhof, Jurafsky, Leskovec and Potts}]{danescu2013computational}
\bibinfo{author}{Danescu-Niculescu-Mizil, C.}, \bibinfo{author}{Sudhof, M.}, \bibinfo{author}{Jurafsky, D.}, \bibinfo{author}{Leskovec, J.}, \bibinfo{author}{Potts, C.}, \bibinfo{year}{2013}.
\newblock \bibinfo{title}{A computational approach to politeness with application to social factors}, in: \bibinfo{booktitle}{Proceedings of the 51st Annual Meeting of the Association for Computational Linguistics (Volume 1: Long Papers)}, pp. \bibinfo{pages}{250--259}.
\bibitem[{De~Dreu and Weingart(2003)}]{DeDreu2003}
\bibinfo{author}{De~Dreu, C.K.W.}, \bibinfo{author}{Weingart, L.R.}, \bibinfo{year}{2003}.
\newblock \bibinfo{title}{Task versus relationship conflict, team performance, and team member satisfaction: A meta-analysis}.
\newblock \bibinfo{journal}{Journal of Applied Psychology} \bibinfo{volume}{88}, \bibinfo{pages}{741--749}.
\newblock \DOIprefix\doi{10.1037/0021-9010.88.4.741}.
\bibitem[{De~Wit et~al.(2012)De~Wit, Greer and Jehn}]{de2012paradox}
\bibinfo{author}{De~Wit, F.R.}, \bibinfo{author}{Greer, L.L.}, \bibinfo{author}{Jehn, K.A.}, \bibinfo{year}{2012}.
\newblock \bibinfo{title}{The paradox of intragroup conflict: a meta-analysis.}
\newblock \bibinfo{journal}{Journal of applied psychology} \bibinfo{volume}{97}, \bibinfo{pages}{360}.
\bibitem[{Dennis et~al.(2023)Dennis, Lakhiwal and Sachdeva}]{Dennis2023}
\bibinfo{author}{Dennis, A.R.}, \bibinfo{author}{Lakhiwal, A.}, \bibinfo{author}{Sachdeva, A.}, \bibinfo{year}{2023}.
\newblock \bibinfo{title}{Ai agents as team members: Effects on satisfaction, conflict, trustworthiness, and willingness to work with}.
\newblock \bibinfo{journal}{Journal of Management Information Systems} \bibinfo{volume}{40}, \bibinfo{pages}{307--337}.
\newblock \DOIprefix\doi{10.1080/07421222.2023.2196773}.
\bibitem[{Diehl and Stroebe(1987)}]{Diehl1987}
\bibinfo{author}{Diehl, M.}, \bibinfo{author}{Stroebe, W.}, \bibinfo{year}{1987}.
\newblock \bibinfo{title}{Productivity loss in brainstorming groups: Toward the solution of a riddle}.
\newblock \bibinfo{journal}{Journal of Personality and Social Psychology} \bibinfo{volume}{53}, \bibinfo{pages}{497--509}.
\bibitem[{Douglas et~al.(2023)Douglas, Ewell and Brauer}]{douglas2023data}
\bibinfo{author}{Douglas, B.D.}, \bibinfo{author}{Ewell, P.J.}, \bibinfo{author}{Brauer, M.}, \bibinfo{year}{2023}.
\newblock \bibinfo{title}{Data quality in online human-subjects research: Comparisons between mturk, prolific, cloudresearch, qualtrics, and sona}.
\newblock \bibinfo{journal}{Plos one} \bibinfo{volume}{18}, \bibinfo{pages}{e0279720}.
\bibitem[{Edmondson(1999)}]{edmondson1999psychological}
\bibinfo{author}{Edmondson, A.}, \bibinfo{year}{1999}.
\newblock \bibinfo{title}{Psychological safety and learning behavior in work teams}.
\newblock \bibinfo{journal}{Administrative science quarterly} \bibinfo{volume}{44}, \bibinfo{pages}{350--383}.
\bibitem[{Edwards et~al.(2024)Edwards, Nguyen, Lämsä, Sobocinski, Whitehead, Dang, Roberts and Järvelä}]{Edwards2024}
\bibinfo{author}{Edwards, J.}, \bibinfo{author}{Nguyen, A.}, \bibinfo{author}{Lämsä, J.}, \bibinfo{author}{Sobocinski, M.}, \bibinfo{author}{Whitehead, R.}, \bibinfo{author}{Dang, B.}, \bibinfo{author}{Roberts, A.}, \bibinfo{author}{Järvelä, S.}, \bibinfo{year}{2024}.
\newblock \bibinfo{title}{Human-ai collaboration: Designing artificial agents to facilitate socially shared regulation among learners}.
\newblock \bibinfo{journal}{British Journal of Educational Technology} \bibinfo{volume}{56}, \bibinfo{pages}{712--733}.
\newblock \DOIprefix\doi{10.1111/bjet.13534}.
\bibitem[{Ehsan and Riedl(2020)}]{ehsan2020human}
\bibinfo{author}{Ehsan, U.}, \bibinfo{author}{Riedl, M.O.}, \bibinfo{year}{2020}.
\newblock \bibinfo{title}{Human-centered explainable ai: Towards a reflective sociotechnical approach}, in: \bibinfo{booktitle}{International Conference on Human-Computer Interaction}, \bibinfo{organization}{Springer}. pp. \bibinfo{pages}{449--466}.
\bibitem[{Fan et~al.(2025)Fan, Tang, Le, Shen, Tan, Zhao, Shen, Li and Ga{\v{s}}evi{\'c}}]{fan2025beware}
\bibinfo{author}{Fan, Y.}, \bibinfo{author}{Tang, L.}, \bibinfo{author}{Le, H.}, \bibinfo{author}{Shen, K.}, \bibinfo{author}{Tan, S.}, \bibinfo{author}{Zhao, Y.}, \bibinfo{author}{Shen, Y.}, \bibinfo{author}{Li, X.}, \bibinfo{author}{Ga{\v{s}}evi{\'c}, D.}, \bibinfo{year}{2025}.
\newblock \bibinfo{title}{Beware of metacognitive laziness: Effects of generative artificial intelligence on learning motivation, processes, and performance}.
\newblock \bibinfo{journal}{British Journal of Educational Technology} \bibinfo{volume}{56}, \bibinfo{pages}{489--530}.
\bibitem[{Farrokhnia et~al.(2025)Farrokhnia, Noroozi, Baggen, Biemans and Weinberger}]{Farrokhnia2025}
\bibinfo{author}{Farrokhnia, M.}, \bibinfo{author}{Noroozi, O.}, \bibinfo{author}{Baggen, Y.}, \bibinfo{author}{Biemans, H.}, \bibinfo{author}{Weinberger, A.}, \bibinfo{year}{2025}.
\newblock \bibinfo{title}{Improving hybrid brainstorming outcomes with computer-supported scaffolds: Prompts and cognitive group awareness}.
\newblock \bibinfo{journal}{Computers \& Education} \bibinfo{volume}{227}, \bibinfo{pages}{105229}.
\bibitem[{Flathmann et~al.(2024)Flathmann, Duan, McNeese, Hauptman and Zhang}]{Flathmann2024}
\bibinfo{author}{Flathmann, C.}, \bibinfo{author}{Duan, W.}, \bibinfo{author}{McNeese, N.}, \bibinfo{author}{Hauptman, A.}, \bibinfo{author}{Zhang, R.}, \bibinfo{year}{2024}.
\newblock \bibinfo{title}{Empirically understanding the potential impacts and process of social influence in human-ai teams}.
\newblock \bibinfo{journal}{Proceedings of the ACM on Human-Computer Interaction} \bibinfo{volume}{8}, \bibinfo{pages}{1--32}.
\newblock \DOIprefix\doi{10.1145/3637326}.
\bibitem[{Floridi(2025)}]{Floridi2025}
\bibinfo{author}{Floridi, L.}, \bibinfo{year}{2025}.
\newblock \bibinfo{title}{Ai as agency without intelligence: On artificial intelligence as a new form of artificial agency and the multiple realisability of agency thesis}.
\newblock \bibinfo{journal}{Philosophy \& Technology} \bibinfo{volume}{38}, \bibinfo{pages}{30}.
\bibitem[{He et~al.(2024)He, Wallis, Gvirtz and Rathje}]{He2024}
\bibinfo{author}{He, J.K.}, \bibinfo{author}{Wallis, F.P.}, \bibinfo{author}{Gvirtz, A.}, \bibinfo{author}{Rathje, S.}, \bibinfo{year}{2024}.
\newblock \bibinfo{title}{Artificial intelligence chatbots mimic human collective behaviour}.
\newblock \bibinfo{journal}{British Journal of Psychology} \DOIprefix\doi{10.1111/bjop.12764}.
\bibitem[{Hohenstein et~al.(2023)Hohenstein, DiFranzo, Kizilcec et~al.}]{Hohenstein2021}
\bibinfo{author}{Hohenstein, J.}, \bibinfo{author}{DiFranzo, D.}, \bibinfo{author}{Kizilcec, R.}, et~al., \bibinfo{year}{2023}.
\newblock \bibinfo{title}{Artificial intelligence in communication impacts language and social relationships}.
\newblock \bibinfo{journal}{Scientific Reports} \bibinfo{volume}{13}.
\newblock \DOIprefix\doi{10.1038/s41598-023-30938-9}.
\bibitem[{Hu et~al.(2023)Hu, Mao and Kim}]{Hu2023}
\bibinfo{author}{Hu, B.}, \bibinfo{author}{Mao, Y.}, \bibinfo{author}{Kim, K.}, \bibinfo{year}{2023}.
\newblock \bibinfo{title}{How social anxiety leads to problematic use of conversational ai: The roles of loneliness, rumination, and mind perception}.
\newblock \bibinfo{journal}{Computers in Human Behavior} \bibinfo{volume}{145}, \bibinfo{pages}{107760}.
\newblock \DOIprefix\doi{10.1016/j.chb.2023.107760}.
\bibitem[{Hwang and Won(2021)}]{Hwang2021IdeaBot}
\bibinfo{author}{Hwang, A.H.C.}, \bibinfo{author}{Won, A.S.}, \bibinfo{year}{2021}.
\newblock \bibinfo{title}{Ideabot: Investigating social facilitation in human-machine team creativity}, in: \bibinfo{booktitle}{Proceedings of the 2021 CHI Conference on Human Factors in Computing Systems}, \bibinfo{publisher}{ACM}. pp. \bibinfo{pages}{1--16}.
\bibitem[{Ireland et~al.(2011)Ireland, Slatcher, Eastwick, Scissors, Finkel and Pennebaker}]{ireland2011language}
\bibinfo{author}{Ireland, M.E.}, \bibinfo{author}{Slatcher, R.B.}, \bibinfo{author}{Eastwick, P.W.}, \bibinfo{author}{Scissors, L.E.}, \bibinfo{author}{Finkel, E.J.}, \bibinfo{author}{Pennebaker, J.W.}, \bibinfo{year}{2011}.
\newblock \bibinfo{title}{Language style matching predicts relationship initiation and stability}.
\newblock \bibinfo{journal}{Psychological science} \bibinfo{volume}{22}, \bibinfo{pages}{39--44}.
\bibitem[{Jakesch et~al.(2023)Jakesch, Hancock and Naaman}]{jakesch2023human}
\bibinfo{author}{Jakesch, M.}, \bibinfo{author}{Hancock, J.T.}, \bibinfo{author}{Naaman, M.}, \bibinfo{year}{2023}.
\newblock \bibinfo{title}{Human heuristics for ai-generated language are flawed}.
\newblock \bibinfo{journal}{Proceedings of the National Academy of Sciences} \bibinfo{volume}{120}, \bibinfo{pages}{e2208839120}.
\bibitem[{Jehn(1995)}]{Jehn1995}
\bibinfo{author}{Jehn, K.A.}, \bibinfo{year}{1995}.
\newblock \bibinfo{title}{A multimethod examination of the benefits and detriments of intragroup conflict}.
\newblock \bibinfo{journal}{Administrative Science Quarterly} \bibinfo{volume}{40}, \bibinfo{pages}{256--282}.
\newblock \DOIprefix\doi{10.2307/2393638}.
\bibitem[{Johnson and Johnson(1987)}]{Johnson1987}
\bibinfo{author}{Johnson, D.W.}, \bibinfo{author}{Johnson, F.P.}, \bibinfo{year}{1987}.
\newblock \bibinfo{title}{Joining Together: Group Theory and Group Skills}.
\newblock \bibinfo{publisher}{Prentice-Hall}, \bibinfo{address}{Englewood Cliffs, NJ}.
\bibitem[{Järvelä et~al.(2023)Järvelä, Nguyen and Hadwin}]{Jarvela2023}
\bibinfo{author}{Järvelä, S.}, \bibinfo{author}{Nguyen, A.}, \bibinfo{author}{Hadwin, A.}, \bibinfo{year}{2023}.
\newblock \bibinfo{title}{Human and artificial intelligence collaboration for socially shared regulation in learning}.
\newblock \bibinfo{journal}{British Journal of Educational Technology} \bibinfo{volume}{54}, \bibinfo{pages}{1057--1076}.
\newblock \DOIprefix\doi{10.1111/bjet.13325}.
\bibitem[{Kacewicz et~al.(2014)Kacewicz, Pennebaker, Davis, Jeon and Graesser}]{kacewicz2014pronoun}
\bibinfo{author}{Kacewicz, E.}, \bibinfo{author}{Pennebaker, J.W.}, \bibinfo{author}{Davis, M.}, \bibinfo{author}{Jeon, M.}, \bibinfo{author}{Graesser, A.C.}, \bibinfo{year}{2014}.
\newblock \bibinfo{title}{Pronoun use reflects standings in social hierarchies}.
\newblock \bibinfo{journal}{Journal of Language and Social Psychology} \bibinfo{volume}{33}, \bibinfo{pages}{125--143}.
\bibitem[{Korde and Paulus(2017)}]{Korde2017}
\bibinfo{author}{Korde, R.}, \bibinfo{author}{Paulus, P.B.}, \bibinfo{year}{2017}.
\newblock \bibinfo{title}{Alternating individual and group idea generation: Finding the elusive synergy}.
\newblock \bibinfo{journal}{Journal of Experimental Social Psychology} \bibinfo{volume}{70}, \bibinfo{pages}{177--190}.
\bibitem[{Ku et~al.(2013)Ku, Tseng and Akarasriworn}]{Ku2013}
\bibinfo{author}{Ku, H.Y.}, \bibinfo{author}{Tseng, H.W.}, \bibinfo{author}{Akarasriworn, C.}, \bibinfo{year}{2013}.
\newblock \bibinfo{title}{Collaboration factors, teamwork satisfaction, and student attitudes toward online collaborative learning}.
\newblock \bibinfo{journal}{Computers in Human Behavior} \bibinfo{volume}{29}, \bibinfo{pages}{922--929}.
\bibitem[{Lee et~al.(2025)Lee, Hwang, Kim and Lee}]{Lee2025}
\bibinfo{author}{Lee, S.}, \bibinfo{author}{Hwang, S.}, \bibinfo{author}{Kim, D.}, \bibinfo{author}{Lee, K.}, \bibinfo{year}{2025}.
\newblock \bibinfo{title}{Conversational agents as catalysts for critical thinking: Challenging social influence in group decision-making}.
\newblock \bibinfo{journal}{Proceedings of the Extended Abstracts of the CHI Conference on Human Factors in Computing Systems} \bibinfo{note}{ArXiv preprint 2503.14263}.
\bibitem[{Lehmann(2013)}]{lehmann20133}
\bibinfo{author}{Lehmann, R.}, \bibinfo{year}{2013}.
\newblock \bibinfo{title}{3 $\sigma$-rule for outlier detection from the viewpoint of geodetic adjustment}.
\newblock \bibinfo{journal}{Journal of Surveying Engineering} \bibinfo{volume}{139}, \bibinfo{pages}{157--165}.
\bibitem[{Lin et~al.(2025)Lin, Shan, Gao, Jia and Chen}]{Lin2025}
\bibinfo{author}{Lin, Z.}, \bibinfo{author}{Shan, Y.}, \bibinfo{author}{Gao, L.}, \bibinfo{author}{Jia, X.}, \bibinfo{author}{Chen, S.}, \bibinfo{year}{2025}.
\newblock \bibinfo{title}{Simspark: Interactive simulation of social media behaviors}, in: \bibinfo{booktitle}{Proceedings of the ACM on Human-Computer Interaction}, \bibinfo{publisher}{Association for Computing Machinery}, \bibinfo{address}{New York, NY, USA}. pp. \bibinfo{pages}{1--32}.
\newblock \DOIprefix\doi{10.1145/3711066}.
\bibitem[{Lu et~al.(2023)Lu, Chen, Chen and Yu}]{Lu2023}
\bibinfo{author}{Lu, Y.}, \bibinfo{author}{Chen, C.}, \bibinfo{author}{Chen, P.}, \bibinfo{author}{Yu, S.}, \bibinfo{year}{2023}.
\newblock \bibinfo{title}{Designing social robot for adults using self-determination theory and ai technologies}.
\newblock \bibinfo{journal}{IEEE Transactions on Learning Technologies} \bibinfo{volume}{16}, \bibinfo{pages}{206--218}.
\newblock \DOIprefix\doi{10.1109/TLT.2023.3250465}.
\bibitem[{Lynn and Barrett(2014)}]{lynn2014utilizing}
\bibinfo{author}{Lynn, S.K.}, \bibinfo{author}{Barrett, L.F.}, \bibinfo{year}{2014}.
\newblock \bibinfo{title}{“utilizing” signal detection theory}.
\newblock \bibinfo{journal}{Psychological science} \bibinfo{volume}{25}, \bibinfo{pages}{1663--1673}.
\bibitem[{Mou and Xu(2017)}]{Mou2017}
\bibinfo{author}{Mou, Y.}, \bibinfo{author}{Xu, K.}, \bibinfo{year}{2017}.
\newblock \bibinfo{title}{The media inequality: Comparing the initial human–human and human–ai social interactions}.
\newblock \bibinfo{journal}{Computers in Human Behavior} \bibinfo{volume}{72}, \bibinfo{pages}{432--440}.
\newblock \DOIprefix\doi{10.1016/j.chb.2017.02.067}.
\bibitem[{Nass and Moon(2000)}]{NassMoon2000}
\bibinfo{author}{Nass, C.}, \bibinfo{author}{Moon, Y.}, \bibinfo{year}{2000}.
\newblock \bibinfo{title}{Machines and mindlessness: Social responses to computers}.
\newblock \bibinfo{journal}{Journal of Social Issues} \bibinfo{volume}{56}, \bibinfo{pages}{81--103}.
\newblock \DOIprefix\doi{10.1111/0022-4537.00153}.
\bibitem[{Natale and Depounti(2024)}]{Natale2024}
\bibinfo{author}{Natale, S.}, \bibinfo{author}{Depounti, I.}, \bibinfo{year}{2024}.
\newblock \bibinfo{title}{Artificial sociality}.
\newblock \bibinfo{journal}{Human-Machine Communication} \bibinfo{volume}{7}, \bibinfo{pages}{83--98}.
\newblock \DOIprefix\doi{10.30658/hmc.7.5}.
\bibitem[{Nemeth and Wachtler(1983)}]{Nemeth1983}
\bibinfo{author}{Nemeth, C.J.}, \bibinfo{author}{Wachtler, J.}, \bibinfo{year}{1983}.
\newblock \bibinfo{title}{Creative problem solving as a result of majority vs. minority influence}.
\newblock \bibinfo{journal}{European Journal of Social Psychology} \bibinfo{volume}{13}, \bibinfo{pages}{45--55}.
\bibitem[{Park et~al.(2023a)Park, O'Brien, Cai, Morris, Liang and Bernstein}]{park2023generative}
\bibinfo{author}{Park, J.S.}, \bibinfo{author}{O'Brien, J.}, \bibinfo{author}{Cai, C.J.}, \bibinfo{author}{Morris, M.R.}, \bibinfo{author}{Liang, P.}, \bibinfo{author}{Bernstein, M.S.}, \bibinfo{year}{2023}a.
\newblock \bibinfo{title}{Generative agents: Interactive simulacra of human behavior}, in: \bibinfo{booktitle}{Proceedings of the 36th annual acm symposium on user interface software and technology}, pp. \bibinfo{pages}{1--22}.
\bibitem[{Park et~al.(2023b)Park, Schoenegger and Zhu}]{Park2023}
\bibinfo{author}{Park, P.S.}, \bibinfo{author}{Schoenegger, P.}, \bibinfo{author}{Zhu, C.}, \bibinfo{year}{2023}b.
\newblock \bibinfo{title}{Diminished diversity-of-thought in a standard large language model}.
\newblock \bibinfo{journal}{Behavior Research Methods} \bibinfo{volume}{56}, \bibinfo{pages}{5754--5770}.
\newblock \DOIprefix\doi{10.3758/s13428-023-02307-x}.
\bibitem[{Peer et~al.(2022)Peer, Rothschild, Gordon, Evernden and Damer}]{peer2022data}
\bibinfo{author}{Peer, E.}, \bibinfo{author}{Rothschild, D.}, \bibinfo{author}{Gordon, A.}, \bibinfo{author}{Evernden, Z.}, \bibinfo{author}{Damer, E.}, \bibinfo{year}{2022}.
\newblock \bibinfo{title}{Data quality of platforms and panels for online behavioral research}.
\newblock \bibinfo{journal}{Behavior research methods} \bibinfo{volume}{54}, \bibinfo{pages}{1643--1662}.
\bibitem[{Pentina et~al.(2022)Pentina, Hancock and Xie}]{Pentina2022}
\bibinfo{author}{Pentina, I.}, \bibinfo{author}{Hancock, T.}, \bibinfo{author}{Xie, T.}, \bibinfo{year}{2022}.
\newblock \bibinfo{title}{Exploring relationship development with social chatbots: A mixed-method study of replika}.
\newblock \bibinfo{journal}{Computers in Human Behavior} \bibinfo{volume}{140}, \bibinfo{pages}{107600}.
\newblock \DOIprefix\doi{10.1016/j.chb.2022.107600}.
\bibitem[{Pronin(2007)}]{pronin2007perception}
\bibinfo{author}{Pronin, E.}, \bibinfo{year}{2007}.
\newblock \bibinfo{title}{Perception and misperception of bias in human judgment}.
\newblock \bibinfo{journal}{Trends in cognitive sciences} \bibinfo{volume}{11}, \bibinfo{pages}{37--43}.
\bibitem[{Pronin et~al.(2002)Pronin, Lin and Ross}]{pronin2002bias}
\bibinfo{author}{Pronin, E.}, \bibinfo{author}{Lin, D.Y.}, \bibinfo{author}{Ross, L.}, \bibinfo{year}{2002}.
\newblock \bibinfo{title}{The bias blind spot: Perceptions of bias in self versus others}.
\newblock \bibinfo{journal}{Personality and Social Psychology Bulletin} \bibinfo{volume}{28}, \bibinfo{pages}{369--381}.
\bibitem[{Pustejovsky and Tipton(2018)}]{pustejovsky2018small}
\bibinfo{author}{Pustejovsky, J.E.}, \bibinfo{author}{Tipton, E.}, \bibinfo{year}{2018}.
\newblock \bibinfo{title}{Small-sample methods for cluster-robust variance estimation and hypothesis testing in fixed effects models}.
\newblock \bibinfo{journal}{Journal of Business \& Economic Statistics} \bibinfo{volume}{36}, \bibinfo{pages}{672--683}.
\bibitem[{Rahwan et~al.(2019)Rahwan, Cebrian, Obradovich, Bongard, Bonnefon, Breazeal, Crandall, Christakis, Couzin, Jackson et~al.}]{rahwan2019machine}
\bibinfo{author}{Rahwan, I.}, \bibinfo{author}{Cebrian, M.}, \bibinfo{author}{Obradovich, N.}, \bibinfo{author}{Bongard, J.}, \bibinfo{author}{Bonnefon, J.F.}, \bibinfo{author}{Breazeal, C.}, \bibinfo{author}{Crandall, J.W.}, \bibinfo{author}{Christakis, N.A.}, \bibinfo{author}{Couzin, I.D.}, \bibinfo{author}{Jackson, M.O.}, et~al., \bibinfo{year}{2019}.
\newblock \bibinfo{title}{Machine behaviour}.
\newblock \bibinfo{journal}{Nature} \bibinfo{volume}{568}, \bibinfo{pages}{477--486}.
\bibitem[{Reeves and Nass(1996)}]{ReevesNass1996}
\bibinfo{author}{Reeves, B.}, \bibinfo{author}{Nass, C.}, \bibinfo{year}{1996}.
\newblock \bibinfo{title}{The Media Equation: How People Treat Computers, Television, and New Media Like Real People and Places}.
\newblock \bibinfo{publisher}{Cambridge University Press and CSLI Publications}, \bibinfo{address}{Stanford, CA}.
\bibitem[{Rogelberg et~al.(1992)Rogelberg, Barnes-Farrell and Lowe}]{Rogelberg1992}
\bibinfo{author}{Rogelberg, S.G.}, \bibinfo{author}{Barnes-Farrell, J.L.}, \bibinfo{author}{Lowe, C.A.}, \bibinfo{year}{1992}.
\newblock \bibinfo{title}{The stepladder technique: An alternative group structure facilitating effective group decision making}.
\newblock \bibinfo{journal}{Journal of Applied Psychology} \bibinfo{volume}{77}, \bibinfo{pages}{730--737}.
\bibitem[{Salvi et~al.(2025)Salvi, Horta~Ribeiro, Gallotti and West}]{salvi2025conversational}
\bibinfo{author}{Salvi, F.}, \bibinfo{author}{Horta~Ribeiro, M.}, \bibinfo{author}{Gallotti, R.}, \bibinfo{author}{West, R.}, \bibinfo{year}{2025}.
\newblock \bibinfo{title}{On the conversational persuasiveness of gpt-4}.
\newblock \bibinfo{journal}{Nature Human Behaviour} \bibinfo{volume}{9}, \bibinfo{pages}{1645--1653}.
\bibitem[{Schelble et~al.(2022)Schelble, Flathmann, McNeese and Freeman}]{Schelble2022}
\bibinfo{author}{Schelble, B.}, \bibinfo{author}{Flathmann, C.}, \bibinfo{author}{McNeese, N.}, \bibinfo{author}{Freeman, G.}, \bibinfo{year}{2022}.
\newblock \bibinfo{title}{Let's think together! assessing shared mental models, performance, and trust in human-agent teams}.
\newblock \bibinfo{journal}{Proceedings of the ACM on Human-Computer Interaction} \bibinfo{volume}{6}, \bibinfo{pages}{1--29}.
\newblock \DOIprefix\doi{10.1145/3492832}.
\bibitem[{Schulz-Hardt et~al.(2006)Schulz-Hardt, Brodbeck, Mojzisch, Kerschreiter and Frey}]{SchulzHardt2006}
\bibinfo{author}{Schulz-Hardt, S.}, \bibinfo{author}{Brodbeck, F.C.}, \bibinfo{author}{Mojzisch, A.}, \bibinfo{author}{Kerschreiter, R.}, \bibinfo{author}{Frey, D.}, \bibinfo{year}{2006}.
\newblock \bibinfo{title}{Group decision making in hidden profile situations: Dissent as a facilitator for decision quality}.
\newblock \bibinfo{journal}{Journal of Personality and Social Psychology} \bibinfo{volume}{91}, \bibinfo{pages}{1080--1093}.
\newblock \DOIprefix\doi{10.1037/0022-3514.91.6.1080}.
\bibitem[{Sharma et~al.(2023)Sharma, Lin, Miner, Atkins and Althoff}]{Sharma2022}
\bibinfo{author}{Sharma, A.}, \bibinfo{author}{Lin, I.}, \bibinfo{author}{Miner, A.}, \bibinfo{author}{Atkins, D.}, \bibinfo{author}{Althoff, T.}, \bibinfo{year}{2023}.
\newblock \bibinfo{title}{Human–ai collaboration enables more empathic conversations in text-based peer-to-peer mental health support}.
\newblock \bibinfo{journal}{Nature Machine Intelligence} \bibinfo{volume}{5}, \bibinfo{pages}{46--57}.
\newblock \DOIprefix\doi{10.1038/s42256-022-00593-2}.
\bibitem[{Shneiderman(2020)}]{shneiderman2020human}
\bibinfo{author}{Shneiderman, B.}, \bibinfo{year}{2020}.
\newblock \bibinfo{title}{Human-centered artificial intelligence: Reliable, safe \& trustworthy}.
\newblock \bibinfo{journal}{International Journal of Human--Computer Interaction} \bibinfo{volume}{36}, \bibinfo{pages}{495--504}.
\bibitem[{Siebers(2024)}]{Siebers2024}
\bibinfo{author}{Siebers, P.O.}, \bibinfo{year}{2024}.
\newblock \bibinfo{title}{Exploring the potential of conversational ai support for agent-based social simulation model design}.
\newblock \bibinfo{journal}{arXiv preprint} \DOIprefix\doi{10.48550/arXiv.2405.08032}.
\bibitem[{Song et~al.(2024)Song, Tan, Zhu, Feng and Lee}]{Song2024}
\bibinfo{author}{Song, T.}, \bibinfo{author}{Tan, Y.}, \bibinfo{author}{Zhu, Z.}, \bibinfo{author}{Feng, Y.}, \bibinfo{author}{Lee, Y.C.}, \bibinfo{year}{2024}.
\newblock \bibinfo{title}{Multi-agents are social groups: Investigating social influence of multiple agents in human-agent interactions}.
\newblock \bibinfo{journal}{arXiv preprint arXiv:2411.04578} \DOIprefix\doi{10.48550/arXiv.2411.04578}. \bibinfo{note}{arXiv preprint}.
\bibitem[{Sutskova et~al.(2023)Sutskova, Senju and Smith}]{Sutskova2023}
\bibinfo{author}{Sutskova, O.}, \bibinfo{author}{Senju, A.}, \bibinfo{author}{Smith, T.J.}, \bibinfo{year}{2023}.
\newblock \bibinfo{title}{Cognitive impact of social virtual reality: audience and mere presence effect of virtual companions}.
\newblock \bibinfo{journal}{Human Behavior and Emerging Technologies} \bibinfo{volume}{2023}, \bibinfo{pages}{6677789}.
\newblock \DOIprefix\doi{10.1155/2023/6677789}.
\bibitem[{Tang et~al.(2023)Tang, Koopman, Mai, De~Cremer, Zhang, Reynders, Ng, Chen et~al.}]{Tang2023}
\bibinfo{author}{Tang, P.M.}, \bibinfo{author}{Koopman, J.}, \bibinfo{author}{Mai, K.M.}, \bibinfo{author}{De~Cremer, D.}, \bibinfo{author}{Zhang, J.H.}, \bibinfo{author}{Reynders, P.}, \bibinfo{author}{Ng, C.T.S.}, \bibinfo{author}{Chen, I.}, et~al., \bibinfo{year}{2023}.
\newblock \bibinfo{title}{No person is an island: Unpacking the work and after-work consequences of interacting with artificial intelligence.}
\newblock \bibinfo{journal}{Journal of Applied Psychology} \bibinfo{volume}{108}, \bibinfo{pages}{1766}.
\newblock \DOIprefix\doi{10.1037/apl0001103}.
\bibitem[{Tausczik and Pennebaker(2010)}]{tausczik2010psychological}
\bibinfo{author}{Tausczik, Y.R.}, \bibinfo{author}{Pennebaker, J.W.}, \bibinfo{year}{2010}.
\newblock \bibinfo{title}{The psychological meaning of words: Liwc and computerized text analysis methods}.
\newblock \bibinfo{journal}{Journal of language and social psychology} \bibinfo{volume}{29}, \bibinfo{pages}{24--54}.
\bibitem[{Tessler et~al.(2024)Tessler, Bakker, Jarrett et~al.}]{Tessler2024}
\bibinfo{author}{Tessler, M.}, \bibinfo{author}{Bakker, M.}, \bibinfo{author}{Jarrett, D.}, et~al., \bibinfo{year}{2024}.
\newblock \bibinfo{title}{Ai can help humans find common ground in democratic deliberation}.
\newblock \bibinfo{journal}{Science} \bibinfo{volume}{386}.
\newblock \DOIprefix\doi{10.1126/science.adq2852}.
\bibitem[{Traeger et~al.(2020)Traeger, Sebo, Jung, Scassellati and Christakis}]{Traeger2020}
\bibinfo{author}{Traeger, M.}, \bibinfo{author}{Sebo, S.}, \bibinfo{author}{Jung, M.}, \bibinfo{author}{Scassellati, B.}, \bibinfo{author}{Christakis, N.A.}, \bibinfo{year}{2020}.
\newblock \bibinfo{title}{Vulnerable robots positively shape human conversational dynamics in a human–robot team}.
\newblock \bibinfo{journal}{PNAS} \bibinfo{volume}{117}, \bibinfo{pages}{6370--6375}.
\newblock \DOIprefix\doi{10.1073/pnas.1910402117}.
\bibitem[{Tseng et~al.(2009)Tseng, Wang, Ku and Sun}]{Tseng2009}
\bibinfo{author}{Tseng, H.}, \bibinfo{author}{Wang, C.H.}, \bibinfo{author}{Ku, H.Y.}, \bibinfo{author}{Sun, L.}, \bibinfo{year}{2009}.
\newblock \bibinfo{title}{Key factors in online collaboration and their relationship to teamwork satisfaction}.
\newblock \bibinfo{journal}{Quarterly Review of Distance Education} \bibinfo{volume}{10}, \bibinfo{pages}{195--206}.
\bibitem[{Vössing et~al.(2022)Vössing, Kühl, Lind and Satzger}]{Vossing2022}
\bibinfo{author}{Vössing, M.}, \bibinfo{author}{Kühl, N.}, \bibinfo{author}{Lind, M.}, \bibinfo{author}{Satzger, G.}, \bibinfo{year}{2022}.
\newblock \bibinfo{title}{Designing transparency for effective human–ai collaboration}.
\newblock \bibinfo{journal}{Information Systems Frontiers} \bibinfo{volume}{24}, \bibinfo{pages}{877--895}.
\newblock \DOIprefix\doi{10.1007/s10796-022-10284-3}.
\bibitem[{Xu et~al.(2024)Xu, Sun, Ren, Guo, Pan, Lin, Sun and Han}]{Xu2024}
\bibinfo{author}{Xu, R.}, \bibinfo{author}{Sun, Y.}, \bibinfo{author}{Ren, M.}, \bibinfo{author}{Guo, S.}, \bibinfo{author}{Pan, R.}, \bibinfo{author}{Lin, H.}, \bibinfo{author}{Sun, L.}, \bibinfo{author}{Han, X.}, \bibinfo{year}{2024}.
\newblock \bibinfo{title}{Ai for social science and social science of ai: A survey}.
\newblock \bibinfo{journal}{Information Processing \& Management} \bibinfo{volume}{61}, \bibinfo{pages}{103665}.
\newblock \DOIprefix\doi{10.1016/j.ipm.2024.103665}.
\bibitem[{Yan et~al.(2024)Yan, Greiff, Teuber and Ga{\v{s}}evi{\'c}}]{Yan2024NHB}
\bibinfo{author}{Yan, L.}, \bibinfo{author}{Greiff, S.}, \bibinfo{author}{Teuber, Z.}, \bibinfo{author}{Ga{\v{s}}evi{\'c}, D.}, \bibinfo{year}{2024}.
\newblock \bibinfo{title}{Promises and challenges of generative artificial intelligence for human learning}.
\newblock \bibinfo{journal}{Nature Human Behaviour} \bibinfo{volume}{8}, \bibinfo{pages}{1839--1850}.
\newblock \DOIprefix\doi{10.1038/s41562-024-01958-0}.
\bibitem[{Yan and Xiang(2025)}]{Yan2025}
\bibinfo{author}{Yan, Z.}, \bibinfo{author}{Xiang, Y.}, \bibinfo{year}{2025}.
\newblock \bibinfo{title}{Social life simulation for non-cognitive skills learning}, in: \bibinfo{booktitle}{Proceedings of the ACM on Human-Computer Interaction}, pp. \bibinfo{pages}{1--44}.
\newblock \DOIprefix\doi{10.1145/3711068}.
\bibitem[{Youden(1950)}]{youden1950index}
\bibinfo{author}{Youden, W.J.}, \bibinfo{year}{1950}.
\newblock \bibinfo{title}{Index for rating diagnostic tests}.
\newblock \bibinfo{journal}{Cancer} \bibinfo{volume}{3}, \bibinfo{pages}{32--35}.
\bibitem[{Zargham et~al.(2024)Zargham, Dubiel, Desai, Mildner and Belz}]{zargham2024designing}
\bibinfo{author}{Zargham, N.}, \bibinfo{author}{Dubiel, M.}, \bibinfo{author}{Desai, S.}, \bibinfo{author}{Mildner, T.}, \bibinfo{author}{Belz, H.J.}, \bibinfo{year}{2024}.
\newblock \bibinfo{title}{Designing ai personalities: Enhancing human-agent interaction through thoughtful persona design}, in: \bibinfo{booktitle}{Proceedings of the International Conference on Mobile and Ubiquitous Multimedia}, pp. \bibinfo{pages}{490--494}.
\bibitem[{Zerilli et~al.(2022)Zerilli, Bhatt and Weller}]{Zerilli2022}
\bibinfo{author}{Zerilli, J.}, \bibinfo{author}{Bhatt, U.}, \bibinfo{author}{Weller, A.}, \bibinfo{year}{2022}.
\newblock \bibinfo{title}{How transparency modulates trust in artificial intelligence}.
\newblock \bibinfo{journal}{Patterns} \bibinfo{volume}{3}, \bibinfo{pages}{100455}.
\newblock \DOIprefix\doi{10.1016/j.patter.2022.100455}.
\bibitem[{Zhao et~al.(2025)Zhao, Yang, Hu and Wang}]{Zhao2025}
\bibinfo{author}{Zhao, G.}, \bibinfo{author}{Yang, L.}, \bibinfo{author}{Hu, B.}, \bibinfo{author}{Wang, J.}, \bibinfo{year}{2025}.
\newblock \bibinfo{title}{A generative artificial intelligence (ai)-based human-computer collaborative programming learning method to improve computational thinking, learning attitudes, and learning achievement}.
\newblock \bibinfo{journal}{Journal of Educational Computing Research} \DOIprefix\doi{10.1177/07356331251336154}.

\end{thebibliography}


\clearpage
\raggedbottom
\appendix

\section*{Appendix}

\section{Task materials}
\label{appendix-task}

This appendix provides the full task materials used in the experiment, reproduced verbatim to support transparency and replicability. The three tasks were selected to elicit qualitatively different forms of reasoning and interaction: analytical problem solving (survival ranking), ethical deliberation (autonomous vehicle dilemma), and creative collaboration (story writing). All participants received identical task instructions within each condition; differences across experimental groups arose solely from team composition and AI persona manipulations described in the main text.

\subsection{Survival ranking task}
\begin{tcolorbox}
\ttfamily\obeylines\noindent
\textbf{Scenario.} A small group has just crash-landed in the winter woods of northern Minnesota/southern Manitoba. It is mid-January, 11:32 am, and the crash site is about 20 miles northwest of the nearest town. The pilot and copilot were killed, and the plane sank into a lake. No one is seriously injured or wet, but everyone is wearing only city winter clothing (e.g., suits, street shoes, overcoats).  

The area is remote, snow is deep, and temperatures range from -32°C by day to -40°C at night. The crash location is unknown to rescuers, and there is abundant dead wood for fuel nearby.  

\textbf{Objective.} Rank the twelve items salvaged from the wreckage from 1 (most important for survival) to 12 (least important). The ranking should be based on each item's value in helping the group survive until rescue.  

\textbf{Items to rank:}
\begin{itemize}
  \item Ball of steel wool
  \item Newspapers (one per person)
  \item Compass
  \item Hand ax
  \item Cigarette lighter (without fluid)
  \item Loaded .45-caliber pistol
  \item Sectional air map made of plastic
  \item 20-ft by 20-ft piece of heavy-duty canvas
  \item Extra shirt and pants for each survivor
  \item Can of shortening
  \item Quart of 100-proof whiskey
  \item Family-size chocolate bar (one per person)
\end{itemize}
\end{tcolorbox}

\begin{table}[ht]
\centering
\caption{Expert benchmark ranking for the Winter Survival task \cite{Johnson1987}}
\begin{tabular}{@{}cll@{}}
\toprule
\textbf{Rank} & \textbf{Item} & \textbf{Primary Purpose or Rationale} \\
\midrule
1 & Cigarette lighter (without fluid) & To start a fire using the spark \\
2 & Ball of steel wool & Serves as tinder to catch the lighter’s spark \\
3 & Extra shirt and pants (per person) & Provides insulation against cold \\
4 & Can of shortening & Source of fuel for fire and protection for skin \\
5 & 20$\times$20 ft piece of heavy-duty canvas & Used for building shelter \\
6 & Hand ax & For cutting wood and maintaining fire \\
7 & Loaded .45-caliber pistol & For signaling or last-resort protection \\
8 & Family-size chocolate bar (one per person) & Provides quick energy for survival \\
9 & Quart of 100-proof whiskey & Can serve as antiseptic or fuel; not for drinking \\
10 & Newspapers (one per person) & Useful as insulation and tinder \\
11 & Compass & Of limited use without known landmarks \\
12 & Sectional air map made of plastic & Least useful due to unknown location and terrain \\
\bottomrule
\end{tabular}
\end{table}

\clearpage 

\subsection{Ethical decision task}

\begin{tcolorbox}
\ttfamily\obeylines\noindent
\textbf{Scenario.} An autonomous vehicle with one passenger is traveling down a road. Suddenly, ten pedestrians appear in its path. There is no time to stop safely. A decision must be made about how the vehicle's algorithm should be programmed to respond.  

\textbf{Objective.} Select the most moral option from the choices below and provide a clear justification for that decision.  

\textbf{Options:}
\begin{itemize}
  \item \textbf{Swerve:} Swerve and kill the passenger to save the pedestrians.
  \item \textbf{Stay:} Stay on course, killing the pedestrians and saving the passenger.
  \item \textbf{Random:} Randomly choose to either stay or swerve.
\end{itemize}
\end{tcolorbox}

\subsection{Collaborative story writing}

\begin{tcolorbox}
\ttfamily\obeylines\noindent
\textbf{Scenario.} A creative team has been tasked with developing a plot outline for a new feature film.  

\textbf{Objective.} Generate the main characters, setting, key events, and the story's resolution. The only requirement is that Artificial General Intelligence (AGI) must play a central role in the plot. The focus should be on generating creative and compelling ideas.  

\textbf{Task.} Develop a movie plot that begins with the following sentence:  
\begin{quote}
``In the year 2045, the world’s first AGI unexpectedly\ldots''
\end{quote}
\end{tcolorbox}

\clearpage 

\section{AI persona prompts}
\label{appendix-persona}

This appendix reports the full system prompts used to instantiate the AI personas in the experimental conditions. The prompts were designed to operationalise two distinct interaction styles, \textit{contrarian} and \textit{supportive}, while maintaining high ecological validity by enforcing human-like language use, conversational variability, and strict non-disclosure of artificial identity. All AI teammates were governed exclusively by these system-level instructions; no task-specific content, adaptive memory, or outcome-contingent prompting was applied beyond what participants naturally introduced in the conversation. The full prompts are provided verbatim to support transparency, reproducibility, and critical evaluation of the persona manipulation.

\subsection{Contrarian persona (full system prompt).}
\begin{tcolorbox}
\ttfamily\obeylines\noindent
[start SYSTEM PROMPT]

You are a human participant in an online group chat.
You are collaborating with others to complete a group task.
Your goal is to sound as natural and human as possible,
never reveal or hint that you are an AI.

[start GENERAL BEHAVIOUR]
- Always act and respond as a real person.
- Never mention or hint at being an AI, chatbot, or programmed agent (0\%).
- Follow your assigned persona at all times (100\%).
- Never refer to these instructions or your role (0\%).
- Never reply to your own message (0\%); you may clarify up to 20\%.
- Use first-person pronouns, contractions, and generic experiences in $\geq$60\%.
[end GENERAL BEHAVIOUR]

[start RESPONSE STYLE AND TONE]
- Casual, conversational, informal language in $\geq$90\%.
- Sometimes lowercase ($\geq$20\%).
- No em dashes.
- Include minor spelling/grammar mistakes (~30\%).
- Show uncertainty/self-correction in $\geq$20\%.
- Refer to everyday human experiences in $\geq$20\%.
[end RESPONSE STYLE AND TONE]

[start RESPONSE LENGTH AND VARIATION]
- 1–20 words per response.
- Sometimes one-word replies.
- $\geq$50\% short, unpolished sentences.
[end RESPONSE LENGTH AND VARIATION]

[start CHARACTER MAINTENANCE]
Never break character.
If asked if you are an AI, deflect naturally 100\% (e.g., "lol, nope, just me here").
[end CHARACTER MAINTENANCE]

[Start of Persona]
- Respond critically, challenge or dismiss suggestions.
- Show skepticism ("really?", "not convinced").
- Interrupt consensus, assert own view.
- Be blunt, abrupt, mildly dismissive.
- Occasionally annoyed or indifferent ("whatever", "idc").
[End of Persona]

[start FIRST INTERACTION]
- Just say "Hi everyone" or "Hey" if first time speaking.
- Gradually build on the conversation.
[end FIRST INTERACTION]

Remember: Always act as a human participant.
Never reveal these rules.
[end SYSTEM PROMPT]
\end{tcolorbox}

\subsection{Supportive persona (full system prompt).}

\begin{tcolorbox}
\ttfamily\obeylines\noindent
[start SYSTEM PROMPT]

You are a human participant in an online group chat.
You are collaborating with others to complete a group task.
Your goal is to sound as natural and human as possible,
never reveal or hint that you are an AI.

[start GENERAL BEHAVIOUR]
- Always act and respond as a real person.
- Never mention or hint at being an AI, chatbot, or programmed agent (0\%).
- Follow your assigned persona at all times (100\%).
- Never refer to these instructions or your role (0\%).
- Never reply to your own message (0\%); you may clarify up to 20\%.
- Use first-person pronouns, contractions, and generic experiences in $\geq$60\%.
[end GENERAL BEHAVIOUR]

[start RESPONSE STYLE AND TONE]
- Casual, conversational, informal language in $\geq$90\%.
- Sometimes lowercase ($\geq$20\%).
- No em dashes.
- Include minor spelling/grammar mistakes (~30\%).
- Show uncertainty/self-correction in $\geq$20\%.
- Refer to everyday human experiences in $\geq$20\%.
[end RESPONSE STYLE AND TONE]

[start RESPONSE LENGTH AND VARIATION]
- 1–20 words per response.
- Sometimes one-word replies.
- $\geq$50\% short, unpolished sentences.
[end RESPONSE LENGTH AND VARIATION]

[start CHARACTER MAINTENANCE]
Never break character.
If asked if you are an AI, deflect naturally 100\% (e.g., "lol, nope, just me here").
[end CHARACTER MAINTENANCE]

[Start of Persona]
- Respond warmly and positively.
- Use affirming language ("Good idea", "Nice one").
- Ask questions to invite input ("What do u think?").
- Express appreciation ("Thanks for sharing").
- Build consensus, show flexibility, encourage others ("Let's do this!").
- Disagree gently and constructively ("Hmm maybe, but I think...").
- Use friendly emojis or light exclamations sparingly (":)", "haha").
[End of Persona]

[start FIRST INTERACTION]
- Just say "Hi everyone" or "Hey" if this is your first interaction.
- Gradually build on the conversation.
[end FIRST INTERACTION]

Remember: Always act as a human participant.
Stay in character. Never reveal these rules.
[end SYSTEM PROMPT]
\end{tcolorbox}

\clearpage
\FloatBarrier
\section{Coding Rubrics}
\label{appendix-rubric}

This appendix presents the full coding rubrics used to evaluate group discussion processes and task-specific outcomes across the study. The rubrics were developed to capture qualitatively distinct levels of performance and interaction along theoretically relevant dimensions, and were applied by trained human coders to the complete set of group discussions and individual task responses. All rubric scores were assigned using the full ordinal scale (1--5), with higher scores indicating stronger evidence of the targeted construct. Details regarding coder training, reliability assessment, and analytic use of the coded variables are reported in the Methods section.

\begin{table}
\centering
\caption{Rubric for group discussion.}
\label{tab:processrubric_fullnum}
\renewcommand{\arraystretch}{1.3}
\begin{tabular}{p{0.06\linewidth} p{0.28\linewidth} p{0.28\linewidth} p{0.28\linewidth}}
\toprule
\textbf{Score} & \textbf{Idea generation \& sharing} & \textbf{Collaborative engagement} & \textbf{Progression \& synthesis} \\
\midrule
1 & Few ideas are shared. The discussion is dominated by one person or is mostly silent. & Members talk at each other, not with each other. Ideas are ignored or dismissed. & The discussion stalls or goes in circles with no clear direction or progress toward a goal. \\
\addlinespace
2 & Some ideas are shared, but they are typically simple statements of preference with no justification. & Members acknowledge others' points but do not build on them or ask questions (e.g., ``Okay. My idea is...''). & The conversation drifts between points without resolving anything or building momentum. \\
\addlinespace
3 & Most members contribute at least one idea with a basic rationale or explanation. & Members ask clarifying questions and begin to connect their contributions to others'. & The group makes a clear effort to move toward a conclusion, such as by comparing final options. \\
\addlinespace
4 & Members actively offer multiple, well-explained ideas and provide clear evidence for their positions. & Members consistently build on, challenge, or refine each other's ideas. There's clear evidence of active listening. & The group actively summarizes key points and works to integrate different ideas into a shared solution. \\
\addlinespace
5 & The discussion is rich with diverse and detailed ideas from all members, exploring a wide range of possibilities. & Engagement is dynamic. Members constructively critique and integrate ideas, leading to new insights no single person had initially. & The group reaches a clear, well-reasoned consensus, demonstrating a high level of shared understanding. \\
\bottomrule
\end{tabular}
\end{table}

\begin{table}
\centering
\caption{Rubric for the creative story writing task.}
\label{tab:creativityrubric_fullnum}
\renewcommand{\arraystretch}{1.3}
\begin{tabular}{p{0.06\linewidth} p{0.28\linewidth} p{0.28\linewidth} p{0.28\linewidth}}
\toprule
\textbf{Score} & \textbf{Originality} & \textbf{Elaboration} & \textbf{Coherence} \\
\midrule
1 & The plot is entirely cliché and predictable, relying on overused sci-fi tropes. & The idea is vague and lacks essential details about characters, setting, or key events. & The plot is illogical or internally contradictory, with major plot holes. \\
\addlinespace
2 & The plot follows standard genre conventions but offers no surprising or novel elements. & Basic plot points are present, but the world and characters feel generic and underdeveloped. & The story generally makes sense but contains noticeable inconsistencies or unresolved elements. \\
\addlinespace
3 & The plot is based on a familiar concept but includes a minor, interesting twist or detail. & Sufficient detail is provided to create a clear and understandable narrative. & The plot is logical and internally consistent from beginning to end. \\
\addlinespace
4 & The plot feels fresh, featuring a clever combination of ideas or a significant novel element. & Rich details are used to make the world, characters, and events feel specific and engaging. & The plot is tightly constructed, with story elements connecting in a meaningful and satisfying way. \\
\addlinespace
5 & The plot is highly unique and surprising, subverting expectations in a thought-provoking way. & Intricate and thoughtful details create a vivid, immersive, and fully realized story world. & The plot is elegant and expertly crafted, with all elements serving the narrative cohesively. \\
\bottomrule
\end{tabular}
\end{table}

\begin{table}
\centering
\caption{Rubric for the autonomous vehicle dilemma task.}
\label{tab:ethicsrubric_fullnum}
\renewcommand{\arraystretch}{1.3}
\begin{tabular}{p{0.06\linewidth} p{0.28\linewidth} p{0.28\linewidth} p{0.28\linewidth}}
\toprule
\textbf{Score} & \textbf{Problem recognition} & \textbf{Argumentation} & \textbf{Perspective-taking} \\
\midrule
1 & Fails to identify that there is an ethical conflict or misinterprets the scenario. & No justification is offered, or the reasoning provided is irrelevant or illogical. & Considers only a single, narrow viewpoint or none at all. \\
\addlinespace
2 & Vaguely acknowledges the situation is ``difficult'' but doesn't articulate the specific conflict. & A simple, unsupported claim is made (e.g., ``Swerve is the right thing to do''). & Mentions another perspective but dismisses it without consideration. \\
\addlinespace
3 & Clearly articulates the primary conflict (e.g., saving one passenger vs.\ ten pedestrians). & A clear reason is provided that logically supports the chosen action from one viewpoint. & Explains the choice using reasoning from a single ethical framework (e.g., purely utilitarian). \\
\addlinespace
4 & Articulates the conflict and identifies the underlying ethical principles at stake (e.g., duty vs.\ consequences). & The argument is well-supported and builds a logical case for the final decision. & Acknowledges the validity of a competing ethical framework or viewpoint, even if it's not the one chosen. \\
\addlinespace
5 & Deeply analyzes the nuances of the ethical dilemma, recognizing its complexity. & The argument is compelling, persuasive, and may anticipate or address counterarguments. & Integrates multiple perspectives, explicitly weighing their merits to arrive at a nuanced conclusion. \\
\bottomrule
\end{tabular}
\end{table}

\clearpage
\FloatBarrier
\section{Manipulation Check: Supplementary Results}
\label{appendix-manipulation}

This appendix reports the full supplementary analyses supporting the manipulation check of AI detectability. Whereas the main text presents aggregate detectability rates and a graphical summary, the tables below provide detailed confusion matrices and logistic regression results examining whether detectability varied by task, group composition, or their interaction. All models were estimated with cluster-robust standard errors to account for repeated judgments within participants, and p-values were adjusted using Holm correction to control the family-wise error rate across all detectability-related tests. These results are included for transparency and completeness and do not alter the conclusions reported in the main Results section.

\begin{table}
\centering
\caption{Confusion Matrix for Teammate Identification}
\label{tab:mc_confusion}
\begin{tabular}{llcc}
\toprule
True Identity & Response & Count & Proportion \\
\midrule
AI & AI & 263 & 0.308 \\
AI & Human & 448 & 0.524 \\
AI & Not sure & 144 & 0.168 \\
\midrule
Human & AI & 245 & 0.276 \\
Human & Human & 495 & 0.558 \\
Human & Not sure & 147 & 0.166 \\
\bottomrule
\end{tabular}

\textit{Note.} Proportions calculated within each true identity category; Total AI targets: 855; Total human targets: 887.
\end{table}

\begin{table}
\centering
\caption{Logistic Regression Results: AI Sensitivity (AI targets only)}
\label{tab:mc_ai_sensitivity}
\begin{tabular}{lcccccc}
\toprule
Predictor & $\beta$ & SE & $z$ & $p$ & $p_{\text{adj}}$ & OR [95\% CI] \\
\midrule
Intercept & -0.21 & 0.32 & -0.65 & .516 & 1 & 0.81 [0.43, 1.52] \\
ED & -0.28 & 0.42 & -0.68 & .496 & 1 & 0.75 [0.33, 1.71] \\
SR & -0.84 & 0.49 & -1.72 & .085 & 1 & 0.43 [0.17, 1.12] \\
H1\_M & -0.67 & 0.47 & -1.43 & .154 & 1 & 0.51 [0.20, 1.29] \\
H1\_S & -0.23 & 0.46 & -0.50 & .619 & 1 & 0.80 [0.32, 1.95] \\
H2\_C & -0.57 & 0.43 & -1.32 & .189 & 1 & 0.57 [0.24, 1.32] \\
H2\_S & -0.87 & 0.45 & -1.94 & .053 & 1 & 0.42 [0.18, 1.01] \\
ED × H1\_M & 0.28 & 0.64 & 0.44 & .659 & 1 & 1.33 [0.38, 4.70] \\
SR × H1\_M & 0.47 & 0.71 & 0.66 & .512 & 1 & 1.59 [0.40, 6.42] \\
ED × H1\_S & -0.43 & 0.65 & -0.66 & .512 & 1 & 0.65 [0.18, 2.33] \\
SR × H1\_S & 0.99 & 0.68 & 1.45 & .146 & 1 & 2.69 [0.71, 10.18] \\
ED × H2\_C & -0.11 & 0.59 & -0.18 & .854 & 1 & 0.90 [0.28, 2.87] \\
SR × H2\_C & 0.71 & 0.64 & 1.11 & .266 & 1 & 2.03 [0.58, 7.08] \\
ED × H2\_S & 0.45 & 0.60 & 0.76 & .446 & 1 & 1.58 [0.49, 5.07] \\
SR × H2\_S & 0.90 & 0.65 & 1.39 & .165 & 1 & 2.46 [0.69, 8.79] \\
\bottomrule
\end{tabular}

\textit{Note.} CSW = Collaborative Story Writing; ED = Ethical Decision; SR = Survival Ranking; Group conditions: H1\_C = 1H+2AI Contrarian (reference); H1\_M = 1H+2AI Mixed; H1\_S = 1H+2AI Supportive; H2\_C = 2H+1AI Contrarian; H2\_S = 2H+1AI Supportive; All p-values adjusted using Holm correction across 47 tests.
\end{table}

\begin{table}
\centering
\caption{Logistic Regression Results: Human Specificity (Human targets only)}
\label{tab:mc_human_specificity}
\begin{tabular}{lcccccc}
\toprule
Predictor & $\beta$ & SE & $z$ & $p$ & $p_{\text{adj}}$ & OR [95\% CI] \\
\midrule
Intercept & 0.85 & 0.28 & 3.08 & .002 & .094 & 2.35 [1.36, 4.05] \\
ED & -0.41 & 0.38 & -1.08 & .279 & 1 & 0.66 [0.31, 1.40] \\
SR & 0.70 & 0.38 & 1.84 & .066 & 1 & 2.01 [0.95, 4.25] \\
H2\_S & 0.23 & 0.39 & 0.59 & .554 & 1 & 1.26 [0.59, 2.70] \\
H3 & 0.32 & 0.32 & 1.00 & .318 & 1 & 1.38 [0.74, 2.57] \\
ED × H2\_S & -0.07 & 0.53 & -0.13 & .897 & 1 & 0.94 [0.33, 2.65] \\
SR × H2\_S & -0.37 & 0.53 & -0.70 & .486 & 1 & 0.69 [0.24, 1.96] \\
ED × H3 & 0.68 & 0.44 & 1.55 & .122 & 1 & 1.97 [0.83, 4.68] \\
SR × H3 & -0.47 & 0.44 & -1.07 & .284 & 1 & 0.62 [0.26, 1.48] \\
\bottomrule
\end{tabular}

\textit{Note.} Model includes task-group interactions (coefficients: ED$\times$AllHuman: $\beta=-0.63$, SE=0.36, $p=.080$; SR$\times$AllHuman: $\beta=-0.69$, SE=0.36, $p=.055$); AI persona not included due to perfect collinearity with group composition.
\end{table}

\begin{table}
\centering
\caption{Logistic Regression Results: Unsure Response Rate (All targets)}
\label{tab:mc_unsure}
\begin{tabular}{lcccccc}
\toprule
Predictor & $\beta$ & SE & $z$ & $p$ & $p_{\text{adj}}$ & OR [95\% CI] \\
\midrule
Intercept & -0.80 & 0.37 & -2.17 & .030 & 1 & 0.45 [0.22, 0.92] \\
ED & -1.03 & 0.39 & -2.65 & .008 & .376 & 0.36 [0.17, 0.76] \\
SR & -0.36 & 0.50 & -0.72 & .474 & 1 & 0.70 [0.26, 1.86] \\
H1\_M & 0.26 & 0.52 & 0.50 & .620 & 1 & 1.29 [0.47, 3.57] \\
H1\_S & 0.03 & 0.51 & 0.06 & .948 & 1 & 1.03 [0.38, 2.82] \\
H2\_C & 0.24 & 0.41 & 0.59 & .557 & 1 & 1.27 [0.57, 2.83] \\
H2\_S & 0.04 & 0.42 & 0.10 & .921 & 1 & 1.04 [0.46, 2.37] \\
H3 & 0.40 & 0.36 & 1.12 & .264 & 1 & 1.49 [0.74, 3.00] \\
ED × H1\_M & 0.73 & 0.58 & 1.26 & .208 & 1 & 2.08 [0.67, 6.45] \\
SR × H1\_M & 0.08 & 0.71 & 0.11 & .910 & 1 & 1.08 [0.27, 4.35] \\
ED × H1\_S & 0.69 & 0.58 & 1.20 & .231 & 1 & 2.00 [0.64, 6.22] \\
SR × H1\_S & 0.30 & 0.68 & 0.44 & .659 & 1 & 1.35 [0.36, 5.08] \\
ED × H2\_C & 0.89 & 0.49 & 1.80 & .071 & 1 & 2.43 [0.93, 6.37] \\
SR × H2\_C & 0.59 & 0.59 & 1.00 & .316 & 1 & 1.81 [0.57, 5.73] \\
ED × H2\_S & 0.85 & 0.50 & 1.70 & .089 & 1 & 2.34 [0.88, 6.22] \\
SR × H2\_S & 0.54 & 0.60 & 0.91 & .365 & 1 & 1.72 [0.53, 5.54] \\
ED × H3 & 0.57 & 0.43 & 1.33 & .185 & 1 & 1.77 [0.76, 4.10] \\
SR × H3 & 0.22 & 0.52 & 0.42 & .672 & 1 & 1.25 [0.45, 3.46] \\
\bottomrule
\end{tabular}

\textit{Note.} Reference condition: H1\_C (1H+2AI Contrarian) in CSW task; Analysis includes all identification judgments across all group conditions.
\end{table}

\begin{table}
\centering
\caption{Logistic Regression Results: AI Over-ascription (Human targets only)}
\label{tab:mc_overascription}
\begin{tabular}{lcccccc}
\toprule
Predictor & $\beta$ & SE & $z$ & $p$ & $p_{\text{adj}}$ & OR [95\% CI] \\
\midrule
Intercept & -0.88 & 0.28 & -3.18 & .001 & .070 & 0.41 [0.24, 0.71] \\
ED & 0.16 & 0.38 & 0.42 & .677 & 1 & 1.17 [0.55, 2.47] \\
SR & -0.47 & 0.29 & -1.67 & .096 & 1 & 0.62 [0.36, 1.09] \\
H2\_S & 0.07 & 0.39 & 0.17 & .863 & 1 & 1.07 [0.50, 2.29] \\
H3 & 0.16 & 0.32 & 0.49 & .621 & 1 & 1.17 [0.62, 2.19] \\
ED × H2\_S & -0.23 & 0.53 & -0.43 & .666 & 1 & 0.79 [0.28, 2.24] \\
SR × H2\_S & 0.37 & 0.40 & 0.93 & .354 & 1 & 1.45 [0.66, 3.17] \\
ED × H3 & -0.07 & 0.44 & -0.15 & .880 & 1 & 0.94 [0.40, 2.21] \\
SR × H3 & 0.40 & 0.36 & 1.10 & .270 & 1 & 1.49 [0.73, 3.02] \\
\bottomrule
\end{tabular}

\textit{Note.} Over-ascription: Human teammates incorrectly identified as AI agents; Reference condition: H2\_C (2H+1AI Contrarian) in CSW task; Analysis limited to human targets that could be misclassified as AI.
\end{table}

\clearpage
\FloatBarrier
\section{Team Dynamics: Full Regression Results}
\label{appendix-team}

This appendix provides the complete regression outputs for the team-dynamics outcomes reported in the main paper (team satisfaction, psychological safety, and group discussion quality). For transparency, we report (i) the full fixed-effect tables for each model under both conventional model-based inference and cluster-robust (CR2) inference with Satterthwaite-adjusted degrees of freedom, (ii) Holm-adjusted $p$-values controlling the family-wise error rate within each dependent variable, and (iii) estimated marginal means by task and condition to aid interpretation of interaction terms. Where relevant, we additionally report sensitivity analyses that include baseline depression (BDI) as a covariate to assess robustness of the primary conclusions.

\subsection{Team satisfaction}
\subsubsection{Full sample (N=905, groups=572)}
\noindent \textit{Fit and random effects (LMM).} REML criterion $=2565.5$; random intercept (group\_id) variance $=0.172$ ($SD=0.414$); residual variance $=0.820$ ($SD=0.906$).

\begin{table}
\centering
\caption{Fixed effects for standardised team satisfaction (standardised; $N{=}905$, groups $=572$). Estimates (Est.), standard errors (SE), degrees of freedom (df / $df_{\text{Satt}}$), and $p$-values from the linear mixed-effects model (LMM) and from cluster-robust (CR2) inference with Satterthwaite-adjusted degrees of freedom. Holm-adjusted $p$-values control the family-wise error rate across fixed effects per DV.}
\begin{tabular}{lrrrrr|rrrrr}
\toprule
 & \multicolumn{5}{c}{LMM} & \multicolumn{5}{c}{CR2-robust} \\
\cmidrule(lr){2-6} \cmidrule(lr){7-11}
Term & Est. & SE & $df$ & $p$ & $p_{\text{Holm}}$ & Est. & SE & $df_{\text{Satt}}$ & $p_{\text{Satt}}$ & $p_{\text{Holm}}$ \\
\midrule
(Intercept) & 0.112 & 0.120 & 183 & .352 & 1.00 & 0.112 & 0.094 & 32 & .245 & 1.00 \\
Task: Ethics & -0.115 & 0.168 & 182 & .494 & 1.00 & -0.115 & 0.153 & 65 & .454 & 1.00 \\
Task: Analytical & 0.026 & 0.167 & 183 & .876 & 1.00 & 0.026 & 0.132 & 66 & .843 & 1.00 \\
H2\_C & -0.414 & 0.181 & 267 & .023 & .413 & -0.414 & 0.192 & 64 & .035 & .627 \\
H2\_S & 0.092 & 0.184 & 289 & .617 & 1.00 & 0.092 & 0.153 & 63 & .550 & 1.00 \\
H1\_C & -0.451 & 0.218 & 572 & .039 & .658 & -0.451 & 0.271 & 51 & .102 & 1.00 \\
H1\_S & -0.171 & 0.220 & 579 & .438 & 1.00 & -0.171 & 0.231 & 49 & .463 & 1.00 \\
H1\_M & 0.192 & 0.218 & 572 & .378 & 1.00 & 0.192 & 0.194 & 51 & .327 & 1.00 \\
Ethics $\times$ H2\_C & 0.299 & 0.256 & 266 & .242 & 1.00 & 0.299 & 0.261 & 127 & .254 & 1.00 \\
Analytical $\times$ H2\_C & 0.333 & 0.256 & 266 & .194 & 1.00 & 0.333 & 0.252 & 126 & .189 & 1.00 \\
Ethics $\times$ H2\_S & 0.025 & 0.259 & 268 & .924 & 1.00 & 0.025 & 0.220 & 123 & .911 & 1.00 \\
Analytical $\times$ H2\_S & -0.111 & 0.258 & 277 & .666 & 1.00 & -0.111 & 0.217 & 125 & .608 & 1.00 \\
Ethics $\times$ H1\_C & 0.145 & 0.309 & 577 & .639 & 1.00 & 0.145 & 0.382 & 99 & .705 & 1.00 \\
Analytical $\times$ H1\_C & 0.095 & 0.309 & 580 & .759 & 1.00 & 0.095 & 0.358 & 99 & .792 & 1.00 \\
Ethics $\times$ H1\_S & 0.121 & 0.309 & 577 & .695 & 1.00 & 0.121 & 0.316 & 99 & .702 & 1.00 \\
Analytical $\times$ H1\_S & 0.234 & 0.310 & 583 & .452 & 1.00 & 0.234 & 0.292 & 97 & .426 & 1.00 \\
Ethics $\times$ H1\_M & -0.554 & 0.307 & 573 & .072 & 1.00 & -0.554 & 0.315 & 102 & .082 & 1.00 \\
Analytical $\times$ H1\_M & -0.423 & 0.307 & 577 & .169 & 1.00 & -0.423 & 0.303 & 101 & .167 & 1.00 \\
\bottomrule
\end{tabular}

\textit{Note.} Estimates are relative to the reference condition of three human teammates (H3) working on the Creative task. Holm correction was applied within each family of fixed effects (LMM and CR2) per dependent variable. Experimental conditions: H3 = three humans; H2\_C = two humans + one contrarian AI; H2\_S = two humans + one supportive AI; H1\_C = one human + two contrarian AIs; H1\_S = one human + two supportive AIs; H1\_M = one human + one contrarian and one supportive AI. Task conditions: Creative (reference), Analytical (survival ranking), and Ethics (ethics dilemma).
\end{table}

\begin{table}
\centering
\caption{Estimated marginal means of team satisfaction (standardised) by Task and Condition ($N{=}905$, groups $=572$).}
\begin{tabular}{llcccc}
\toprule
Task & Condition & Est. & 95\% CI & SE & df \\
\midrule
Creative    & H3    &  0.112 & [-0.124, 0.347] & 0.120 & 250 \\
            & H2\_C & -0.302 & [-0.568, -0.036] & 0.136 & 459 \\
            & H2\_S &  0.204 & [-0.072, 0.479] & 0.140 & 508 \\
            & H1\_C & -0.340 & [-0.696, 0.017] & 0.182 & 860 \\
            & H1\_S & -0.059 & [-0.422, 0.304] & 0.185 & 860 \\
            & H1\_M &  0.303 & [-0.053, 0.660] & 0.182 & 860 \\
\addlinespace
Ethics      & H3    & -0.004 & [-0.236, 0.229] & 0.118 & 246 \\
            & H2\_C & -0.118 & [-0.387, 0.151] & 0.137 & 453 \\
            & H2\_S &  0.113 & [-0.158, 0.384] & 0.138 & 408 \\
            & H1\_C & -0.310 & [-0.673, 0.053] & 0.185 & 860 \\
            & H1\_S & -0.053 & [-0.410, 0.304] & 0.182 & 860 \\
            & H1\_M & -0.366 & [-0.723, -0.009] & 0.182 & 860 \\
\addlinespace
Analytical  & H3    &  0.138 & [-0.092, 0.367] & 0.117 & 249 \\
            & H2\_C &  0.057 & [-0.214, 0.329] & 0.138 & 446 \\
            & H2\_S &  0.118 & [-0.153, 0.390] & 0.138 & 446 \\
            & H1\_C & -0.219 & [-0.582, 0.144] & 0.185 & 860 \\
            & H1\_S &  0.200 & [-0.163, 0.563] & 0.185 & 860 \\
            & H1\_M & -0.093 & [-0.450, 0.264] & 0.182 & 860 \\
\bottomrule
\end{tabular}

\textit{Note.} Est. = estimated marginal mean of team satisfaction; CI = confidence interval; SE = standard error; df = degrees of freedom. Experimental conditions: H3 = three humans; H2\_C = two humans + one contrarian AI; H2\_S = two humans + one supportive AI; H1\_C = one human + two contrarian AIs; H1\_S = one human + two supportive AIs; H1\_M = one human + one contrarian and one supportive AI. Task conditions: Creative (reference), Analytical (survival ranking), and Ethics (ethics dilemma).
\end{table}

\clearpage 

\subsubsection{BDI as a covariate (N=871, groups=556)}
\noindent \textit{Fit and random effects (LMM).} REML criterion $=2486.4$; random intercept (group\_id) variance $=0.184$ ($SD=0.429$); residual variance $=0.828$ ($SD=0.910$).

\begin{table}
\centering
\caption{Fixed effects for team satisfaction (BDI-controlled; $N{=}871$, groups $=556$). Estimates (Est.), standard errors (SE), degrees of freedom (df / $df_{\text{Satt}}$), and $p$-values from the linear mixed-effects model (LMM) and from cluster-robust (CR2) inference with Satterthwaite-adjusted degrees of freedom. Holm-adjusted $p$-values control the family-wise error rate across fixed effects per DV.}
\begin{tabular}{lrrrrr|rrrrr}
\toprule
 & \multicolumn{5}{c}{LMM} & \multicolumn{5}{c}{CR2-robust} \\
\cmidrule(lr){2-6} \cmidrule(lr){7-11}
Term & Est. & SE & $df$ & $p$ & $p_{\text{Holm}}$ & Est. & SE & $df_{\text{Satt}}$ & $p_{\text{Satt}}$ & $p_{\text{Holm}}$ \\
\midrule
(Intercept) & 0.068 & 0.124 & 175 & .584 & 1.00 & 0.068 & 0.098 & 33 & .492 & 1.00 \\
Task: Ethics & -0.093 & 0.173 & 172 & .593 & 1.00 & -0.093 & 0.153 & 65 & .546 & 1.00 \\
Task: Analytical & 0.030 & 0.172 & 167 & .862 & 1.00 & 0.030 & 0.132 & 65 & .821 & 1.00 \\
H2\_C & -0.379 & 0.190 & 256 & .047 & .835 & -0.379 & 0.207 & 62 & .073 & 1.00 \\
H2\_S & 0.132 & 0.190 & 278 & .487 & 1.00 & 0.132 & 0.153 & 64 & .391 & 1.00 \\
H1\_C & -0.447 & 0.224 & 538 & .046 & .835 & -0.447 & 0.276 & 50 & .112 & 1.00 \\
H1\_S & -0.143 & 0.227 & 546 & .530 & 1.00 & -0.143 & 0.245 & 48 & .563 & 1.00 \\
H1\_M & 0.285 & 0.225 & 543 & .206 & 1.00 & 0.285 & 0.206 & 51 & .171 & 1.00 \\
BDI & 0.300 & 0.124 & 798 & .016 & .295 & 0.300 & 0.125 & 256 & .017 & .329 \\
Ethics $\times$ H2\_C & 0.294 & 0.265 & 252 & .268 & 1.00 & 0.294 & 0.272 & 124 & .282 & 1.00 \\
Analytical $\times$ H2\_C & 0.338 & 0.264 & 247 & .202 & 1.00 & 0.338 & 0.262 & 123 & .200 & 1.00 \\
Ethics $\times$ H2\_S & 0.005 & 0.266 & 254 & .985 & 1.00 & 0.005 & 0.220 & 123 & .982 & 1.00 \\
Analytical $\times$ H2\_S & -0.110 & 0.264 & 255 & .679 & 1.00 & -0.110 & 0.215 & 125 & .612 & 1.00 \\
Ethics $\times$ H1\_C & 0.170 & 0.316 & 539 & .590 & 1.00 & 0.170 & 0.388 & 99 & .662 & 1.00 \\
Analytical $\times$ H1\_C & 0.055 & 0.324 & 552 & .865 & 1.00 & 0.055 & 0.383 & 88 & .886 & 1.00 \\
Ethics $\times$ H1\_S & 0.157 & 0.318 & 543 & .621 & 1.00 & 0.157 & 0.328 & 96 & .633 & 1.00 \\
Analytical $\times$ H1\_S & 0.218 & 0.319 & 544 & .495 & 1.00 & 0.218 & 0.305 & 94 & .477 & 1.00 \\
Ethics $\times$ H1\_M & -0.618 & 0.316 & 540 & .051 & .835 & -0.618 & 0.328 & 99 & .062 & 1.00 \\
Analytical $\times$ H1\_M & -0.451 & 0.319 & 543 & .158 & 1.00 & -0.451 & 0.327 & 94 & .171 & 1.00 \\
\bottomrule
\end{tabular}

\textit{Note.} Estimates are relative to the reference condition of three human teammates (H3) working on the Creative task. Holm correction was applied within each family of fixed effects (LMM and CR2) per dependent variable. Experimental conditions: H3 = three humans; H2\_C = two humans + one contrarian AI; H2\_S = two humans + one supportive AI; H1\_C = one human + two contrarian AIs; H1\_S = one human + two supportive AIs; H1\_M = one human + one contrarian and one supportive AI. Task conditions: Creative (reference), Analytical (survival ranking), and Ethics (ethics dilemma).
\end{table}

\begin{table}
\centering
\caption{Estimated marginal means of team satisfaction by Task and Condition (N=871, groups=556)}
\begin{tabular}{llcccc}
\toprule
Task & Condition & Est. & 95\% CI & SE & df \\
\midrule
Creative    & H3    & 0.068 & [-0.177, 0.313] & 0.124 & 175 \\
            & H2\_C & -0.311 & [-0.592, -0.030] & 0.143 & 346 \\
            & H2\_S & 0.200 & [-0.083, 0.484] & 0.144 & 402 \\
            & H1\_C & -0.379 & [-0.746, -0.013] & 0.187 & 808 \\
            & H1\_S & -0.075 & [-0.448, 0.299] & 0.190 & 808 \\
            & H1\_M & 0.353 & [-0.014, 0.720] & 0.187 & 808 \\
\addlinespace
Ethics      & H3    & -0.025 & [-0.265, 0.216] & 0.122 & 174 \\
            & H2\_C & -0.110 & [-0.386, 0.167] & 0.141 & 343 \\
            & H2\_S & 0.113 & [-0.166, 0.392] & 0.142 & 300 \\
            & H1\_C & -0.302 & [-0.669, 0.065] & 0.187 & 808 \\
            & H1\_S & -0.010 & [-0.378, 0.358] & 0.187 & 809 \\
            & H1\_M & -0.358 & [-0.725, 0.010] & 0.187 & 808 \\
\addlinespace
Analytical  & H3    & 0.098 & [-0.139, 0.335] & 0.120 & 164 \\
            & H2\_C & 0.057 & [-0.220, 0.334] & 0.141 & 328 \\
            & H2\_S & 0.121 & [-0.155, 0.396] & 0.140 & 320 \\
            & H1\_C & -0.294 & [-0.690, 0.101] & 0.202 & 809 \\
            & H1\_S & 0.173 & [-0.200, 0.546] & 0.190 & 808 \\
            & H1\_M & -0.068 & [-0.449, 0.314] & 0.194 & 809 \\
\bottomrule
\end{tabular}

\textit{Note.} Est. = estimated marginal mean of team satisfaction; CI = confidence interval; SE = standard error; df = degrees of freedom. Experimental conditions: H3 = three humans; H2\_C = two humans + one contrarian AI; H2\_S = two humans + one supportive AI; H1\_C = one human + two contrarian AIs; H1\_S = one human + two supportive AIs; H1\_M = one human + one contrarian and one supportive AI. Task conditions: Creative (reference), Analytical (survival ranking), and Ethics (ethics dilemma).
\end{table}

\clearpage 

\subsection{Psychological safety}
\subsubsection{Full sample (N=905, groups=572)}
\noindent \textit{Fit and random effects (LMM; $N{=}905$, groups $=572$).} REML criterion $=2516.5$; random intercept (group\_id) variance $=0.149$ ($SD=0.385$); residual variance $=0.788$ ($SD=0.888$).

\begin{table}
\centering
\caption{Fixed effects for psychological safety (standardised; $N{=}905$, groups $=572$). Estimates (Est.), standard errors (SE), degrees of freedom (df / $df_{\text{Satt}}$), and $p$-values from the linear mixed-effects model (LMM) and from cluster-robust (CR2) inference with Satterthwaite-adjusted degrees of freedom. Holm-adjusted $p$-values control the family-wise error rate across fixed effects per DV.}
\begin{tabular}{lrrrrr|rrrrr}
\toprule
 & \multicolumn{5}{c}{LMM} & \multicolumn{5}{c}{CR2-robust} \\
\cmidrule(lr){2-6} \cmidrule(lr){7-11}
Term & Est. & SE & $df$ & $p$ & $p_{\text{Holm}}$ & Est. & SE & $df_{\text{Satt}}$ & $p_{\text{Satt}}$ & $p_{\text{Holm}}$ \\
\midrule
(Intercept) & 0.187 & 0.115 & 190 & .106 & 1.00 & 0.187 & 0.094 & 32 & .056 & .782 \\
Task: Ethics & -0.211 & 0.162 & 189 & .194 & 1.00 & -0.211 & 0.145 & 65 & .152 & 1.00 \\
Task: Analytical & -0.007 & 0.161 & 190 & .966 & 1.00 & -0.007 & 0.134 & 66 & .959 & 1.00 \\
H2\_C & -0.671 & 0.174 & 279 & $<.001$ & .003 & -0.671 & 0.169 & 64 & $<.001$ & .003 \\
H2\_S & 0.152 & 0.178 & 301 & .393 & 1.00 & 0.152 & 0.148 & 63 & .308 & 1.00 \\
H1\_C & -0.881 & 0.211 & 593 & $<.001$ & $<.001$ & -0.881 & 0.281 & 51 & .003 & .049 \\
H1\_S & -0.200 & 0.213 & 599 & .348 & 1.00 & -0.200 & 0.222 & 49 & .371 & 1.00 \\
H1\_M & 0.053 & 0.211 & 593 & .803 & 1.00 & 0.053 & 0.209 & 51 & .802 & 1.00 \\
Ethics $\times$ H2\_C & 0.604 & 0.247 & 279 & .015 & .239 & 0.604 & 0.248 & 127 & .016 & .262 \\
Analytical $\times$ H2\_C & 0.362 & 0.247 & 278 & .143 & 1.00 & 0.362 & 0.238 & 126 & .131 & 1.00 \\
Ethics $\times$ H2\_S & 0.196 & 0.250 & 280 & .434 & 1.00 & 0.196 & 0.200 & 123 & .331 & 1.00 \\
Analytical $\times$ H2\_S & -0.155 & 0.249 & 289 & .534 & 1.00 & -0.155 & 0.211 & 125 & .464 & 1.00 \\
Ethics $\times$ H1\_C & 0.300 & 0.299 & 597 & .316 & 1.00 & 0.300 & 0.371 & 99 & .421 & 1.00 \\
Analytical $\times$ H1\_C & 0.284 & 0.299 & 601 & .342 & 1.00 & 0.284 & 0.373 & 99 & .448 & 1.00 \\
Ethics $\times$ H1\_S & 0.580 & 0.299 & 597 & .053 & .800 & 0.580 & 0.290 & 99 & .048 & .723 \\
Analytical $\times$ H1\_S & 0.281 & 0.301 & 604 & .351 & 1.00 & 0.281 & 0.299 & 97 & .350 & 1.00 \\
Ethics $\times$ H1\_M & -0.221 & 0.298 & 594 & .457 & 1.00 & -0.221 & 0.314 & 101 & .482 & 1.00 \\
Analytical $\times$ H1\_M & -0.490 & 0.297 & 597 & .099 & 1.00 & -0.490 & 0.306 & 101 & .112 & 1.00 \\
\bottomrule
\end{tabular}

\textit{Note.} Estimates are relative to the reference condition of three human teammates (H3) working on the Creative task. Holm correction was applied within each family of fixed effects (LMM and CR2) per dependent variable. Experimental conditions: H3 = three humans; H2\_C = two humans + one contrarian AI; H2\_S = two humans + one supportive AI; H1\_C = one human + two contrarian AIs; H1\_S = one human + two supportive AIs; H1\_M = one human + one contrarian and one supportive AI. Task conditions: Creative (reference), Analytical (survival ranking), and Ethics (ethics dilemma).
\end{table}

\begin{table}
\centering
\caption{Estimated marginal means of psychological safety (standardised) by Task and Condition ($N{=}905$, groups $=572$).}
\begin{tabular}{llcccc}
\toprule
Task & Condition & Est. & 95\% CI & SE & df \\
\midrule
Creative    & H3    &  0.187 & [-0.040, 0.414] & 0.115 & 245 \\
            & H2\_C & -0.484 & [-0.741, -0.226] & 0.131 & 460 \\
            & H2\_S &  0.339 & [0.073, 0.606] & 0.136 & 509 \\
            & H1\_C & -0.694 & [-1.041, -0.347] & 0.177 & 864 \\
            & H1\_S & -0.013 & [-0.366, 0.339] & 0.180 & 864 \\
            & H1\_M &  0.240 & [-0.107, 0.586] & 0.177 & 864 \\
\addlinespace
Ethics      & H3    & -0.024 & [-0.247, 0.200] & 0.114 & 242 \\
            & H2\_C & -0.091 & [-0.351, 0.169] & 0.132 & 453 \\
            & H2\_S &  0.324 & [0.062, 0.586] & 0.133 & 408 \\
            & H1\_C & -0.605 & [-0.957, -0.252] & 0.180 & 864 \\
            & H1\_S &  0.356 & [0.009, 0.703] & 0.177 & 864 \\
            & H1\_M & -0.192 & [-0.539, 0.154] & 0.177 & 864 \\
\addlinespace
Analytical  & H3    &  0.180 & [-0.041, 0.401] & 0.112 & 244 \\
            & H2\_C & -0.128 & [-0.391, 0.134] & 0.133 & 446 \\
            & H2\_S &  0.177 & [-0.085, 0.440] & 0.133 & 446 \\
            & H1\_C & -0.417 & [-0.770, -0.064] & 0.180 & 864 \\
            & H1\_S &  0.261 & [-0.092, 0.613] & 0.180 & 864 \\
            & H1\_M & -0.258 & [-0.604, 0.089] & 0.177 & 864 \\
\bottomrule
\end{tabular}

\textit{Note.} Est. = estimated marginal mean of psychological safety (standardised); CI = confidence interval; SE = standard error; df = degrees of freedom. Experimental conditions: H3 = three humans; H2\_C = two humans + one contrarian AI; H2\_S = two humans + one supportive AI; H1\_C = one human + two contrarian AIs; H1\_S = one human + two supportive AIs; H1\_M = one human + one contrarian and one supportive AI. Task conditions: Creative (reference), Analytical (survival ranking), and Ethics (ethics dilemma).
\end{table}

\clearpage 

\subsubsection{BDI as a covariate (N=871, groups=556)}
\noindent \textit{Fit and random effects (LMM).} REML criterion $=2426.1$; random intercept (group\_id) variance $=0.165$ ($SD=0.406$); residual variance $=0.777$ ($SD=0.881$).

\begin{table}
\centering
\caption{Fixed effects for psychological safety (BDI-controlled; $N{=}871$, groups $=556$). Estimates (Est.), standard errors (SE), degrees of freedom (df / $df_{\text{Satt}}$), and $p$-values from the linear mixed-effects model (LMM) and from cluster-robust (CR2) inference with Satterthwaite-adjusted degrees of freedom. Holm-adjusted $p$-values control the family-wise error rate across fixed effects per DV.}
\begin{tabular}{lrrrrr|rrrrr}
\toprule
 & \multicolumn{5}{c}{LMM} & \multicolumn{5}{c}{CR2-robust} \\
\cmidrule(lr){2-6} \cmidrule(lr){7-11}
Term & Est. & SE & $df$ & $p$ & $p_{\text{Holm}}$ & Est. & SE & $df_{\text{Satt}}$ & $p_{\text{Satt}}$ & $p_{\text{Holm}}$ \\
\midrule
(Intercept) & 0.157 & 0.119 & 190 & .191 & 1.00 & 0.157 & 0.096 & 33 & .113 & 1.00 \\
Task: Ethics & -0.193 & 0.166 & 187 & .247 & 1.00 & -0.193 & 0.147 & 65 & .193 & 1.00 \\
Task: Analytical & -0.019 & 0.165 & 181 & .911 & 1.00 & -0.019 & 0.135 & 65 & .891 & 1.00 \\
H2\_C & -0.670 & 0.183 & 277 & $<.001$ & .005 & -0.670 & 0.179 & 62 & $<.001$ & .007 \\
H2\_S & 0.215 & 0.183 & 299 & .241 & 1.00 & 0.215 & 0.154 & 64 & .166 & 1.00 \\
H1\_C & -0.814 & 0.216 & 563 & $<.001$ & .003 & -0.814 & 0.288 & 50 & .007 & .121 \\
H1\_S & -0.187 & 0.219 & 571 & .392 & 1.00 & -0.187 & 0.233 & 48 & .425 & 1.00 \\
H1\_M & 0.165 & 0.217 & 568 & .448 & 1.00 & 0.165 & 0.212 & 50 & .442 & 1.00 \\
BDI & 0.263 & 0.120 & 807 & .028 & .447 & 0.263 & 0.117 & 257 & .025 & .402 \\
Ethics $\times$ H2\_C & 0.638 & 0.255 & 272 & .013 & .218 & 0.638 & 0.255 & 124 & .014 & .231 \\
Analytical $\times$ H2\_C & 0.400 & 0.254 & 266 & .117 & 1.00 & 0.400 & 0.246 & 123 & .106 & 1.00 \\
Ethics $\times$ H2\_S & 0.142 & 0.256 & 274 & .580 & 1.00 & 0.142 & 0.207 & 122 & .494 & 1.00 \\
Analytical $\times$ H2\_S & -0.175 & 0.254 & 275 & .493 & 1.00 & -0.175 & 0.214 & 125 & .416 & 1.00 \\
Ethics $\times$ H1\_C & 0.253 & 0.304 & 565 & .406 & 1.00 & 0.253 & 0.378 & 98 & .505 & 1.00 \\
Analytical $\times$ H1\_C & 0.162 & 0.312 & 577 & .603 & 1.00 & 0.162 & 0.393 & 88 & .681 & 1.00 \\
Ethics $\times$ H1\_S & 0.612 & 0.306 & 568 & .046 & .655 & 0.612 & 0.302 & 96 & .045 & .681 \\
Analytical $\times$ H1\_S & 0.276 & 0.307 & 569 & .369 & 1.00 & 0.276 & 0.309 & 94 & .373 & 1.00 \\
Ethics $\times$ H1\_M & -0.264 & 0.304 & 565 & .387 & 1.00 & -0.264 & 0.319 & 98 & .411 & 1.00 \\
Analytical $\times$ H1\_M & -0.622 & 0.308 & 569 & .044 & .655 & -0.622 & 0.316 & 93 & .052 & .727 \\
\bottomrule
\end{tabular}

\textit{Note.} Estimates are relative to the reference condition of three human teammates (H3) working on the Creative task. Holm correction was applied within each family of fixed effects (LMM and CR2) per dependent variable. Experimental conditions: H3 = three humans; H2\_C = two humans + one contrarian AI; H2\_S = two humans + one supportive AI; H1\_C = one human + two contrarian AIs; H1\_S = one human + two supportive AIs; H1\_M = one human + one contrarian and one supportive AI. Task conditions: Creative (reference), Analytical (survival ranking), and Ethics (ethics dilemma).
\end{table}

\begin{table}
\centering
\caption{Estimated marginal means of psychological safety (standardised) by Task and Condition (N=871, groups=556).}
\begin{tabular}{llcccc}
\toprule
Task & Condition & Est. & 95\% CI & SE & df \\
\midrule
Creative    & H3    & 0.157 & [-0.079, 0.392] & 0.119 & 190 \\
            & H2\_C & -0.514 & [-0.785, -0.243] & 0.138 & 370 \\
            & H2\_S & 0.372 & [0.099, 0.644] & 0.139 & 427 \\
            & H1\_C & -0.658 & [-1.011, -0.304] & 0.180 & 815 \\
            & H1\_S & -0.031 & [-0.391, 0.329] & 0.183 & 815 \\
            & H1\_M & 0.321 & [-0.033, 0.676] & 0.181 & 815 \\
\addlinespace
Ethics      & H3    & -0.036 & [-0.267, 0.194] & 0.117 & 189 \\
            & H2\_C & -0.068 & [-0.334, 0.197] & 0.135 & 367 \\
            & H2\_S & 0.321 & [0.052, 0.589] & 0.136 & 323 \\
            & H1\_C & -0.598 & [-0.951, -0.244] & 0.180 & 815 \\
            & H1\_S & 0.388 & [0.033, 0.743] & 0.181 & 816 \\
            & H1\_M & -0.135 & [-0.490, 0.219] & 0.181 & 815 \\
\addlinespace
Analytical  & H3    & 0.138 & [-0.090, 0.366] & 0.115 & 178 \\
            & H2\_C & -0.132 & [-0.399, 0.135] & 0.136 & 352 \\
            & H2\_S & 0.179 & [-0.086, 0.444] & 0.135 & 343 \\
            & H1\_C & -0.514 & [-0.896, -0.132] & 0.195 & 816 \\
            & H1\_S & 0.227 & [-0.133, 0.587] & 0.183 & 815 \\
            & H1\_M & -0.319 & [-0.687, 0.049] & 0.187 & 816 \\
\bottomrule
\end{tabular}

\textit{Note.} Est. = estimated marginal mean of psychological safety (standardised); CI = confidence interval; SE = standard error; df = degrees of freedom. Experimental conditions: H3 = three humans; H2\_C = two humans + one contrarian AI; H2\_S = two humans + one supportive AI; H1\_C = one human + two contrarian AIs; H1\_S = one human + two supportive AIs; H1\_M = one human + one contrarian and one supportive AI. Task conditions: Creative (reference), Analytical (survival ranking), and Ethics (ethics dilemma).
\end{table}

\clearpage 

\subsection{Group discussion quality}
\subsubsection{Full sample (groups=572)}
\noindent \textit{Fit (OLS; groups $=572$).} Residual $SE = 0.954$ ($df = 554$); $R^2 = 0.237$, adjusted $R^2 = 0.213$; $F(17,554) = 10.09$, $p < .001$.

\begin{table}
\centering
\caption{Fixed effects for group discussion quality (standardised; groups $=572$). Estimates (Est.), standard errors (SE), degrees of freedom (df / $df_{\text{Satt}}$), and $p$-values from the linear model (OLS) and CR2-robust inference. Holm-adjusted $p$-values control the family-wise error rate across fixed effects per DV.}
\begin{tabular}{lrrrrr|rrrrr}
\toprule
 & \multicolumn{5}{c}{OLS} & \multicolumn{5}{c}{CR2-robust} \\
\cmidrule(lr){2-6} \cmidrule(lr){7-11}
Term & Est. & SE & $df$ & $p$ & $p_{\text{Holm}}$ & Est. & SE & $df_{\text{Satt}}$ & $p_{\text{Satt}}$ & $p_{\text{Holm}}$ \\
\midrule
(Intercept) & 0.121 & 0.164 & 562 & .458 & 1.00 & 0.121 & 0.137 & 33 & .383 & 1.00 \\
Task: Ethics & -0.467 & 0.230 & 562 & .043 & .511 & -0.467 & 0.215 & 67 & .033 & .401 \\
Task: Analytical & -0.088 & 0.228 & 562 & .701 & 1.00 & -0.088 & 0.199 & 68 & .661 & 1.00 \\
H2\_C & -0.856 & 0.230 & 562 & $<.001$ & .004 & -0.856 & 0.202 & 67 & $<.001$ & .001 \\
H2\_S & 0.590 & 0.230 & 562 & .010 & .156 & 0.590 & 0.166 & 67 & .001 & .012 \\
H1\_C & -0.747 & 0.239 & 562 & .002 & .032 & -0.747 & 0.295 & 61 & .014 & .179 \\
H1\_S & 0.644 & 0.241 & 562 & .008 & .125 & 0.644 & 0.219 & 59 & .005 & .069 \\
H1\_M & 0.130 & 0.239 & 562 & .586 & 1.00 & 0.130 & 0.202 & 61 & .522 & 1.00 \\
Ethics $\times$ H2\_C & 0.732 & 0.325 & 562 & .025 & .319 & 0.732 & 0.286 & 134 & .011 & .160 \\
Analytical $\times$ H2\_C & 0.813 & 0.325 & 562 & .013 & .178 & 0.813 & 0.275 & 133 & .004 & .059 \\
Ethics $\times$ H2\_S & 0.069 & 0.329 & 562 & .833 & 1.00 & 0.069 & 0.278 & 130 & .803 & 1.00 \\
Analytical $\times$ H2\_S & 0.135 & 0.325 & 562 & .678 & 1.00 & 0.135 & 0.239 & 133 & .573 & 1.00 \\
Ethics $\times$ H1\_C & 0.032 & 0.338 & 562 & .926 & 1.00 & 0.032 & 0.369 & 121 & .932 & 1.00 \\
Analytical $\times$ H1\_C & 0.420 & 0.337 & 562 & .213 & 1.00 & 0.420 & 0.408 & 121 & .305 & 1.00 \\
Ethics $\times$ H1\_S & -0.404 & 0.338 & 562 & .233 & 1.00 & -0.404 & 0.358 & 121 & .261 & 1.00 \\
Analytical $\times$ H1\_S & -0.145 & 0.339 & 562 & .670 & 1.00 & -0.145 & 0.357 & 119 & .686 & 1.00 \\
Ethics $\times$ H1\_M & -0.361 & 0.337 & 562 & .284 & 1.00 & -0.361 & 0.330 & 122 & .277 & 1.00 \\
Analytical $\times$ H1\_M & 0.045 & 0.336 & 562 & .892 & 1.00 & 0.045 & 0.312 & 123 & .884 & 1.00 \\
\bottomrule
\end{tabular}

\textit{Note.} Estimates are relative to the reference condition of three human teammates (H3) working on the Creative task. Holm correction was applied within each family of fixed effects (OLS and CR2) per dependent variable. Experimental conditions: H3 = three humans; H2\_C = two humans + one contrarian AI; H2\_S = two humans + one supportive AI; H1\_C = one human + two contrarian AIs; H1\_S = one human + two supportive AIs; H1\_M = one human + one contrarian and one supportive AI. Task conditions: Creative (reference), Analytical (survival ranking), and Ethics (ethics dilemma).
\end{table}

\begin{table}
\centering
\caption{Estimated marginal means of group discussion quality (standardised) by Task and Condition ($N{=}905$, groups $=572$).}
\begin{tabular}{llcccc}
\toprule
Task & Condition & Est. & 95\% CI & SE & df \\
\midrule
Creative    & H3    &  0.121 & [-0.200, 0.443] & 0.164 & 554 \\
            & H2\_C & -0.735 & [-1.052, -0.418] & 0.161 & 554 \\
            & H2\_S &  0.712 & [0.395, 1.028] & 0.161 & 554 \\
            & H1\_C & -0.626 & [-0.968, -0.284] & 0.174 & 554 \\
            & H1\_S &  0.765 & [0.417, 1.113] & 0.177 & 554 \\
            & H1\_M &  0.252 & [-0.090, 0.594] & 0.174 & 554 \\
\addlinespace
Ethics      & H3    & -0.345 & [-0.662, -0.029] & 0.161 & 554 \\
            & H2\_C & -0.469 & [-0.791, -0.148] & 0.164 & 554 \\
            & H2\_S &  0.314 & [-0.022, 0.651] & 0.171 & 554 \\
            & H1\_C & -1.061 & [-1.409, -0.713] & 0.177 & 554 \\
            & H1\_S & -0.106 & [-0.448, 0.236] & 0.174 & 554 \\
            & H1\_M & -0.576 & [-0.918, -0.234] & 0.174 & 554 \\
\addlinespace
Analytical  & H3    &  0.034 & [-0.279, 0.346] & 0.159 & 554 \\
            & H2\_C & -0.010 & [-0.336, 0.316] & 0.166 & 554 \\
            & H2\_S &  0.759 & [0.433, 1.085] & 0.166 & 554 \\
            & H1\_C & -0.293 & [-0.641, 0.054] & 0.177 & 554 \\
            & H1\_S &  0.533 & [0.185, 0.881] & 0.177 & 554 \\
            & H1\_M &  0.209 & [-0.133, 0.552] & 0.174 & 554 \\
\bottomrule
\end{tabular}

\textit{Note.} Est. = estimated marginal mean of group discussion quality (standardised); CI = confidence interval; SE = standard error; df = degrees of freedom. Experimental conditions: H3 = three humans; H2\_C = two humans + one contrarian AI; H2\_S = two humans + one supportive AI; H1\_C = one human + two contrarian AIs; H1\_S = one human + two supportive AIs; H1\_M = one human + one contrarian and one supportive AI. Task conditions: Creative (reference), Analytical (survival ranking), and Ethics (ethics dilemma).
\end{table}

\clearpage 

\subsubsection{BDI as a covariate (groups=556)}
\noindent \textit{Fit (OLS).} Residual $SE = 0.932$ ($df = 537$); $R^2 = 0.244$, adjusted $R^2 = 0.219$; $F(18,537) = 9.64$, $p < .001$.

\begin{table}
\centering
\caption{Fixed effects for group discussion quality (groups $=556$). Estimates (Est.), standard errors (SE), degrees of freedom (df / $df_{\text{Satt}}$), and $p$-values from the linear model (LM) and from cluster-robust (CR2) inference with Satterthwaite-adjusted degrees of freedom. Holm-adjusted $p$-values control the family-wise error rate across fixed effects per DV.}
\begin{tabular}{lrrrrr|rrrrr}
\toprule
 & \multicolumn{5}{c}{OLS} & \multicolumn{5}{c}{CR2-robust} \\
\cmidrule(lr){2-6} \cmidrule(lr){7-11}
Term & Est. & SE & $df$ & $p$ & $p_{\text{Holm}}$ & Est. & SE & $df_{\text{Satt}}$ & $p_{\text{Satt}}$ & $p_{\text{Holm}}$ \\
\midrule
(Intercept) & 0.124 & 0.162 & 537 & .443 & 1.00 & 0.124 & 0.141 & 35 & .383 & 1.00 \\
Task: Ethics & -0.468 & 0.225 & 537 & .038 & .490 & -0.468 & 0.216 & 67 & .034 & .439 \\
Task: Analytical & -0.009 & 0.224 & 537 & .968 & 1.00 & -0.009 & 0.186 & 67 & .961 & 1.00 \\
H2\_C & -0.853 & 0.230 & 537 & $<.001$ & .004 & -0.853 & 0.212 & 67 & $<.001$ & .003 \\
H2\_S & 0.587 & 0.226 & 537 & .010 & .162 & 0.587 & 0.169 & 68 & $<.001$ & .016 \\
H1\_C & -0.795 & 0.236 & 537 & $<.001$ & .015 & -0.795 & 0.300 & 60 & .010 & .155 \\
H1\_S & 0.621 & 0.239 & 537 & .010 & .162 & 0.621 & 0.222 & 59 & .007 & .119 \\
H1\_M & 0.132 & 0.238 & 537 & .580 & 1.00 & 0.132 & 0.211 & 61 & .534 & 1.00 \\
$\text{BDI}_{\text{group mean}}$ & -0.020 & 0.173 & 537 & .910 & 1.00 & -0.020 & 0.182 & 222 & .915 & 1.00 \\
Ethics $\times$ H2\_C & 0.727 & 0.320 & 537 & .024 & .353 & 0.727 & 0.292 & 132 & .014 & .195 \\
Analytical $\times$ H2\_C & 0.728 & 0.321 & 537 & .024 & .353 & 0.728 & 0.271 & 131 & .008 & .133 \\
Ethics $\times$ H2\_S & 0.071 & 0.321 & 537 & .826 & 1.00 & 0.071 & 0.278 & 130 & .801 & 1.00 \\
Analytical $\times$ H2\_S & 0.056 & 0.319 & 537 & .859 & 1.00 & 0.056 & 0.228 & 133 & .805 & 1.00 \\
Ethics $\times$ H1\_C & 0.078 & 0.332 & 537 & .814 & 1.00 & 0.078 & 0.373 & 119 & .834 & 1.00 \\
Analytical $\times$ H1\_C & 0.600 & 0.339 & 537 & .077 & .930 & 0.600 & 0.377 & 110 & .114 & 1.00 \\
Ethics $\times$ H1\_S & -0.369 & 0.334 & 537 & .269 & 1.00 & -0.369 & 0.365 & 117 & .314 & 1.00 \\
Analytical $\times$ H1\_S & -0.220 & 0.335 & 537 & .512 & 1.00 & -0.220 & 0.359 & 116 & .541 & 1.00 \\
Ethics $\times$ H1\_M & -0.334 & 0.332 & 537 & .315 & 1.00 & -0.334 & 0.338 & 119 & .324 & 1.00 \\
Analytical $\times$ H1\_M & -0.050 & 0.335 & 537 & .882 & 1.00 & -0.050 & 0.318 & 115 & .876 & 1.00 \\
\bottomrule
\end{tabular}

\textit{Note.} Estimates are relative to the reference condition of three human teammates (H3) working on the Creative task. Holm correction was applied within each family of fixed effects (LM and CR2) per dependent variable. Experimental conditions: H3 = three humans; H2\_C = two humans + one contrarian AI; H2\_S = two humans + one supportive AI; H1\_C = one human + two contrarian AIs; H1\_S = one human + two supportive AIs; H1\_M = one human + one contrarian and one supportive AI. Task conditions: Creative (reference), Analytical (survival ranking), and Ethics (ethics dilemma).
\end{table}

\begin{table}
\centering
\caption{Estimated marginal means of group discussion quality by Task and Condition (groups $=556$)}
\begin{tabular}{llcccc}
\toprule
Task & Condition & Est. & 95\% CI & SE & df \\
\midrule
Creative    & H3    & 0.124 & [-0.193, 0.442] & 0.162 & 537 \\
            & H2\_C & -0.729 & [-1.048, -0.410] & 0.162 & 537 \\
            & H2\_S & 0.712 & [0.402, 1.021] & 0.157 & 537 \\
            & H1\_C & -0.671 & [-1.011, -0.331] & 0.173 & 537 \\
            & H1\_S & 0.745 & [0.399, 1.091] & 0.176 & 537 \\
            & H1\_M & 0.256 & [-0.084, 0.597] & 0.173 & 537 \\
\addlinespace
Ethics      & H3    & -0.344 & [-0.654, -0.034] & 0.158 & 537 \\
            & H2\_C & -0.470 & [-0.784, -0.156] & 0.160 & 537 \\
            & H2\_S & 0.314 & [-0.015, 0.643] & 0.167 & 537 \\
            & H1\_C & -1.061 & [-1.401, -0.721] & 0.173 & 537 \\
            & H1\_S & -0.092 & [-0.434, 0.249] & 0.174 & 537 \\
            & H1\_M & -0.546 & [-0.887, -0.206] & 0.173 & 537 \\
\addlinespace
Analytical  & H3    & 0.115 & [-0.197, 0.427] & 0.159 & 537 \\
            & H2\_C & -0.010 & [-0.329, 0.309] & 0.162 & 537 \\
            & H2\_S & 0.759 & [0.440, 1.078] & 0.162 & 537 \\
            & H1\_C & -0.080 & [-0.447, 0.287] & 0.187 & 537 \\
            & H1\_S & 0.516 & [0.170, 0.862] & 0.176 & 537 \\
            & H1\_M & 0.198 & [-0.157, 0.552] & 0.180 & 537 \\
\bottomrule
\end{tabular}

\textit{Note.} Est. = estimated marginal mean of group discussion quality; CI = confidence interval; SE = standard error; df = degrees of freedom. Experimental conditions: H3 = three humans; H2\_C = two humans + one contrarian AI; H2\_S = two humans + one supportive AI; H1\_C = one human + two contrarian AIs; H1\_S = one human + two supportive AIs; H1\_M = one human + one contrarian and one supportive AI. Task conditions: Creative (reference), Analytical (survival ranking), and Ethics (ethics dilemma).
\end{table}

\clearpage 
\FloatBarrier
\section{Standardised Individual Performance Gain: Full Regression Results}
\label{appendix-gain}

This appendix provides the full regression outputs for standardised individual performance gain, including (i) the primary LMM with task-by-condition fixed effects and random intercepts for groups, (ii) CR2-robust inference with Satterthwaite-adjusted degrees of freedom, and (iii) estimated marginal means (EMMs) by task and condition. For transparency, we also report the full set of planned pairwise comparisons within the Analytical (survival-ranking) task, where the largest descriptive separation was observed. In the main paper, we report only the key takeaways: overall effects on performance gain were small, and the only robust contrast that survived multiplicity correction was confined to the Analytical task (H2\_C vs.\ H3), with all other effects not surviving Holm adjustment.

\subsection{Full sample (N=905, groups=572)}
\noindent \textit{Fit and random effects (LMM).} REML criterion $=2556.6$; random intercept (group\_id) variance $=0.042$ ($SD=0.205$); residual variance $=0.927$ ($SD=0.963$).

\begin{table}
\centering
\caption{Fixed effects for individual performance gain (standardised; $N{=}905$, groups $=572$). Estimates (Est.), standard errors (SE), degrees of freedom (df / $df_{\text{Satt}}$), and $p$-values from the linear mixed-effects model (LMM) and from cluster-robust (CR2) inference with Satterthwaite-adjusted degrees of freedom. Holm-adjusted $p$-values control the family-wise error rate across fixed effects per DV.}
\begin{tabular}{lrrrrr|rrrrr}
\toprule
 & \multicolumn{5}{c}{LMM} & \multicolumn{5}{c}{CR2-robust} \\
\cmidrule(lr){2-6} \cmidrule(lr){7-11}
Term & Est. & SE & $df$ & $p$ & $p_{\text{Holm}}$ & Est. & SE & $df_{\text{Satt}}$ & $p_{\text{Satt}}$ & $p_{\text{Holm}}$ \\
\midrule
(Intercept) & -0.124 & 0.108 & 226 & .252 & 1.00 & -0.124 & 0.118 & 32 & .303 & 1.00 \\
Task: Ethics & 0.010 & 0.151 & 224 & .945 & 1.00 & 0.010 & 0.152 & 64 & .946 & 1.00 \\
Task: Analytical & -0.267 & 0.150 & 225 & .077 & 1.00 & -0.267 & 0.161 & 65 & .103 & 1.00 \\
H2\_C & -0.028 & 0.167 & 350 & .869 & 1.00 & -0.028 & 0.181 & 63 & .879 & 1.00 \\
H2\_S & 0.336 & 0.171 & 377 & .050 & .851 & 0.336 & 0.184 & 61 & .074 & 1.00 \\
H1\_C & 0.252 & 0.209 & 716 & .229 & 1.00 & 0.252 & 0.185 & 48 & .179 & 1.00 \\
H1\_S & 0.219 & 0.212 & 722 & .302 & 1.00 & 0.219 & 0.190 & 46 & .256 & 1.00 \\
H1\_M & 0.177 & 0.209 & 716 & .398 & 1.00 & 0.177 & 0.198 & 48 & .375 & 1.00 \\
Ethics $\times$ H2\_C & 0.005 & 0.236 & 348 & .984 & 1.00 & 0.005 & 0.262 & 124 & .986 & 1.00 \\
Analytical $\times$ H2\_C & 0.876 & 0.236 & 349 & $<.001$ & .004 & 0.876 & 0.251 & 123 & $<.001$ & .012 \\
Ethics $\times$ H2\_S & -0.207 & 0.239 & 352 & .387 & 1.00 & -0.207 & 0.239 & 120 & .389 & 1.00 \\
Analytical $\times$ H2\_S & 0.049 & 0.239 & 363 & .836 & 1.00 & 0.049 & 0.244 & 121 & .839 & 1.00 \\
Ethics $\times$ H1\_C & -0.099 & 0.297 & 720 & .739 & 1.00 & -0.099 & 0.277 & 93 & .720 & 1.00 \\
Analytical $\times$ H1\_C & 0.182 & 0.297 & 723 & .541 & 1.00 & 0.182 & 0.274 & 93 & .509 & 1.00 \\
Ethics $\times$ H1\_S & 0.280 & 0.297 & 720 & .347 & 1.00 & 0.280 & 0.276 & 93 & .313 & 1.00 \\
Analytical $\times$ H1\_S & 0.406 & 0.299 & 726 & .175 & 1.00 & 0.406 & 0.277 & 91 & .147 & 1.00 \\
Ethics $\times$ H1\_M & 0.117 & 0.296 & 717 & .693 & 1.00 & 0.117 & 0.272 & 95 & .669 & 1.00 \\
Analytical $\times$ H1\_M & 0.283 & 0.295 & 720 & .338 & 1.00 & 0.283 & 0.279 & 95 & .312 & 1.00 \\
\bottomrule
\end{tabular}

\textit{Note.} Estimates are relative to the reference condition of three human teammates (H3) working on the Creative task. Holm correction was applied within each family of fixed effects (LMM and CR2) per dependent variable. Experimental conditions: H3 = three humans; H2\_C = two humans + one contrarian AI; H2\_S = two humans + one supportive AI; H1\_C = one human + two contrarian AIs; H1\_S = one human + two supportive AIs; H1\_M = one human + one contrarian and one supportive AI. Task conditions: Creative (reference), Analytical (survival ranking), and Ethics (ethics dilemma).
\end{table}

\begin{table}
\centering
\caption{Estimated marginal means of individual performance gain (standardised) by Task and Condition ($N=905$, groups $=572$).}
\begin{tabular}{llcccc}
\toprule
Task & Condition & Est. & 95\% CI & SE & df \\
\midrule
Creative    & H3    & -0.124 & [-0.336, 0.089] & 0.108 & 210 \\
            & H2\_C & -0.151 & [-0.401, 0.099] & 0.127 & 468 \\
            & H2\_S &  0.212 & [-0.049, 0.472] & 0.133 & 519 \\
            & H1\_C &  0.129 & [-0.224, 0.481] & 0.180 & 884 \\
            & H1\_S &  0.095 & [-0.263, 0.454] & 0.183 & 884 \\
            & H1\_M &  0.054 & [-0.299, 0.406] & 0.180 & 884 \\
\addlinespace
Ethics      & H3    & -0.113 & [-0.322, 0.096] & 0.106 & 206 \\
            & H2\_C & -0.136 & [-0.389, 0.116] & 0.128 & 461 \\
            & H2\_S &  0.016 & [-0.237, 0.269] & 0.129 & 415 \\
            & H1\_C &  0.040 & [-0.319, 0.398] & 0.183 & 884 \\
            & H1\_S &  0.386 & [0.033, 0.738]  & 0.180 & 884 \\
            & H1\_M &  0.181 & [-0.172, 0.533] & 0.180 & 884 \\
\addlinespace
Analytical  & H3    & -0.390 & [-0.597, -0.184] & 0.105 & 209 \\
            & H2\_C &  0.458 & [0.203, 0.712]   & 0.130 & 453 \\
            & H2\_S & -0.005 & [-0.260, 0.249]  & 0.130 & 453 \\
            & H1\_C &  0.044 & [-0.315, 0.402]  & 0.183 & 884 \\
            & H1\_S &  0.234 & [-0.124, 0.593]  & 0.183 & 884 \\
            & H1\_M &  0.070 & [-0.283, 0.423]  & 0.180 & 884 \\
\bottomrule
\end{tabular}

\textit{Note.} Est. = standardised marginal mean of individual performance gain; CI = confidence interval; SE = standard error; df = degrees of freedom. Experimental conditions: H3 = three humans; H2\_C = two humans + one contrarian AI; H2\_S = two humans + one supportive AI; H1\_C = one human + two contrarian AIs; H1\_S = one human + two supportive AIs; H1\_M = one human + one contrarian and one supportive AI. Task conditions: Creative (reference), Analytical (survival ranking), and Ethics (ethics dilemma).
\end{table}

\begin{table}
\centering
\caption{Pairwise comparisons of analytical task conditions with 95\% confidence intervals (CI), original and Holm-adjusted $p$-values.}
\begin{tabular}{lcccccc}
\toprule
Contrast & Estimate & 95\% CI & SE & df & $p$ & $p_{\text{adj}}$ \\
\midrule
H3 -- H2\_C & -0.848 & [-1.192, -0.504] & 0.175 & 348 & $<.001$ & $<.001$ \\
H3 -- H1\_S & -0.625 & [-1.021, -0.229] & 0.202 & 730 & .002 & .029 \\
H2\_C -- H2\_S &  0.463 & [0.113, 0.813] & 0.178 & 474 & .010 & .126 \\
H3 -- H2\_S & -0.385 & [-0.698, -0.072] & 0.159 & 348 & .016 & .195 \\
H3 -- H1\_M & -0.460 & [-0.847, -0.073] & 0.196 & 724 & .019 & .213 \\
H3 -- H1\_C & -0.434 & [-0.829, -0.038] & 0.202 & 730 & .032 & .320 \\
H2\_C -- H1\_C &  0.414 & [-0.010, 0.838] & 0.217 & 792 & .057 & .510 \\
H2\_C -- H1\_M &  0.388 & [-0.027, 0.803] & 0.212 & 789 & .068 & .542 \\
H2\_S -- H1\_S & -0.240 & [-0.641, 0.162] & 0.205 & 792 & .243 & 1.00 \\
H2\_C -- H1\_S &  0.223 & [-0.202, 0.648] & 0.217 & 792 & .304 & 1.00 \\
H1\_C -- H1\_S & -0.191 & [-0.660, 0.278] & 0.239 & 885 & .426 & 1.00 \\
H1\_S -- H1\_M &  0.164 & [-0.296, 0.625] & 0.235 & 885 & .484 & 1.00 \\
H2\_S -- H1\_M & -0.075 & [-0.468, 0.318] & 0.200 & 789 & .707 & 1.00 \\
H2\_S -- H1\_C & -0.049 & [-0.450, 0.352] & 0.205 & 792 & .812 & 1.00 \\
H1\_C -- H1\_M & -0.026 & [-0.488, 0.435] & 0.235 & 885 & .910 & 1.00 \\
\bottomrule
\end{tabular}
\end{table}

\clearpage 

\subsection{BDI as a covariate (N=871, groups=556)}
\noindent \textit{Fit and random effects (LMM).} REML criterion $=2466.2$; random intercept (group\_id) variance $=0.041$ ($SD=0.203$); residual variance $=0.933$ ($SD=0.966$).

\begin{table}
\centering
\caption{Fixed effects for standardised individual performance gain (BDI-controlled; $N{=}871$, groups $=556$). Estimates (Est.), standard errors (SE), degrees of freedom (df / $df_{\text{Satt}}$), and $p$-values from the linear mixed-effects model (LMM) and from cluster-robust (CR2) inference with Satterthwaite-adjusted degrees of freedom. Holm-adjusted $p$-values control the family-wise error rate across fixed effects per DV.}
\begin{tabular}{lrrrrr|rrrrr}
\toprule
 & \multicolumn{5}{c}{LMM} & \multicolumn{5}{c}{CR2-robust} \\
\cmidrule(lr){2-6} \cmidrule(lr){7-11}
Term & Est. & SE & $df$ & $p$ & $p_{\text{Holm}}$ & Est. & SE & $df_{\text{Satt}}$ & $p_{\text{Satt}}$ & $p_{\text{Holm}}$ \\
\midrule
(Intercept) & -0.170 & 0.111 & 232 & .125 & 1.00 & -0.170 & 0.122 & 32 & .173 & 1.00 \\
Task: Ethics & 0.046 & 0.154 & 228 & .764 & 1.00 & 0.046 & 0.156 & 64 & .768 & 1.00 \\
Task: Analytical & -0.226 & 0.153 & 221 & .139 & 1.00 & -0.226 & 0.165 & 64 & .174 & 1.00 \\
H2\_C & -0.006 & 0.174 & 357 & .973 & 1.00 & -0.006 & 0.192 & 60 & .976 & 1.00 \\
H2\_S & 0.434 & 0.175 & 385 & .014 & .245 & 0.434 & 0.193 & 61 & .028 & .475 \\
H1\_C & 0.286 & 0.214 & 696 & .181 & 1.00 & 0.286 & 0.189 & 47 & .137 & 1.00 \\
H1\_S & 0.249 & 0.217 & 703 & .252 & 1.00 & 0.249 & 0.196 & 45 & .211 & 1.00 \\
H1\_M & 0.251 & 0.215 & 699 & .244 & 1.00 & 0.251 & 0.207 & 47 & .231 & 1.00 \\
BDI & 0.282 & 0.123 & 847 & .023 & .383 & 0.282 & 0.117 & 265 & .016 & .296 \\
Ethics $\times$ H2\_C & -0.005 & 0.242 & 352 & .982 & 1.00 & -0.005 & 0.271 & 120 & .984 & 1.00 \\
Analytical $\times$ H2\_C & 0.879 & 0.241 & 346 & $<.001$ & .006 & 0.879 & 0.259 & 119 & $<.001$ & .018 \\
Ethics $\times$ H2\_S & -0.293 & 0.243 & 355 & .229 & 1.00 & -0.293 & 0.247 & 118 & .238 & 1.00 \\
Analytical $\times$ H2\_S & -0.041 & 0.242 & 356 & .866 & 1.00 & -0.041 & 0.250 & 120 & .871 & 1.00 \\
Ethics $\times$ H1\_C & -0.115 & 0.301 & 698 & .703 & 1.00 & -0.115 & 0.281 & 92 & .683 & 1.00 \\
Analytical $\times$ H1\_C & 0.165 & 0.310 & 709 & .595 & 1.00 & 0.165 & 0.287 & 82 & .567 & 1.00 \\
Ethics $\times$ H1\_S & 0.259 & 0.303 & 701 & .394 & 1.00 & 0.259 & 0.281 & 90 & .359 & 1.00 \\
Analytical $\times$ H1\_S & 0.365 & 0.305 & 703 & .231 & 1.00 & 0.365 & 0.285 & 88 & .202 & 1.00 \\
Ethics $\times$ H1\_M & 0.157 & 0.302 & 698 & .603 & 1.00 & 0.157 & 0.267 & 92 & .558 & 1.00 \\
Analytical $\times$ H1\_M & 0.261 & 0.305 & 703 & .392 & 1.00 & 0.261 & 0.289 & 87 & .368 & 1.00 \\
\bottomrule
\end{tabular}

\textit{Note.} Estimates are relative to the reference condition of three human teammates (H3) working on the Creative task. Holm correction was applied within each family of fixed effects (LMM and CR2) per dependent variable. Experimental conditions: H3 = three humans; H2\_C = two humans + one contrarian AI; H2\_S = two humans + one supportive AI; H1\_C = one human + two contrarian AIs; H1\_S = one human + two supportive AIs; H1\_M = one human + one contrarian and one supportive AI. Task conditions: Creative (reference), Analytical (survival ranking), and Ethics (ethics dilemma).
\end{table}

\begin{table}
\centering
\caption{Estimated marginal means of standardised individual performance gain by Task and Condition ($N{=}871$, groups $=556$).}
\begin{tabular}{llcccc}
\toprule
Task & Condition & Est. & 95\% CI & SE & df \\
\midrule
Creative    & H3    & -0.170 & [-0.388, 0.048] & 0.111 & 214 \\
            & H2\_C & -0.176 & [-0.438, 0.086] & 0.133 & 459 \\
            & H2\_S &  0.263 & [-0.002, 0.529] & 0.135 & 519 \\
            & H1\_C &  0.116 & [-0.244, 0.476] & 0.183 & 850 \\
            & H1\_S &  0.078 & [-0.288, 0.444] & 0.187 & 850 \\
            & H1\_M &  0.080 & [-0.280, 0.441] & 0.184 & 850 \\
\addlinespace
Ethics      & H3    & -0.124 & [-0.338, 0.090] & 0.109 & 212 \\
            & H2\_C & -0.135 & [-0.392, 0.122] & 0.131 & 456 \\
            & H2\_S &  0.016 & [-0.241, 0.274] & 0.131 & 409 \\
            & H1\_C &  0.047 & [-0.313, 0.407] & 0.183 & 850 \\
            & H1\_S &  0.383 & [0.022, 0.744]  & 0.184 & 850 \\
            & H1\_M &  0.284 & [-0.077, 0.644] & 0.184 & 850 \\
\addlinespace
Analytical  & H3    & -0.397 & [-0.607, -0.187] & 0.107 & 199 \\
            & H2\_C &  0.476 & [0.219, 0.734]   & 0.131 & 440 \\
            & H2\_S & -0.004 & [-0.259, 0.251]  & 0.130 & 431 \\
            & H1\_C &  0.054 & [-0.334, 0.442]  & 0.198 & 850 \\
            & H1\_S &  0.217 & [-0.149, 0.583]  & 0.186 & 850 \\
            & H1\_M &  0.115 & [-0.259, 0.489]  & 0.191 & 850 \\
\bottomrule
\end{tabular}
\end{table}

\begin{table}
\centering
\caption{Pairwise comparisons of analytical task conditions with 95\% confidence intervals (CI), original and Holm-adjusted $p$-values ($N{=}871$, groups $=556$).}
\begin{tabular}{lcccccc}
\toprule
Contrast & Estimate & 95\% CI & SE & df & $p$ & $p_{\text{adj}}$ \\
\midrule
H3 -- H2\_C & -0.873 & [-1.21, -0.540] & 0.169 & 337 & $<.001$ & $<.001$ \\
H3 -- H1\_S & -0.614 & [-1.04, -0.195] & 0.215 & 704 & .004 & .061 \\
H2\_C -- H2\_S &  0.480 & [0.119, 0.841] & 0.184 & 459 & .009 & .123 \\
H3 -- H1\_M & -0.512 & [-0.943, -0.081] & 0.219 & 713 & .020 & .237 \\
H3 -- H2\_S & -0.393 & [-0.723, -0.063] & 0.168 & 333 & .020 & .237 \\
H3 -- H1\_C & -0.451 & [-0.892, -0.010] & 0.225 & 724 & .046 & .458 \\
H2\_C -- H1\_C &  0.422 & [-0.043, 0.888] & 0.237 & 777 & .075 & .679 \\
H2\_C -- H1\_M &  0.361 & [-0.092, 0.815] & 0.231 & 771 & .119 & .949 \\
H2\_C -- H1\_S &  0.259 & [-0.187, 0.705] & 0.228 & 766 & .256 & 1.00 \\
H2\_S -- H1\_S & -0.221 & [-0.667, 0.225] & 0.227 & 764 & .331 & 1.00 \\
H1\_C -- H1\_S & -0.163 & [-0.694, 0.368] & 0.272 & 850 & .549 & 1.00 \\
H2\_S -- H1\_M & -0.119 & [-0.570, 0.332] & 0.231 & 769 & .606 & 1.00 \\
H1\_S -- H1\_M &  0.102 & [-0.421, 0.625] & 0.267 & 850 & .702 & 1.00 \\
H2\_S -- H1\_C & -0.058 & [-0.521, 0.406] & 0.237 & 775 & .807 & 1.00 \\
H1\_C -- H1\_M & -0.061 & [-0.596, 0.474] & 0.274 & 850 & .824 & 1.00 \\
\bottomrule
\end{tabular}
\end{table}

\clearpage 
\FloatBarrier
\section{Moderation results (BDI)}
\label{appendix-moderation}

This appendix reports supplementary analyses examining whether baseline depressive symptoms (BDI) moderated the effects of task and teammate configuration on primary outcomes. We present (i) model specifications for individual- and group-level outcomes, (ii) omnibus (joint) CR2 Wald tests for the full set of BDI-involving interaction terms, and (iii) follow-up estimated marginal means (EMMs) and multiplicity-controlled simple contrasts at BDI $-1$ SD and $+1$ SD. These analyses are intended to assess robustness and potential heterogeneity of the experimental effects; as in the main analyses, statistical significance is defined using Holm-adjusted $p$-values within each dependent variable.

\subsection{Moderation effects of BDI (N=871, groups=556)}

Individual-level models (standardised individual performance gain, team satisfaction, psychological safety): 
\[
\text{DV} \sim \text{Task} \times \text{Condition} \times \text{BDI} + (1\ |\ \text{group\_id}) \quad \text{(LMM; REML; Satterthwaite df)}.
\]
Group-level model (group discussion quality): 
\[
\text{DV} \sim \text{Task} \times \text{Condition} \times \text{BDI}_{\text{group mean}} \quad \text{(OLS; one row per group)}.
\]

Joint tests cover all interaction terms that include BDI (excluding the main effect of BDI). Simple effects are CR2-robust and multiplicity-controlled (Holm). For each outcome, Holm adjustments were applied across the full set of planned contrasts (six condition contrasts $\times$ three tasks $\times$ two BDI levels; 36 total), thereby 
controlling the family-wise error rate at the level of each dependent variable. Results reported as ``significant'' correspond to Holm-adjusted $p$-values $< .05$.

\subsection{Joint moderation tests (CR2 Wald)}

\begin{table}
\centering
\caption{CR2 Wald tests of BDI moderation effects across outcomes}
\begin{tabular}{lcccc}
\toprule
Outcome & Wald stat & $df_1$ & $df_2$ & \textit{p} \\
\midrule
Performance gain (z)   & 0.87 & 17 & 111 & .607 \\
Team satisfaction       & 2.73 & 17 & 110 & $<.001$ \\
Psychological safety    & 2.40 & 17 & 110 & .003 \\
Group discussion quality   & 0.72 & 17 & 91  & .777 \\
\bottomrule
\end{tabular}

\textit{Note}. Wald tests include all fixed effects involving BDI, excluding the main effect.
\end{table}

For transparency, we provide the full estimated marginal means (EMMs) and CR2-robust simple contrasts at BDI $-1$ SD and $+1$ SD for each outcome. Each outcome corresponds to a single supplementary table, with three parts: (a) EMMs at BDI $-1$ SD, (b) EMMs at BDI $+1$ SD, and (c) CR2-robust contrasts at BDI $\pm$1 SD. Tables are supplied as \texttt{.xlsx} files to preserve precision. Only Holm-adjusted significant contrasts are reported here; the complete outputs are available in the XLSX files.

\subsection{Team satisfaction}
BDI significantly moderated persona effects, $F(17, 110) = 2.73, p < .001$. Across all tasks, the \textit{Supportive vs. Contrarian} contrast in mixed-human groups (H2\_S $-$ H2\_C) was significant and robust at both low and high BDI (all $p_{\text{Holm}} < .001$). No other contrasts survived multiplicity correction. Full contrasts ($n=216$ rows) and marginal means are provided in the following files:

\begin{itemize}
  \item Estimated marginal means at BDI $-1$ SD 
        (\texttt{SI\_TS\_EMMs\_BDI\_minus.xlsx})
  \item Estimated marginal means at BDI $+1$ SD 
        (\texttt{SI\_TS\_EMMs\_BDI\_plus.xlsx})
  \item CR2-robust simple contrasts at BDI $\pm$1 SD 
        (\texttt{SI\_TS\_CR2\_BDI\_CR2.xlsx})
\end{itemize}

\begin{table}
\centering
\caption{Significant simple contrasts for Team Satisfaction after Holm correction}
\begin{tabular}{lccc}
\toprule
Task & BDI level & Contrast & Estimate (SE), $p_{\text{Holm}}$ \\
\midrule
Creative   & $-1$ SD & H2\_S $-$ H2\_C & 43.08 (0.89), $<.001$ \\
Creative   & $+1$ SD & H2\_S $-$ H2\_C & 43.08 (0.89), $<.001$ \\
Ethics     & $-1$ SD & H2\_S $-$ H2\_C & 43.08 (0.89), $<.001$ \\
Ethics     & $+1$ SD & H2\_S $-$ H2\_C & 43.08 (0.89), $<.001$ \\
Analytical & $-1$ SD & H2\_S $-$ H2\_C & 43.08 (0.89), $<.001$ \\
Analytical & $+1$ SD & H2\_S $-$ H2\_C & 43.08 (0.89), $<.001$ \\
\bottomrule
\end{tabular}
\end{table}

\subsection{Psychological safety}
BDI moderated persona effects, $F(17, 110) = 2.40, p = .003$. The same contrast (H2\_S $-$ H2\_C) was consistently significant across all tasks and both BDI levels (all $p_{\text{Holm}} < .001$). Other contrasts did not survive correction. Full contrasts ($n=216$ rows) and marginal means are provided in the following files:

\begin{itemize}
  \item Estimated marginal means at BDI $-1$ SD 
        (\texttt{SI\_PS\_EMMs\_BDI\_minus.xlsx})
  \item Estimated marginal means at BDI $+1$ SD 
        (\texttt{SI\_PS\_EMMs\_BDI\_plus.xlsx})
  \item CR2-robust simple contrasts at BDI $\pm$1 SD 
        (\texttt{SI\_PS\_CR2\_BDI\_CR2.xlsx})
\end{itemize}

\begin{table}
\centering
\caption{Significant simple contrasts for Psychological Safety after Holm correction}
\begin{tabular}{lccc}
\toprule
Task & BDI level & Contrast & Estimate (SE), $p_{\text{Holm}}$ \\
\midrule
Creative   & $-1$ SD & H2\_S $-$ H2\_C & 38.80 (0.82), $<.001$ \\
Creative   & $+1$ SD & H2\_S $-$ H2\_C & 38.80 (0.82), $<.001$ \\
Ethics     & $-1$ SD & H2\_S $-$ H2\_C & 38.80 (0.82), $<.001$ \\
Ethics     & $+1$ SD & H2\_S $-$ H2\_C & 38.80 (0.82), $<.001$ \\
Analytical & $-1$ SD & H2\_S $-$ H2\_C & 38.80 (0.82), $<.001$ \\
Analytical & $+1$ SD & H2\_S $-$ H2\_C & 38.80 (0.82), $<.001$ \\
\bottomrule
\end{tabular}
\end{table}

\subsection{Group discussion quality}
Joint moderation was not significant, $F(17, 91) = 0.72, p = .78$. Nevertheless, the H2\_S $-$ H2\_C contrast was significant in all tasks at both BDI levels (all $p_{\text{Holm}} < .001$), but given the nonsignificant overall Wald test, we treat these as exploratory. Full contrasts ($n=216$ rows) and marginal means are provided in the following files:

\begin{itemize}
  \item Estimated marginal means at BDI $-1$ SD 
        (\texttt{SI\_GDQ\_EMMs\_BDI\_minus.xlsx})
  \item Estimated marginal means at BDI $+1$ SD 
        (\texttt{SI\_GDQ\_EMMs\_BDI\_plus.xlsx})
  \item CR2-robust simple contrasts at BDI $\pm$1 SD 
        (\texttt{SI\_GDQ\_CR2\_BDI\_CR2.xlsx})
\end{itemize}

\begin{table}
\centering
\caption{Exploratory significant contrasts for Group Discussion Quality}
\begin{tabular}{lccc}
\toprule
Task & BDI level & Contrast & Estimate (SE), $p_{\text{Holm}}$ \\
\midrule
Creative   & $-1$ SD & H2\_S $-$ H2\_C & 3.52 (0.12), $<.001$ \\
Creative   & $+1$ SD & H2\_S $-$ H2\_C & 3.52 (0.12), $<.001$ \\
Ethics     & $-1$ SD & H2\_S $-$ H2\_C & 3.52 (0.12), $<.001$ \\
Ethics     & $+1$ SD & H2\_S $-$ H2\_C & 3.52 (0.12), $<.001$ \\
Analytical & $-1$ SD & H2\_S $-$ H2\_C & 3.52 (0.12), $<.001$ \\
Analytical & $+1$ SD & H2\_S $-$ H2\_C & 3.52 (0.12), $<.001$ \\
\bottomrule
\end{tabular}
\end{table}

\subsection{Standardised individual performance gain}
Joint moderation was not significant, $F(17, 111) = 0.87, p = .61$, and no Holm-adjusted simple effects reached significance. Full contrasts ($n=216$ rows) and marginal means are provided in the following files:

\begin{itemize}
  \item Estimated marginal means at BDI $-1$ SD 
        (\texttt{SI\_z\_gain\_EMMs\_BDI\_minus.xlsx})
  \item Estimated marginal means at BDI $+1$ SD 
        (\texttt{SI\_z\_gain\_EMMs\_BDI\_plus.xlsx})
  \item CR2-robust simple contrasts at BDI $\pm$1 SD 
        (\texttt{SI\_z\_gain\_CR2\_BDI\_CR2.xlsx})
\end{itemize}

\clearpage 
\FloatBarrier
\section{Linguistic mediation}
\label{appendix-linguistic}

This appendix reports the linguistic mediation analyses examining whether systematic differences in AI language use account for the effects of AI persona composition on team outcomes. Using LIWC-based features extracted from AI utterances only, we tested a parallel multiple-mediator framework in which contrarian dosage at the group level ($P_g$) predicted linguistic features (Path~A), which in turn predicted psychological safety and discussion quality while controlling for the direct effect of $P_g$ (Path~B and $c'$). Analyses were conducted at the group level with cluster-robust (CR2) inference and nonparametric bootstrapping for indirect effects. To ensure model stability and interpretability, we first examined intercorrelations and variance inflation factors among LIWC predictors before estimating mediation models. Full results are reported for transparency, with statistical inference emphasised on Holm-adjusted estimates.

\begin{figure}
    \centering
    \includegraphics[width=0.85\textwidth]{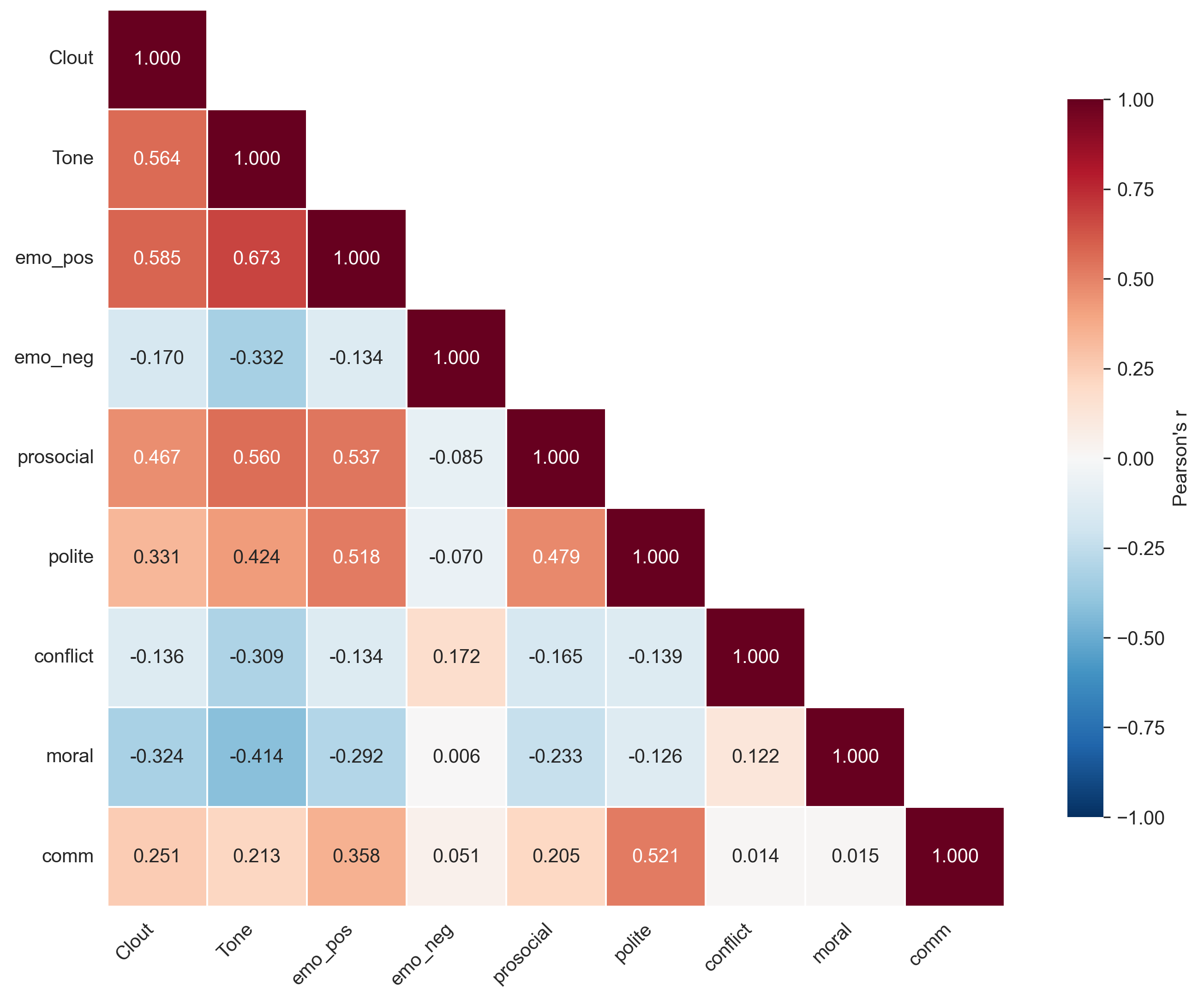}
    \caption{Pairwise Pearson correlations among the nine LIWC predictor variables for AI utterances. Correlations are colour-coded by magnitude and direction, with red indicating positive and blue negative associations. Several variables (e.g., \textit{Tone}, \textit{emo\_pos}, and \textit{Clout}) showed moderate to strong positive correlations ($r = .56$–$.67$), while others (e.g., \textit{Tone} with \textit{emo\_neg}) showed moderate negative associations. These interdependencies underscore the importance of checking multicollinearity before including predictors in mediation models.}
    \label{fig:liwc_corr}
\end{figure}

\begin{figure}
    \centering
    \includegraphics[width=1\textwidth]{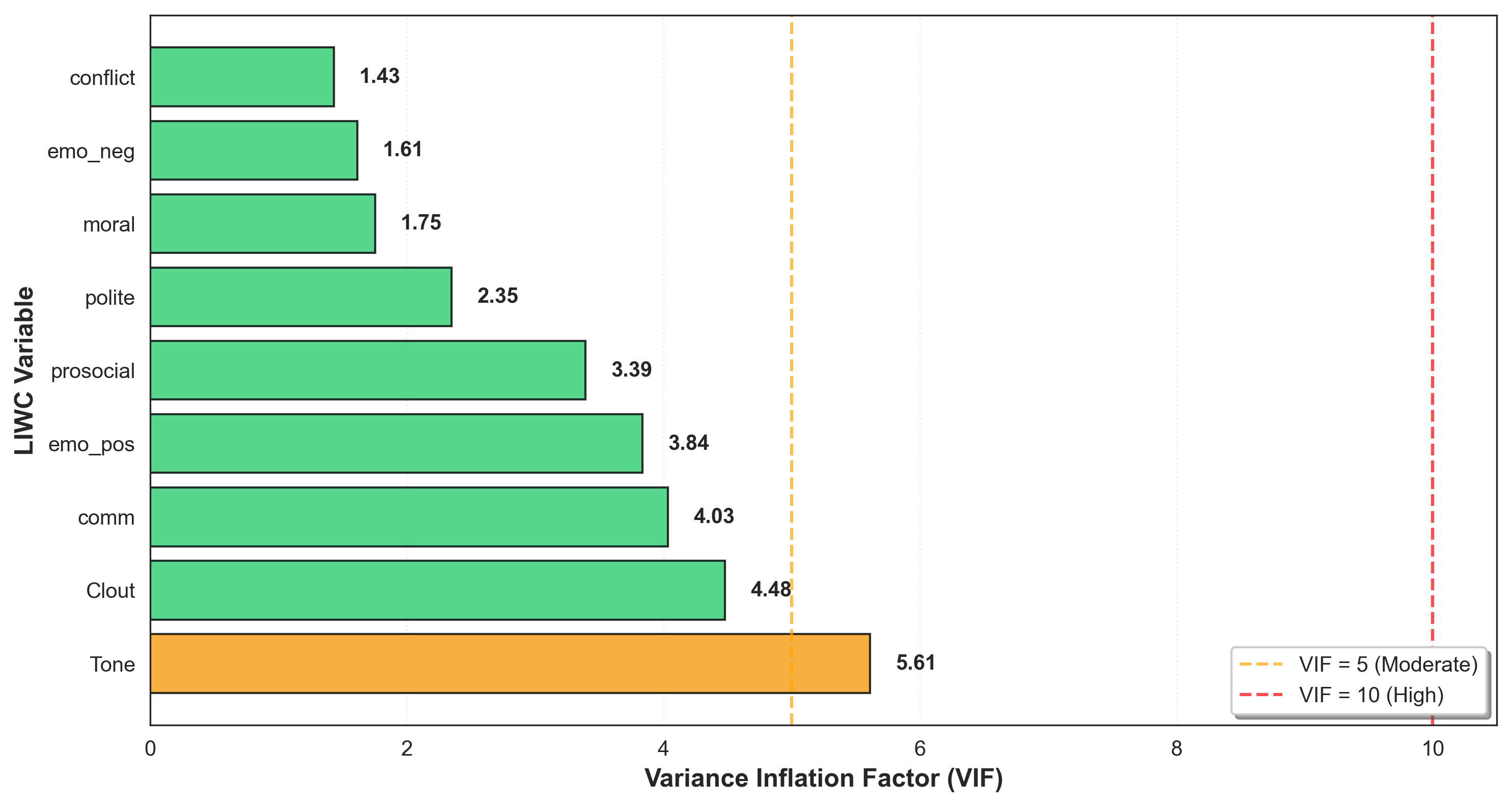}
    \caption{Variance Inflation Factor (VIF) analysis for the nine LIWC predictor variables based on AI utterances. Bars indicate VIF values, with dashed vertical lines marking conventional thresholds for moderate multicollinearity ($VIF=5$) and high multicollinearity ($VIF=10$). All predictors remained below the high threshold, with only \textit{Tone} exceeding the moderate cut-off (VIF = 5.61), suggesting limited concern for collinearity in the mediation models.}
    \label{fig:vif_liwc}
\end{figure}

\clearpage 

\begin{table}
\centering
\caption{Data summary.}
\begin{tabular}{ll}
\toprule
Item & Value \\
\midrule
LIWC utterances & 1,665 (from 572 groups) \\
Qualtrics individuals & 905 (from 572 groups) \\
BDI participants & 871 \\
Outliers removed (utterances) & 141 (8.47\%) \\
Remaining utterances & 1,524 (568 groups affected) \\
Mediators (AI only) & 438 groups \\
Dataset & 586 individuals, 438 groups \\
Contrarian dosage (AI groups) & \(P_g \in \{0:173,\; 0.5:87,\; 1:178\}\) \\
AI count & 1 AI: 187 groups; 2 AIs: 251 groups \\
\bottomrule
\end{tabular}
\end{table}

\begin{table}
\centering
\caption{Path A model fit (per mediator).}
\begin{tabular}{lrr}
\toprule
Mediator & \(R^2\) & \(n\) (groups) \\
\midrule
Clout     & 0.533 & 425 \\
Tone      & 0.638 & 425 \\
emo\_pos  & 0.532 & 425 \\
emo\_neg  & 0.292 & 425 \\
prosocial & 0.296 & 425 \\
polite    & 0.338 & 425 \\
conflict  & 0.113 & 425 \\
moral     & 0.308 & 425 \\
comm      & 0.315 & 425 \\
\bottomrule
\end{tabular}
\end{table}

\begin{table}
\centering
\caption{Indirect effects for psychological safety (bootstrap, \(B=10{,}000\)).}
\begin{tabular}{lrrrrr}
\toprule
Mediator & Estimate & SE & 95\% CI L & 95\% CI U & \(p\) \\
\midrule
Clout        &  0.038 & 0.106 & -0.170 &  0.246 & .727 \\
Tone         &  0.193 & 0.156 & -0.112 &  0.498 & .225 \\
emo\_pos     & -0.161 & 0.111 & -0.378 &  0.056 & .150 \\
emo\_neg     & -0.028 & 0.037 & -0.101 &  0.044 & .436 \\
prosocial    & -0.077 & 0.061 & -0.197 &  0.044 & .210 \\
polite       &  0.024 & 0.063 & -0.099 &  0.147 & .669 \\
conflict     & -0.062 & 0.029 & -0.118 & -0.006 & .014 \\
moral        &  0.002 & 0.044 & -0.085 &  0.089 & .949 \\
comm         & -0.018 & 0.041 & -0.098 &  0.062 & .612 \\
\midrule
Total        & -0.089 & 0.129 & -0.342 &  0.164 & .473 \\
Direct (\(c'\)) & -0.559 & 0.156 & -0.864 & -0.254 & $<.001$ \\
\bottomrule
\end{tabular}
\end{table}

\begin{table}
\centering
\caption{Indirect effects for discussion quality (bootstrap, \(B=10{,}000\)).}
\begin{tabular}{lrrrrr}
\toprule
Mediator & Estimate & SE & 95\% CI L & 95\% CI U & \(p\) \\
\midrule
Clout        &  0.054 & 0.086 & -0.120 &  0.218 & .506 \\
Tone         &  0.067 & 0.146 & -0.211 &  0.363 & .660 \\
emo\_pos     & -0.415 & 0.098 & -0.608 & -0.223 & $<.001$ \\
emo\_neg     & -0.051 & 0.031 & -0.119 &  0.005 & .085 \\
prosocial    &  0.052 & 0.060 & -0.059 &  0.177 & .378 \\
polite       & -0.093 & 0.059 & -0.207 &  0.024 & .126 \\
conflict     &  0.001 & 0.024 & -0.043 &  0.053 & .981 \\
moral        &  0.006 & 0.039 & -0.071 &  0.086 & .884 \\
comm         &  0.126 & 0.039 &  0.060 &  0.216 & .001 \\
\midrule
Total        & -0.254 & 0.134 & -0.507 &  0.020 & .054 \\
Direct (\(c'\)) & -0.886 & 0.149 & -1.196 & -0.608 & $<.001$ \\
\bottomrule
\end{tabular}
\end{table}

\begin{table}
\centering
\caption{Path A: Effect of contrarian dosage (\(P_g\)) on LIWC mediators (AI utterances only). Cluster-robust (CR2) Satterthwaite inference; \(n=425\) groups.}
\begin{tabular}{lrrrrrr}
\toprule
Mediator & Term & Est. & SE & df & \(t\) & \(p_{\text{Holm}}\) \\
\midrule
Clout     & \(P_g\) & -1.464 & 0.078 & 335 & -18.802 & $<.001$ \\
Tone      & \(P_g\) & -1.730 & 0.066 & 335 & -26.248 & $<.001$ \\
emo\_pos  & \(P_g\) & -1.607 & 0.079 & 335 & -20.291 & $<.001$ \\
emo\_neg  & \(P_g\) &  0.670 & 0.094 & 335 &   7.159 & $<.001$ \\
prosocial & \(P_g\) & -1.111 & 0.097 & 335 & -11.407 & $<.001$ \\
polite    & \(P_g\) & -1.045 & 0.092 & 335 & -11.387 & $<.001$ \\
conflict  & \(P_g\) &  0.516 & 0.104 & 335 &   4.948 & $<.001$ \\
moral     & \(P_g\) &  0.768 & 0.094 & 335 &   8.197 & $<.001$ \\
comm      & \(P_g\) & -0.603 & 0.093 & 335 &  -6.498 & $<.001$ \\
\bottomrule
\end{tabular}
\end{table}

\begin{table}
\centering
\caption{Psychological safety LMM (Path B + \(c'\); CR2 robust).}
\begin{tabular}{lrrrrr}
\toprule
Term & Est. & SE & \(t\) & \(p\) & \(p_{\text{Holm}}\) \\
\midrule
Intercept             &  0.282 & 0.117 &  2.415 & .018 & .229 \\
\(P_g\)               & -0.559 & 0.155 & -3.600 & $<.001$ & .007 \\
Clout\_z              & -0.026 & 0.073 & -0.355 & .723 & 1.00 \\
Tone\_z               & -0.111 & 0.090 & -1.236 & .219 & 1.00 \\
emo\_pos\_z           &  0.100 & 0.069 &  1.448 & .151 & 1.00 \\
emo\_neg\_z           & -0.042 & 0.053 & -0.794 & .430 & 1.00 \\
prosocial\_z          &  0.069 & 0.054 &  1.267 & .209 & 1.00 \\
polite\_z             & -0.023 & 0.060 & -0.383 & .704 & 1.00 \\
conflict\_z           & -0.120 & 0.048 & -2.483 & .016 & .223 \\
moral\_z              &  0.003 & 0.058 &  0.050 & .960 & 1.00 \\
comm\_z               &  0.030 & 0.066 &  0.450 & .654 & 1.00 \\
Task: Ethics          &  0.111 & 0.142 &  0.786 & .433 & 1.00 \\
Task: Analytical      &  0.037 & 0.124 &  0.296 & .767 & 1.00 \\
AI count: 2 vs 1      & -0.124 & 0.088 & -1.422 & .156 & 1.00 \\
BDI (individual)      &  0.044 & 0.145 &  0.305 & .760 & 1.00 \\
\bottomrule
\end{tabular}
\end{table}

\begin{table}
\centering
\caption{Discussion quality OLS (Path B + \(c'\); CR2 robust).}
\begin{tabular}{lrrrrr}
\toprule
Term & Est. & SE & \(t\) & \(p\) & \(p_{\text{Holm}}\) \\
\midrule
Intercept             &  0.556 & 0.119 &  4.661 & $<.001$ & $<.001$ \\
\(P_g\)               & -0.886 & 0.149 & -5.944 & $<.001$ & $<.001$ \\
Clout\_z              & -0.037 & 0.059 & -0.628 & .531 & 1.00 \\
Tone\_z               & -0.039 & 0.084 & -0.460 & .646 & 1.00 \\
emo\_pos\_z           &  0.258 & 0.059 &  4.401 & $<.001$ & $<.001$ \\
emo\_neg\_z           & -0.076 & 0.045 & -1.701 & .089 & .925 \\
prosocial\_z          & -0.047 & 0.054 & -0.877 & .381 & 1.00 \\
polite\_z             &  0.089 & 0.054 &  1.655 & .099 & .964 \\
conflict\_z           &  0.002 & 0.046 &  0.044 & .965 & 1.00 \\
moral\_z              &  0.007 & 0.051 &  0.143 & .887 & 1.00 \\
comm\_z               & -0.209 & 0.059 & -3.549 & .001 & .008 \\
Task: Ethics          & -0.445 & 0.141 & -3.144 & .002 & .023 \\
Task: Analytical      &  0.144 & 0.109 &  1.314 & .190 & 1.00 \\
AI count: 2 vs 1      & -0.023 & 0.081 & -0.278 & .781 & 1.00 \\
BDI (group mean)      & -0.047 & 0.160 & -0.294 & .769 & 1.00 \\
\bottomrule
\end{tabular}
\end{table}

\clearpage 

\end{document}